\documentclass[12pt]{article}
\usepackage{graphicx,amsmath,amssymb,cite}

\newlength{\dinwidth}
\newlength{\dinmargin}
\begin{document}
  \titlepage
  \begin{flushright}
    IPPP/09/33 \\
    DCPT/09/66 \\
    Cavendish-HEP-09/06 \\
    25th August 2009 \\
  \end{flushright}
  
  \vspace*{0.5cm}
  
  \begin{center}
    {\Large \bf Uncertainties on $\alpha_S$ in global PDF analyses and \\[1ex] implications for predicted hadronic cross sections}

    \vspace*{1cm}
    \textsc{A.D. Martin$^a$, W.J. Stirling$^b$, R.S. Thorne$^c$ and G. Watt$^a$} \\
    
    \vspace*{0.5cm}$^a$ Institute for Particle Physics Phenomenology, University of Durham, DH1 3LE, UK \\
    $^b$ Cavendish Laboratory, University of Cambridge, CB3 0HE, UK \\
    $^c$ Department of Physics and Astronomy, University College London, WC1E 6BT, UK
  \end{center}
  
  \vspace*{0.5cm}
  
  \begin{abstract}
    We determine the uncertainty on the strong coupling $\alpha_S$ due to the experimental errors on the data fitted in global analysis of hard-scattering data, within the standard framework of leading-twist fixed-order collinear factorisation in the $\overline{\rm MS}$ scheme, finding that $\alpha_S(M_Z^2) = 0.1202^{+0.0012}_{-0.0015}$ at next-to-leading order (NLO) and $\alpha_S(M_Z^2) = 0.1171^{+0.0014}_{-0.0014}$ at next-to-next-to-leading order (NNLO).  We do not address in detail the issue of the additional theory uncertainty on $\alpha_S(M_Z^2)$, but an estimate is $\pm0.003$ at NLO and at most $\pm0.002$ at NNLO.  We investigate the interplay between uncertainties on $\alpha_S$ and uncertainties on parton distribution functions (PDFs).  We show, for the first time, how both these sources of uncertainty can be accounted for simultaneously in calculations of cross sections, and we provide eigenvector PDF sets with different fixed $\alpha_S$ values to allow further studies by the general user.  We illustrate the application of these PDF sets by calculating cross sections for $W$, $Z$, Higgs boson and inclusive jet production at the Tevatron and LHC.
  \end{abstract}

\newpage

\section{Introduction}
There has been a steady improvement in the precision and the variety of the data for deep-inelastic and related hard-scattering processes.  Global parton analyses of these data allow the partonic structure of the proton to be quantified with ever-increasing accuracy.  Analyses are now possible at next-to-next-to-leading order (NNLO) in the running strong coupling $\alpha_S(Q^2)$.  Indeed, the accuracy is such that the theoretical formalism needs to consider effects at the level of $1\%$.  One area which needs careful treatment at this level of accuracy is the strong coupling itself.  This applies both to the definition of the coupling, and to the uncertainties associated with it.  The definition of $\alpha_S$ is not unique and different definitions can give noticeable differences in the results.  In the recent ``MSTW 2008'' global fit~\cite{Martin:2009iq} we changed the definition of the strong coupling compared to that used in previous ``MRST'' analyses.  We discuss the technical details of this change in detail in the Appendix.  However, the main emphasis of this article is the issue of the uncertainty on parton distribution functions (PDFs) and derived quantities such as cross sections, arising from the uncertainty in the strong coupling $\alpha_S$.  This is a non-trivial, and sometimes significant, source of uncertainty which is often not considered when extracting PDFs from global fits or when calculating uncertainties on hadronic cross sections.\footnote{See, for example, Refs.~\cite{Pumplin:2005rh,Chekanov:2006yc} for analyses where the uncertainty on $\alpha_S$ \emph{is} included in the theoretical uncertainty on predicted cross sections.}  The interplay of the PDFs, their uncertainties and $\alpha_S$ is both interesting in itself and important for making precise cross section predictions, and in this paper we give the first comprehensive account of this subject.

\section{Parton distributions and the strong coupling}
The recent MSTW analysis~\cite{Martin:2009iq} (using an improved definition of $\alpha_S$) was based on the obtained ``best-fit'' values of the strong coupling, i.e.~the values obtained by minimisation of a global goodness-of-fit quantity, $\chi^2_{\rm global}$, simultaneously with the PDF parameters at an input scale $Q_0^2=1$~GeV$^2$.  In the NLO and NNLO global PDF fits, the values were found to be $\alpha_S(M_Z^2)=0.1202$ and $0.1171$, respectively.  In addition to the best-fit PDFs and corresponding best-fit $\alpha_S$ values, we also determined a set of eigenvector (``error'') PDF sets, designed to span the variation in the parton distributions allowed by the uncertainties on the data included in the global fit.  We introduced and used a new ``dynamic tolerance'' method for determining 68$\%$ or 90$\%$ confidence-level (C.L.) error PDFs, which is an extension of an earlier method introduced in the CTEQ6 analysis~\cite{Stump:2001gu,Pumplin:2001ct,Pumplin:2002vw}.  Then the prediction and corresponding uncertainty for a physical quantity $F$ that depends on the PDFs, such as a cross section at a hadron--hadron collider, is given by $F(S_0) \pm \Delta F_{\textsc{pdf}}$ where\footnote{In practice, we use slightly more precise formulae, given in Eqs.~(51,52) of Ref.~\cite{Martin:2009iq}, leading to asymmetric PDF uncertainties; see also Eqs.~\eqref{eq:Fp} and \eqref{eq:Fm} below.}
\begin{equation} \label{eq:symmunc}
  \Delta F_{\textsc{pdf}} = \frac{1}{2}\sqrt{\sum_{k=1}^n \left[F(S_k^+)-F(S_k^-)\right]^2}.
\end{equation}
Here, $S_0$ is the central PDF set, and $S_k^{\pm}$ ($k=1,\ldots,n$) are the eigenvector PDF sets, with $n=20$ in the MSTW 2008 analysis~\cite{Martin:2009iq}.  To illustrate our method and its results, we used as examples in Ref.~\cite{Martin:2009iq} the total cross sections for $W^\pm$ and $Z$ production at the Tevatron and LHC, i.e.~we computed the theoretical predictions for the total cross sections and their corresponding PDF uncertainties, $\sigma\pm\Delta\sigma_{\textsc{pdf}}$.

However, the PDF sets obtained in this previous study are only defined for the particular values of $\alpha_S(M_Z^2)$ found by the best fit.  If the user has a preferred (different) value of $\alpha_S(M_Z^2)$ then PDFs determined with that particular value in the global analysis should strictly be used.  Hence, in this paper we provide best-fit PDFs for a range of values of $\alpha_S(M_Z^2)$.  This is a straightforward matter of repetition.  However, when considering \emph{uncertainties} on PDFs and on physical quantities the process is more involved.  The {\it true} theoretical uncertainty on a predicted cross section should also include a contribution from the allowed variation of $\alpha_S$ about its best-fit value.  This is particularly important for cross sections that at leading-order are proportional to a power of the coupling, for example, $\sigma_{\rm Higgs}\propto \alpha_S^2$ for production via gluon--gluon fusion at the Tevatron and LHC.  A na\"ive way of doing this would be to define the overall theory error as $\Delta\sigma_{\rm th}^2 = \Delta\sigma_{\textsc{pdf}}^2 + \Delta\sigma_{\alpha_S}^2$, where $\Delta\sigma_{\alpha_S}$ is the cross section variation when $\alpha_S$ is allowed to vary within some range, determined, for example, by its world-average error, for a fixed set of PDFs.  However, it is not consistent to use different values of $\alpha_S$ in the partonic cross section describing the hard subprocess and in the PDF evolution.  Moreover, in a global PDF analysis, there are non-negligible correlations between the PDFs and the value of the strong coupling.  For example, the evolution of the structure function $F_2$ at small $x$ is driven by the combination $\alpha_S\,g$, and so a large value of the coupling will correspond, in principle, to a smaller gluon distribution at low $x$.  This is indeed observed in practice.

This PDF--$\alpha_S$ correlation has been known for some time, and indeed early global analyses derived a series of best-fit PDFs for a set of discrete $\alpha_S$ values spanning a given range, for example, $\alpha_S(M_Z^2) = 0.105, 0.110, \ldots, 0.130$~\cite{Martin:1995gx}.  Other relevant studies have been made by the CTEQ group~\cite{Huston:2005jm,Pumplin:2005rh}.  However, given that the precision of the data used in global analyses has increased significantly in recent years, with a corresponding reduction in the error on both the PDFs and on $\alpha_S$, a more sophisticated approach is now required, that is, one in which (i) the allowed variation of $\alpha_S$ is determined by the data, and (ii) the correlation between the variation in $\alpha_S$ and in the PDFs is fully taken into account.  The latter would be very difficult to achieve if $\alpha_S(M_Z^2)$ and its uncertainty were taken as externally determined quantities rather than obtained directly from the fit.

In this paper we investigate the interplay between uncertainties on $\alpha_S$ and uncertainties on parton distributions in the framework of the recent MSTW 2008 analysis~\cite{Martin:2009iq}.  We show how both these sources of uncertainty can be properly accounted for simultaneously in calculations of cross sections.  The remainder of this paper is organised as follows.  In Section~\ref{sec:desc} we discuss how the description of each data set included in the global fit depends on the value of $\alpha_S$.  In Section~\ref{sec:unc} we determine the $\pm 1\sigma$ (that is, the 68$\%$ C.L.) and 90$\%$ C.L.~uncertainties on $\alpha_S(M_Z^2)$ as determined in the NLO and NNLO global analyses.\footnote{Note that we do not consider LO fits in the current analysis, since the overall quality of the LO global fit of $\chi^2_{\rm global} = 3066$ for 2598 data points is significantly worse than those at NLO and NNLO, and therefore there is no statistically meaningful constraint on $\alpha_S$ at this order.  As discussed in Ref.~\cite{Martin:2009iq}, the large best-fit value of $\alpha_S(M_Z^2)=0.13939$ obtained is an attempt to mimic missing higher-order terms.}  Moreover, we identify which particular data sets give the strongest constraints on the value of $\alpha_S(M_Z^2)$.  In Section~\ref{sec:eigen} we derive eigenvector PDF sets with $\alpha_S(M_Z^2)$ fixed at the limits of the 68\% (and also 90\%) C.L.~uncertainty region.  In Section~\ref{sec:cross} we illustrate our results using predictions for the $W^\pm$, $Z$, Higgs and inclusive jet cross sections at the Tevatron and LHC, i.e.~we derive uncertainties on these cross section predictions which take both the allowed variation of $\alpha_S$ and the PDFs fully into account.  We conclude in Section~\ref{sec:conclusions}.  Finally, we devote an Appendix to a discussion of the definition of $\alpha_S$ used in our recent MSTW 2008 analysis~\cite{Martin:2009iq}, and to a comparison with the earlier form used in the MRST analyses.  The importance of a consistent definition is especially important at NNLO, where, in the $\overline{\rm MS}$ scheme, the coupling $\alpha_S(Q^2)$ is discontinuous, albeit by a very small amount, as the scale $Q^2$ increases through the heavy flavour thresholds.

\section{Description of data sets as a function of $\alpha_S$} \label{sec:desc}
The MSTW 2008 global fit~\cite{Martin:2009iq} used a wide variety of data from both fixed-target experiments and the HERA $ep$ and Tevatron $p\bar{p}$ colliders.  Neutral-current structure functions ($F_2$ and $F_L$) were included from fixed-target lepton--nucleon scattering experiments (BCDMS~\cite{Benvenuti:1989rh,Benvenuti:1989fm}, NMC~\cite{Arneodo:1996qe,Arneodo:1996kd}, E665~\cite{Adams:1996gu} and SLAC~\cite{Whitlow:1991uw,Whitlow:1990dr,Whitlow:1990gk}), low-mass Drell--Yan cross sections from the E866/NuSea experiment~\cite{Webb:2003bj,Towell:2001nh}, and charged-current structure functions ($F_2$ and $xF_3$) and dimuon cross sections from neutrino--nucleon scattering experiments (CCFR/NuTeV~\cite{Goncharov:2001qe,Tzanov:2005kr} and CHORUS~\cite{Onengut:2005kv}).  From the HERA experiments, H1 and ZEUS, data were included on neutral- and charged-current reduced cross sections ($\sigma_r^{\rm NC}$ and $\sigma_r^{\rm CC}$)~\cite{Lobodzinska:2003yd,Adloff:2000qk,Adloff:2000qj,Adloff:2003uh,Breitweg:1998dz,Chekanov:2001qu,Chekanov:2002ej,Chekanov:2003yv,Chekanov:2003vw}, the charm structure function ($F_2^{\rm charm}$)~\cite{Adloff:1996xq,Adloff:2001zj,Aktas:2005iw,Aktas:2004az,Breitweg:1999ad,Chekanov:2003rb,Chekanov:2007ch}, and inclusive jet production in deep-inelastic scattering~\cite{Aktas:2007pb,Chekanov:2002be,Chekanov:2006xr}.  From the Tevatron experiments, CDF and D{\O}, Run II data were included on inclusive jet production~\cite{Abazov:2008hu,Abulencia:2007ez}, the lepton charge asymmetry from $W$ decays~\cite{Abazov:2007pm,Acosta:2005ud} and the $Z$ rapidity distribution~\cite{Abazov:2007jy,Han:2008}.  A more detailed description of the treatment of each of these data sets can be found in Ref.~\cite{Martin:2009iq}.

The definition of the goodness-of-fit quantity, $\chi^2_n$, for each data set $n$ is given in Section~5.2 of Ref.~\cite{Martin:2009iq}.  Note that the normalisation of each data set in the global analysis is taken as a fitted free parameter to allow for uncertainties in the measured luminosities, with a \emph{quartic} $\chi^2_n$ penalty term given by Eq.~(37) of Ref.~\cite{Martin:2009iq} rather than the more usual quadratic penalty term, in order to bind the fitted data set normalisations more closely to their nominal values of 1, and to recognise that the normalisation uncertainties may well not be Gaussian in nature.  For most data sets, particularly the older ones where correlations are unavailable or statistical uncertainties often dominate, the statistical and systematic (other than normalisation) uncertainties are simply added in quadrature in the definition of $\chi^2_n$.  However, for the data on inclusive jet production, where at the Tevatron the systematic errors are often many times larger than the statistical errors, and the CDF $Z$ rapidity distribution, the full information on correlated systematic uncertainties is included by defining $\chi^2_n$ as in Eq.~(38) of Ref.~\cite{Martin:2009iq}, i.e.~the data points are allowed to shift by the correlated systematic uncertainties in order to give the best fit, with a quadratic penalty term to limit large shifts by more than the experimental errors.

\begin{figure}
  \centering
  \includegraphics[width=0.9\textwidth]{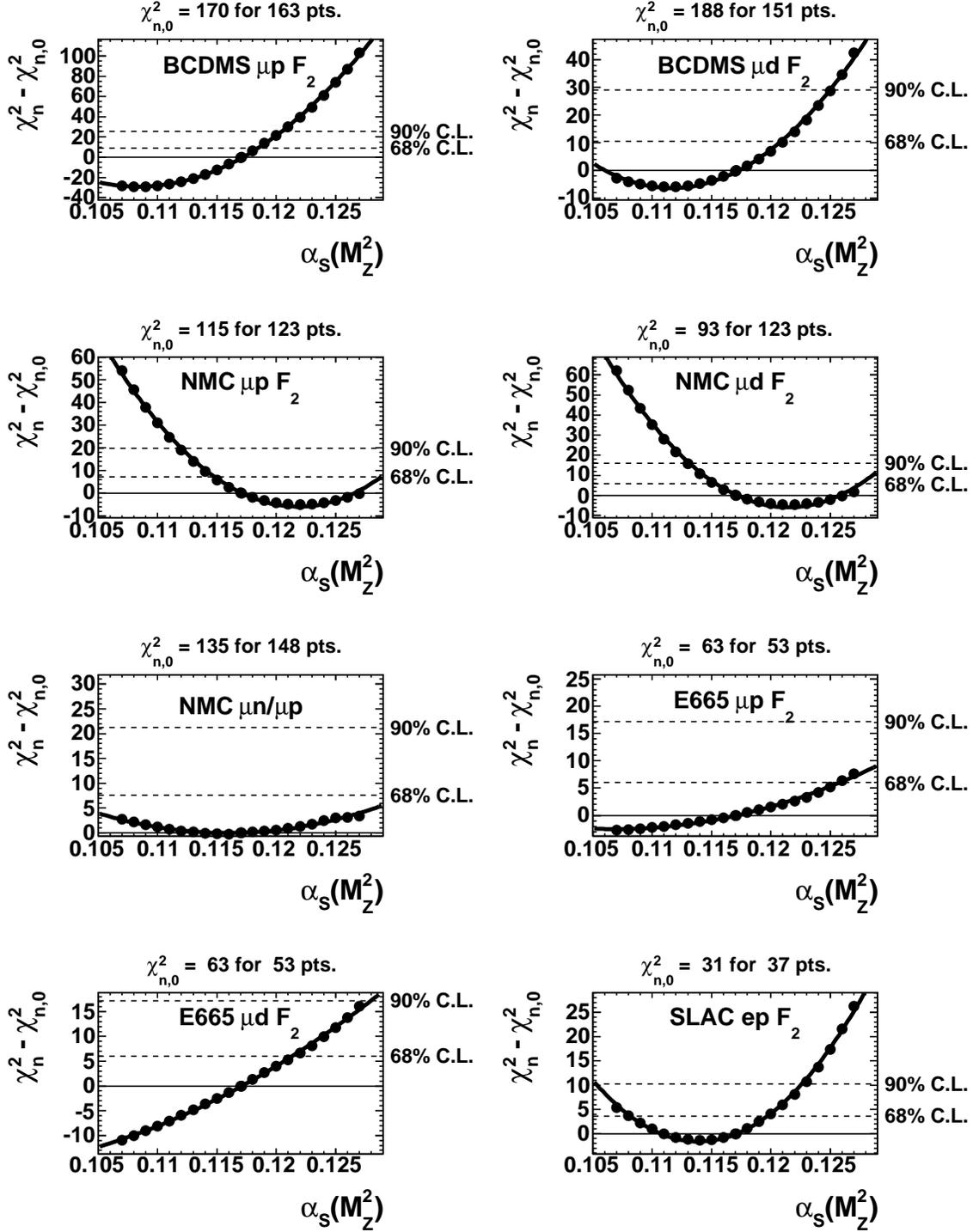}
  \caption{$\chi_n^2$ profiles for the first subset of data sets $n$ in the NNLO fit, when varying $\alpha_S(M_Z^2)$.}
  \label{fig:nnloscanasmzA}
\end{figure}
\begin{figure}
  \centering
  \includegraphics[width=0.9\textwidth]{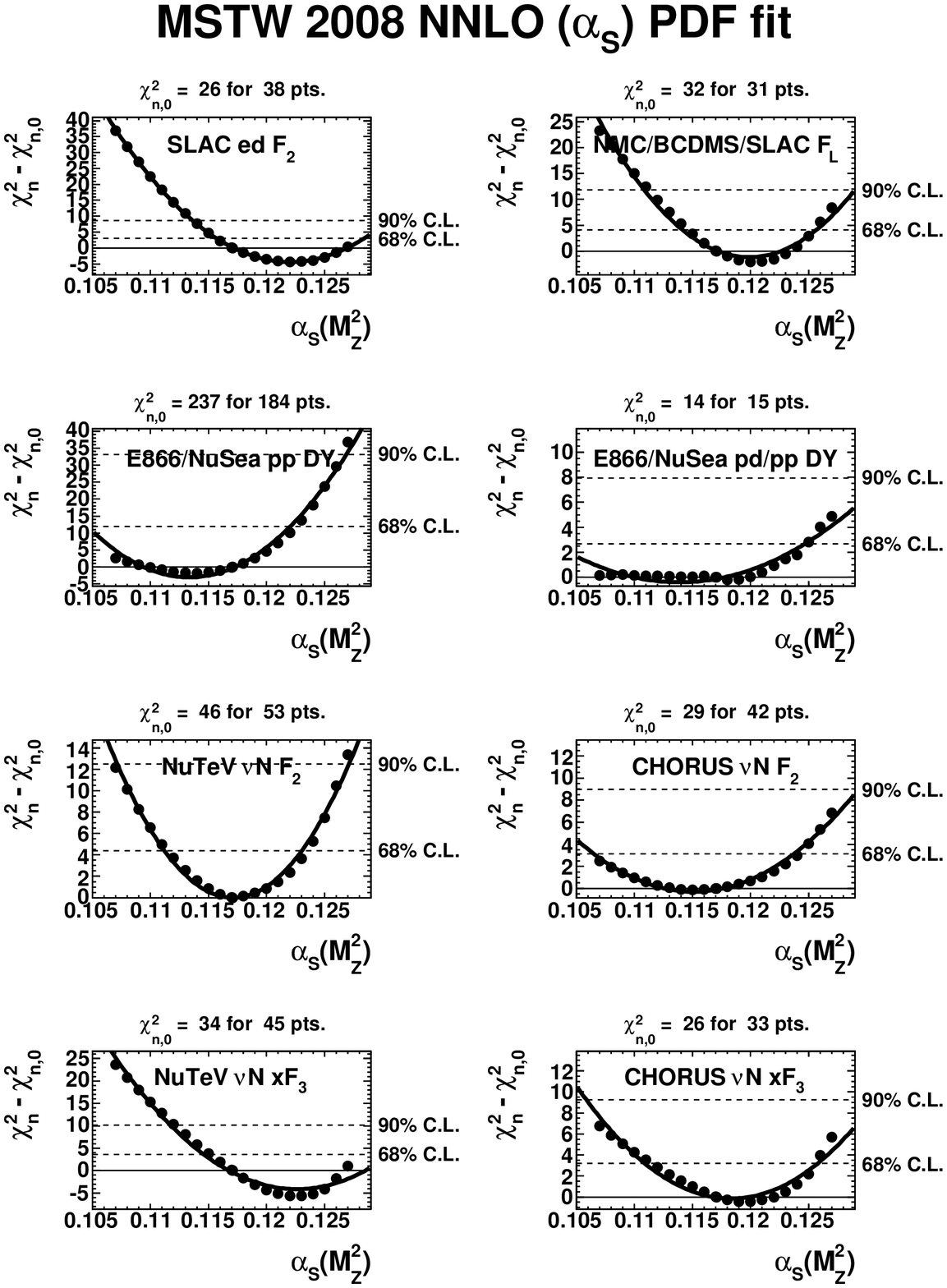}
  \caption{$\chi_n^2$ profiles for the second subset of data sets $n$ in the NNLO fit, when varying $\alpha_S(M_Z^2)$.}
  \label{fig:nnloscanasmzB}
\end{figure}
\begin{figure}
  \centering
  \includegraphics[width=0.9\textwidth]{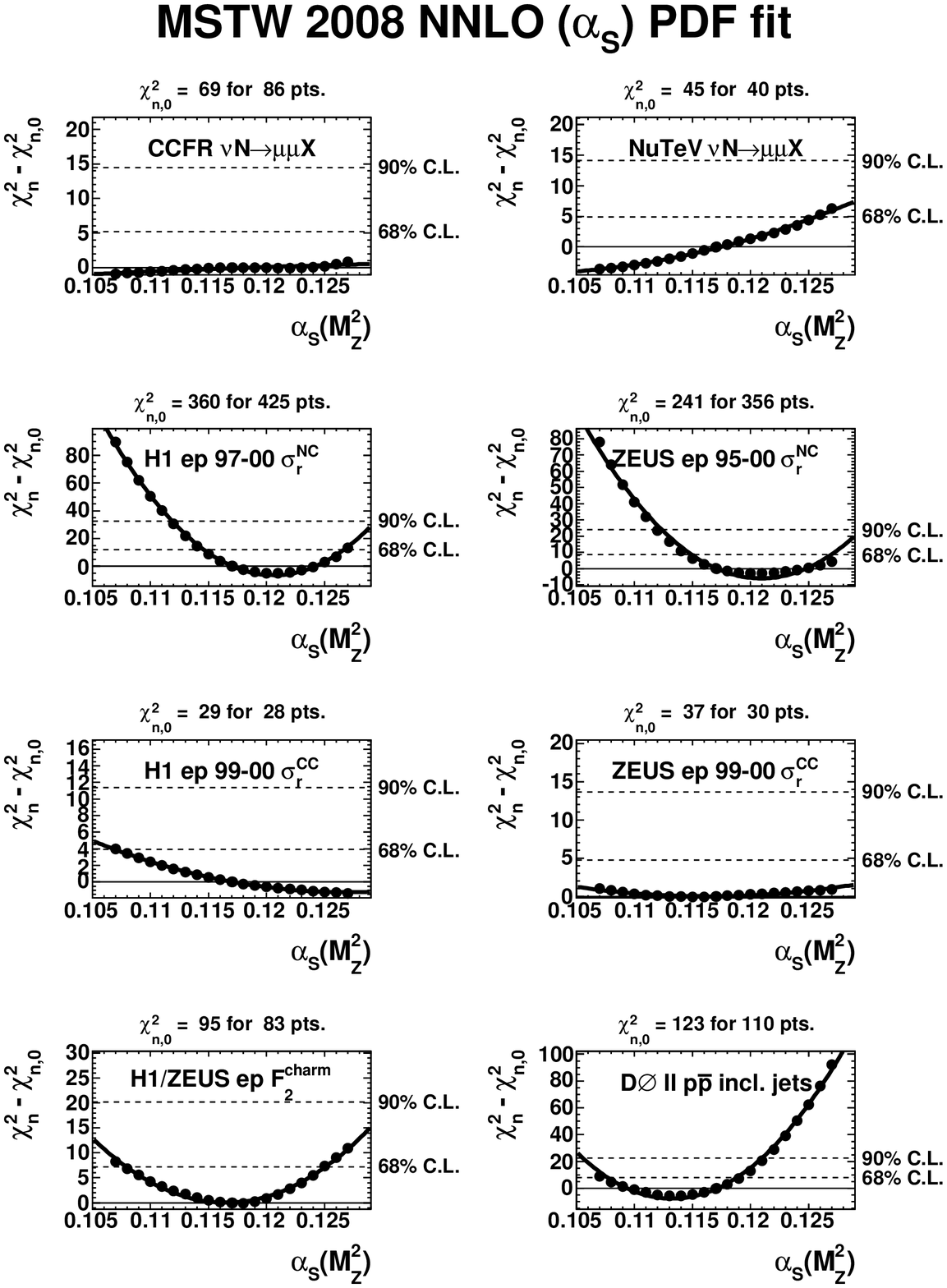}
  \caption{$\chi_n^2$ profiles for the third subset of data sets $n$ in the NNLO fit, when varying $\alpha_S(M_Z^2)$.}
  \label{fig:nnloscanasmzC}
\end{figure}
\begin{figure}
  \centering
  \includegraphics[width=0.9\textwidth]{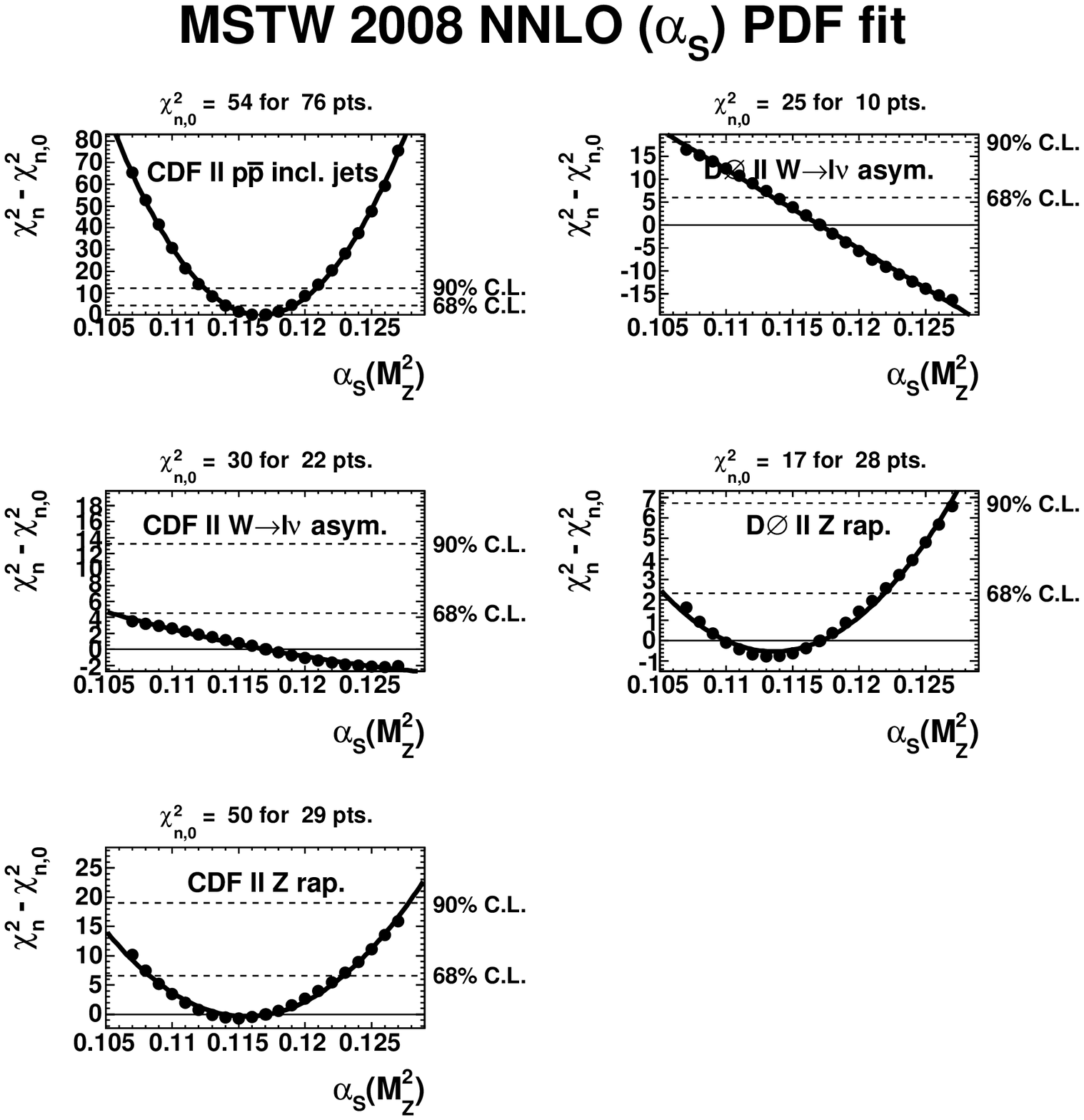}
  \caption{$\chi_n^2$ profiles for the fourth subset of data sets $n$ in the NNLO fit, when varying $\alpha_S(M_Z^2)$.}
  \label{fig:nnloscanasmzD}
\end{figure}
We plot the $\chi^2_n$ profiles for each data set $n$ as the difference from the value at the \emph{global} minimum, $\chi^2_{n,0}$, when varying $\alpha_S(M_Z^2)$; see Figs.~\ref{fig:nnloscanasmzA}--\ref{fig:nnloscanasmzD} for the NNLO $\chi^2_n$ profiles for each of the 29 different types of data.  The points ($\bullet$) in Figs.~\ref{fig:nnloscanasmzA}--\ref{fig:nnloscanasmzD} are generated for fixed values of $\alpha_S(M_Z^2)$ between 0.107 and 0.127 in steps of 0.001.  These points are then fitted to a quadratic function of $\alpha_S(M_Z^2)$ shown by the solid curves.  The horizontal dashed lines in the plots indicate the 68\% and 90\% C.L.~limits for each data set, determined according to a ``hypothesis-testing'' criterion which we will describe below in Section~\ref{sec:unc}.

We see from Figs.~\ref{fig:nnloscanasmzA}--\ref{fig:nnloscanasmzD} that for most data sets the variation of $\chi^2_n$ with respect to $\alpha_S(M_Z^2)$ is indeed approximately quadratic, with minima in the vicinity of the value of $\alpha_S(M_Z^2)$ determined by the global fit.  However, a few data sets have minima which are significantly displaced from the global best-fit $\alpha_S(M_Z^2)$ values of 0.1171 at NNLO (or 0.1202 at NLO).  For example, the BCDMS data on $F_2^{\mu\{p,d\}}$ prefer $\alpha_S(M_Z^2)$ values around 0.110.  With the global best-fit $\alpha_S$, $F_2$ in the region $0.3\lesssim x\lesssim 0.7$ falls too quickly with increasing $Q^2$ for the BCDMS data, particularly for low $Q^2$ (and hence low inelasticity $y$), and so a lower $\alpha_S$ is preferred since it gives a flatter $Q^2$ slope.\footnote{The low $y$ data points from BCDMS are strongly affected by the energy scale uncertainty of the scattered muon.  It has been advocated to impose a cut of $y>0.3$ on the BCDMS data, which caused $\alpha_S(M_Z^2)$ to increase by about 0.004 in a fit to only BCDMS data and by about 0.002 in a combined fit to H1 and BCDMS data~\cite{Adloff:2000qk}.}  By contrast, the NMC $F_2^{\mu\{p,d\}}$ data sets prefer $\alpha_S$ values somewhat larger than the global average, since the increase of these data with $Q^2$ is quicker than the default fit at NLO for $x<0.1$, so a bigger $\alpha_S$ increases the evolution speed (as do NNLO corrections, hence this problem is reduced at NNLO).  Peculiarly, the E665 data, particularly $F_2^{\mu d}$, show the opposite trend to the NMC data in the same region of $x$, and therefore the E665 data prefer low $\alpha_S$ values at NNLO (with the NLO fit giving a better description).  Note also from Fig.~\ref{fig:nnloscanasmzC} that within the full global fit the H1 and ZEUS $\sigma_r^{\rm NC}$ data (mostly at small $x$) have minima at slightly high values of $\alpha_S(M_Z^2)\approx 0.121$, implying that slightly stronger evolution than with the overall best fit is preferred.  A similar relative feature is found at NLO.  The slight tension between the $Q^2$ slopes of the structure functions from various experiments and the overall best-fit theory predictions can be seen clearly in Fig.~24 of Ref.~\cite{Martin:2009iq} (see also Figs.~3--6 of Ref.~\cite{Martin:2001es}).

We find that the preference of the inclusive HERA data for high $\alpha_S$ values holds also when we perform a NLO fit to these data alone, which at first sight seems to contradict previous findings by H1~\cite{Adloff:2000qk} and ZEUS~\cite{Chekanov:2005nn} indicating a preference for low values of 0.115 and 0.110, respectively, although each with a large uncertainty, when fitting only to inclusive data.  Relatively low values of $\alpha_S$ are also preferred by the inclusive HERA 96/97 data sets in the CTEQ global analysis~(see Fig.~4 of Ref.~\cite{Pumplin:2005rh}).  However, the low-$x$ gluon distribution is strongly anticorrelated with $\alpha_S$.  The input gluon parameterisation in the H1~\cite{Adloff:2000qk}, ZEUS~\cite{Chekanov:2005nn} and CTEQ~\cite{Pumplin:2005rh} analyses was assumed to have the form $xg\sim x^{\delta_g}$ at low $x$, whereas the more flexible MSTW~\cite{Martin:2009iq} (and MRST~\cite{Martin:2001es}) input gluon parameterisation is effectively a sum of two powers at low $x$, where the second term is required by the fit to be negative, at the input scale $Q_0^2 = 1$~GeV$^2$.  We find that a NLO fit to only inclusive HERA data using a restricted input gluon parameterisation, with the second term omitted, gives $\alpha_S(M_Z^2)\simeq 0.110$ (with a large uncertainty).  Repeating the \emph{global} fits gives $\alpha_S(M_Z^2)=0.1175$ at NLO and $\alpha_S(M_Z^2)=0.1157$ at NNLO.  However, the global $\chi^2$ is worse by 63 at NLO and by 80 at NNLO, so the more flexible low-$x$ gluon parameterisation is clearly required (particularly for the PDF uncertainties, see Section 6.5 of Ref.~\cite{Martin:2009iq}).  Since the CTEQ PDFs are input at the slightly higher $Q_0^2=1.69$~GeV$^2$, and other sets have even higher input scales, the effect of only having a single power for the gluon parameterisation will be reduced compared to our above example. However, the general feature of a reduction in $\alpha_S(M_Z^2)$ due to a restricted parameterisation is very likely to persist to a greater or lesser extent.

The data set that exhibits particularly anomalous behaviour, where $\chi_n^2$ is significantly reduced for larger $\alpha_S(M_Z^2)$, is the D{\O} Run II $W\to\mu\nu$ charge asymmetry; see Fig.~\ref{fig:nnloscanasmzD}.  We have already pointed out the difficulties of fitting these data in the standard NLO and NNLO global analyses (see Section~11.1 and Fig.~44 in Ref.~\cite{Martin:2009iq}), with both fits yielding a $\chi_{n,0}^2$ of 25 for 10 data points.  We also noted that the asymmetry is sensitive to the separation into valence and sea quarks, particularly at lower lepton $p_T$.  Indeed, this explains the behaviour of $\chi_n^2$ with $\alpha_S(M_Z^2)$: as $\alpha_S(M_Z^2)$ increases, the valence quarks evolve more rapidly at high $x$ than the sea quarks, and the $W$ asymmetry is reduced.  For $\alpha_S(M_Z^2)\simeq 0.126$ at NNLO (or $\simeq 0.129$ at NLO) the decrease in $\chi_n^2$ is approximately 15, and a good fit is obtained.  However, such large $\alpha_S(M_Z^2)$ values are completely inconsistent with most of the other data sets in the global analysis (and with other determinations of $\alpha_S$).

\begin{figure}
  \centering
  \begin{tabular}{c|c}
    {\bf MSTW 2008 NLO ($\alpha_S$)} & {\bf MSTW 2008 NNLO ($\alpha_S$)} \\[1cm]
    \includegraphics[width=0.5\textwidth]{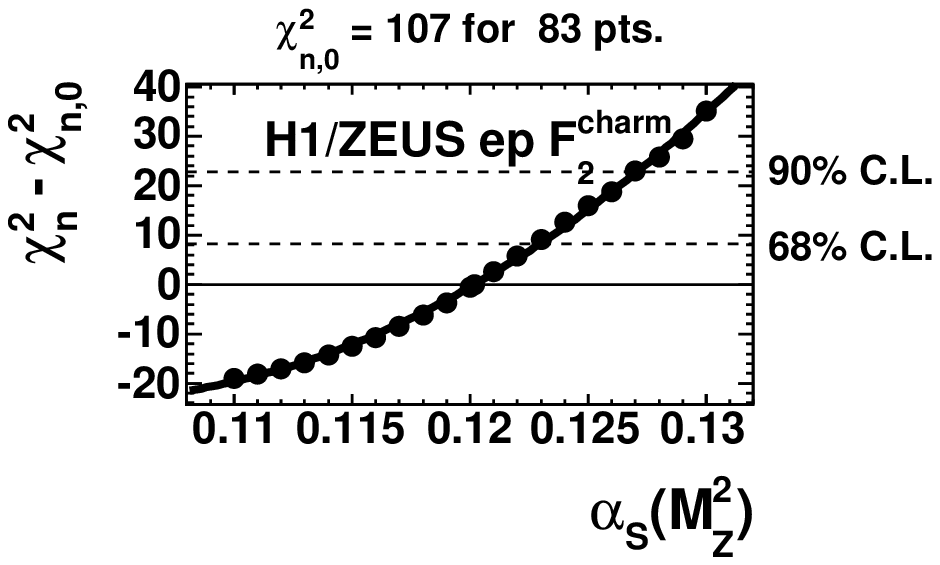} & \includegraphics[width=0.5\textwidth]{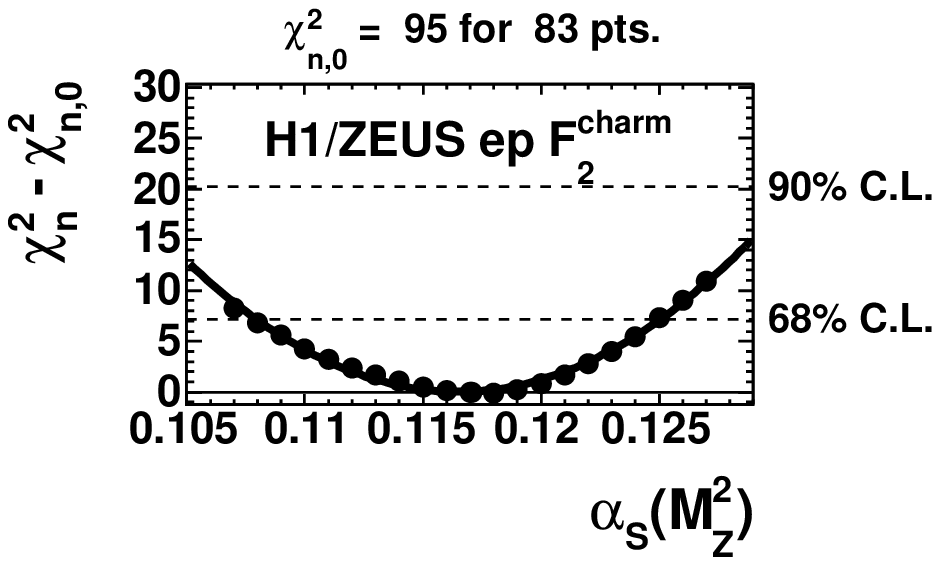}\\
    \includegraphics[width=0.5\textwidth]{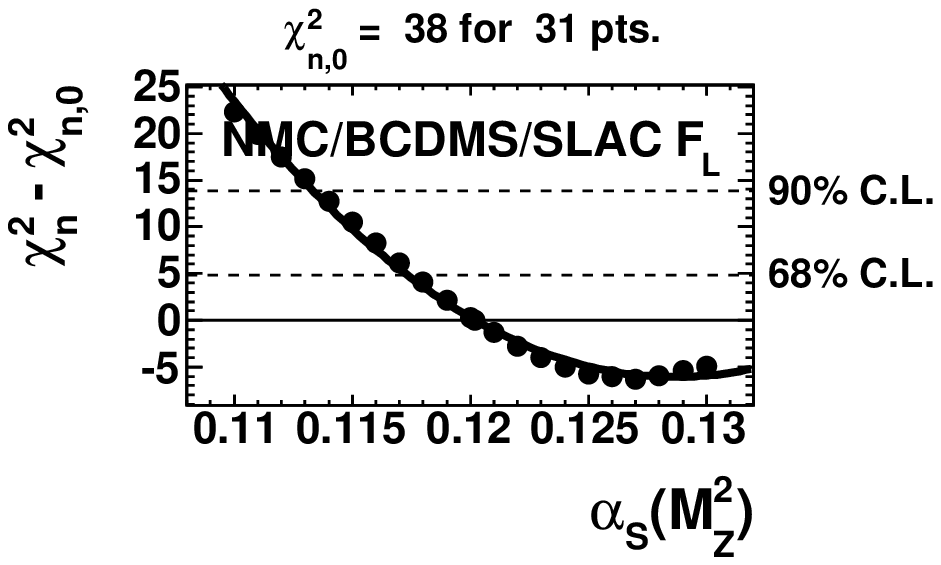} & \includegraphics[width=0.5\textwidth]{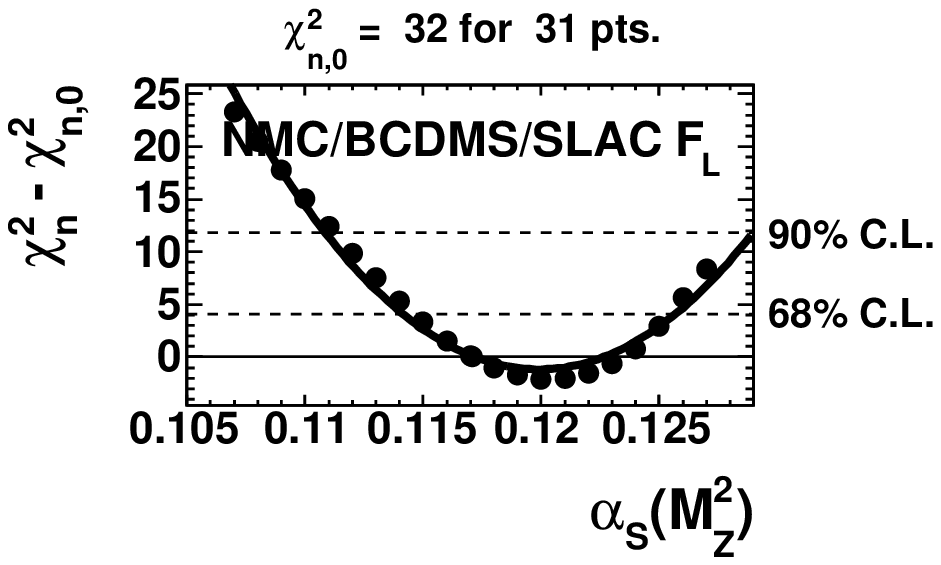}\\
    \includegraphics[width=0.5\textwidth]{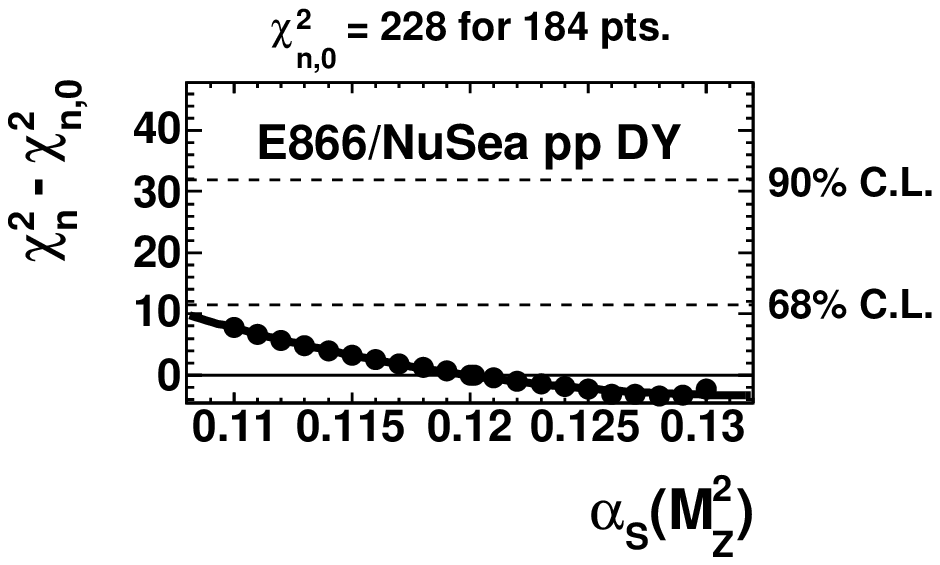} & \includegraphics[width=0.5\textwidth]{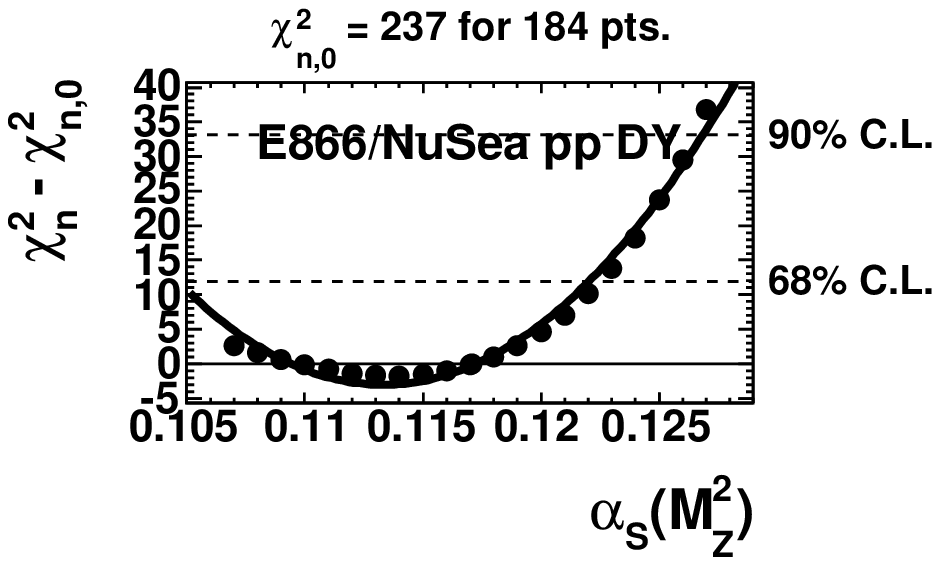}\\
  \end{tabular}
  \caption{Comparison of selected $\chi_n^2$ profiles in the NLO (left) and NNLO (right) fits.}
  \label{fig:scanasmz_selected}
\end{figure}
The NLO $\chi^2_n$ profiles are similar to those at NNLO for most data sets.  In Fig.~\ref{fig:scanasmz_selected} we compare the NLO and NNLO $\chi_n^2$ profiles for those data sets where there are notable differences between the two fits.  Specifically, the $\chi_n^2$ profiles for the H1 and ZEUS $F_2^{\rm charm}$ data, the NMC/BCDMS/SLAC $F_L$ data and the E866/NuSea $pp$ Drell--Yan cross sections are clearly more quadratic at NNLO than at NLO, with minima closer to the best-fit values.  This indicates a strong preference for the NNLO description, which is not so apparent if only the global best-fit values $\chi_{n,0}^2$ are known.  Let us give the reasons for this, taking each of these data sets in turn.  The $\chi_n^2$ minimum for the $F_2^{\rm charm}$ data at NNLO is consistent with the global average value of $\alpha_S(M_Z^2)$, whereas a minimum is not even visible in the NLO plot.  It was shown in Ref.~\cite{Thorne:2006qt} that at NNLO the value of $\partial F_2^{\rm charm}/\partial\ln Q^2$ is lowered compared to the NLO value.  The preference of the $F_2^{\rm charm}$ data for this is borne out by comparing both the values of $\chi_{n,0}^2$ at each order and the shape of the profile, i.e.~the NLO fit tries to flatten the slope with a lower coupling.  The NNLO coefficient functions for $F_L(x,Q^2)$~\cite{Moch:2004xu,Vermaseren:2005qc} are positive and significant, and similarly the NLO fit tries to mimic these with a higher value of $\alpha_S(M_Z^2)$.

The explanation of the different NLO and NNLO profiles for the E866/NuSea $pp$ Drell--Yan cross sections, seen in Fig.~\ref{fig:scanasmz_selected}, is more complicated.  The positive NNLO correction~\cite{Anastasiou:2003ds} requires an increased E866/NuSea normalisation (1.09 at NNLO compared to 1.01 at NLO, cf.~the 1-$\sigma$ normalisation uncertainty of 6.5\%), with a correspondingly larger $\chi^2_n$ penalty (3.2 at NNLO compared to practically zero at NLO).  When varying $\alpha_S(M_Z^2)$ in the range shown in Fig.~\ref{fig:scanasmz_selected}, the $\chi^2_n$ penalty term due to the fitted normalisation of the Drell--Yan data set is negligible at NLO ($\lesssim 1$ unit), while it increases dramatically at NNLO when increasing $\alpha_S(M_Z^2)$ from the best-fit value, reaching 22 units when $\alpha_S(M_Z^2)=0.127$, i.e.~a sizeable proportion of the total increase in $\chi^2_n$ relative to the best-fit $\alpha_S$.  The Drell--Yan data slightly prefer a higher value of $\alpha_S$ at NLO than the global best-fit value, while at NNLO a lower value of $\alpha_S$ is preferred mainly in order to reduce the $\chi^2_n$ penalty term due to the fitted normalisation.

The complete NNLO corrections are not yet known for inclusive jet production in deep-inelastic scattering or in hadron--hadron collisions.  For the Tevatron data, we use the approximation to the NNLO corrections obtained from threshold resummation~\cite{Kidonakis:2000gi}, which is included in the \textsc{fastnlo} package~\cite{Kluge:2006xs}.  No such approximation is available for the HERA data, and so the HERA data on inclusive jet production are simply omitted from the NNLO global fit.\footnote{Recent progress in the NNLO calculation of inclusive jet production in deep-inelastic scattering has been made by the derivation of the relevant two-loop QCD helicity amplitudes~\cite{Gehrmann:2009vu}.}  More discussion of these issues is given in Ref.~\cite{Martin:2009iq}.

\begin{figure}
  (a)\\
  \includegraphics[width=0.5\textwidth]{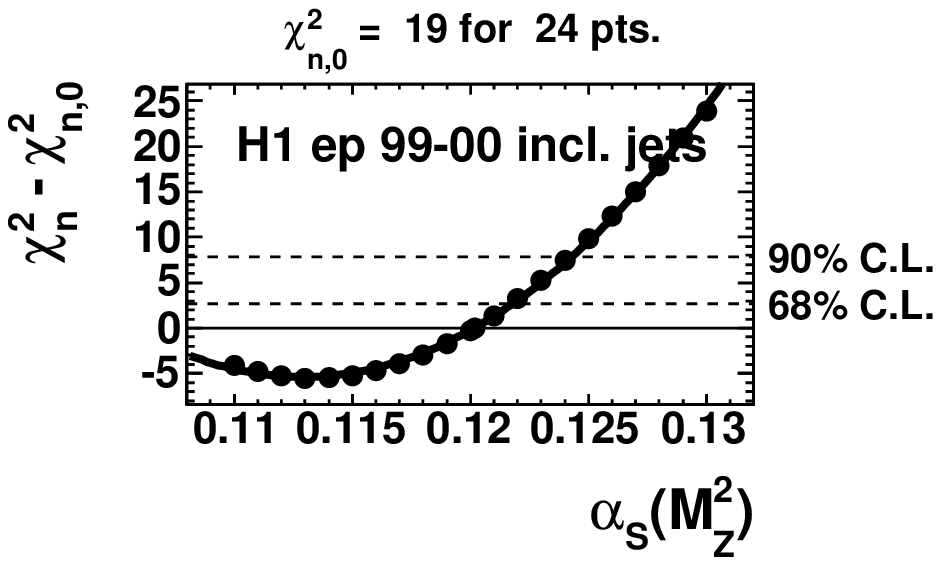}%
  \includegraphics[width=0.5\textwidth]{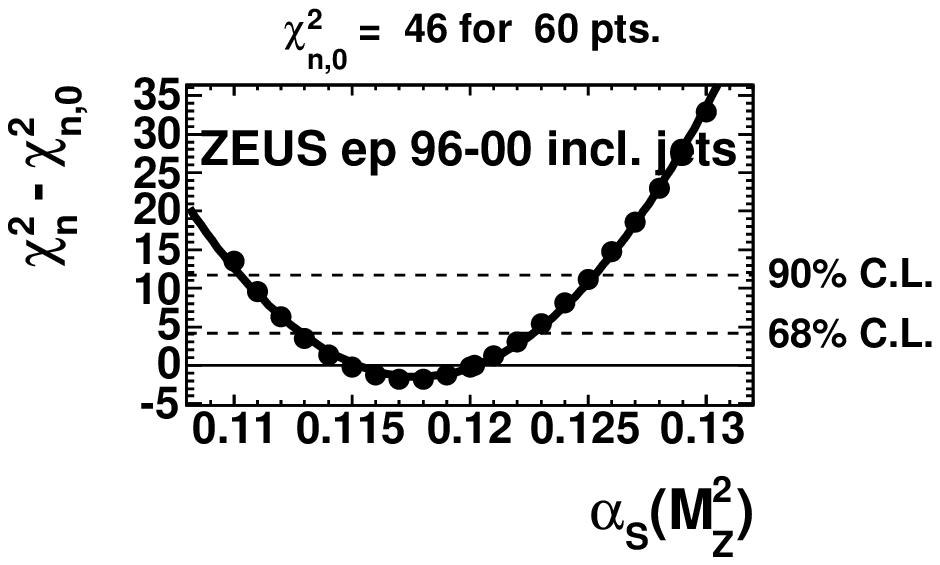}\\[3mm]
  (b)\\
  \includegraphics[width=0.5\textwidth]{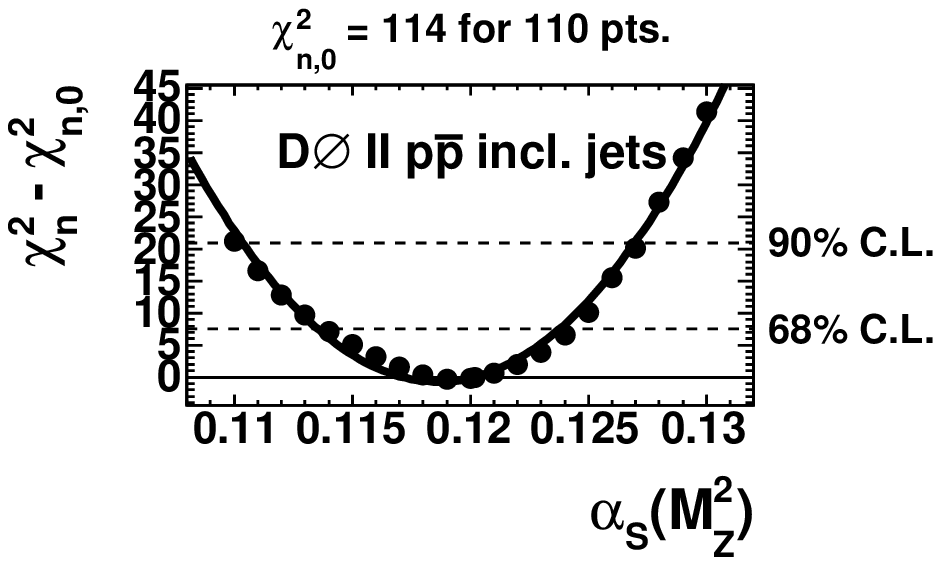}%
  \includegraphics[width=0.5\textwidth]{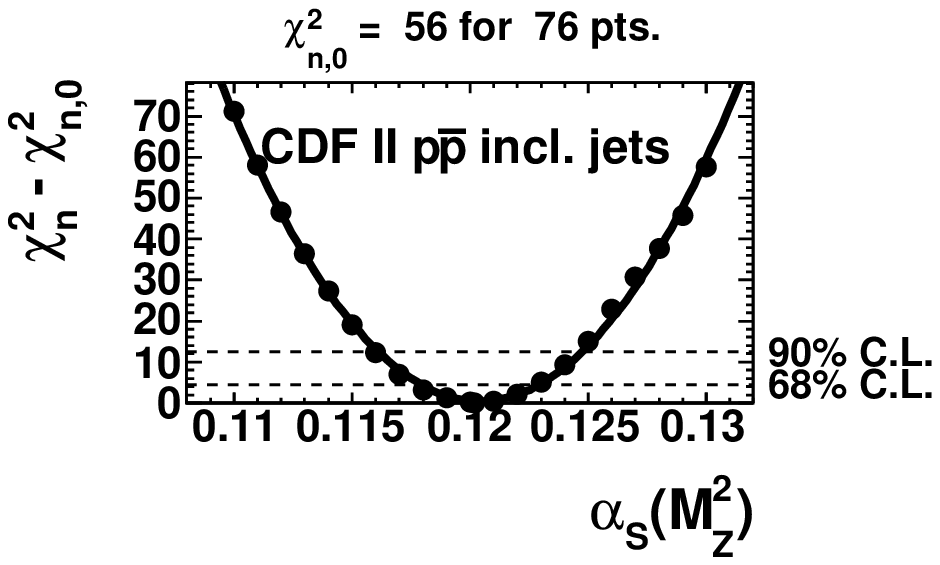}
  \caption{$\chi_n^2$ profiles for the (a)~HERA and (b)~Tevatron inclusive jet production data at NLO.}
  \label{fig:nloscanasmzNLOjets}
\end{figure}
For completeness, in Fig.~\ref{fig:nloscanasmzNLOjets}(a) we show the $\chi_n^2$ profiles for the HERA inclusive jet production data in the NLO fit, since these H1 and ZEUS data sets are not included in the NNLO fit.  Although the impact on the high-$x$ gluon distribution is diluted\footnote{In a restricted fit comprising only ZEUS data on $\sigma_r^{\rm NC}$ and $\sigma_r^{\rm CC}$, the addition of jet cross sections from deep-inelastic scattering and photoproduction was found to significantly decrease the mid- to high-$x$ gluon uncertainty without changing the central value~\cite{Chekanov:2005nn}.  However, within a \emph{global} fit~\cite{Martin:2009iq} constraints are placed on the high-$x$ gluon not only by the Tevatron inclusive jet data, but also by fixed-target structure functions, such that the impact of HERA inclusive jet cross sections is reduced.} in a \emph{global} parton analysis, it is seen from Fig.~\ref{fig:nloscanasmzNLOjets} that these data sets are quite sensitive to the value of $\alpha_S$, with the H1 data preferring $\alpha_S(M_Z^2)\simeq 0.113$ and the ZEUS data preferring $\alpha_S(M_Z^2)\simeq 0.118$.  We also show in Fig.~\ref{fig:nloscanasmzNLOjets}(b) the corresponding NLO plots for the Tevatron inclusive jet production data: the preferred values of $\alpha_S(M_Z^2)$ are 0.120 for the CDF data ($k_T$ jet algorithm) and 0.119 for the D{\O} data (cone jet algorithm), which are very consistent with the best-fit values.

We provide public PDF grids (without uncertainties) for fits with $\alpha_S(M_Z^2)$ in the range 0.110 to 0.130 (at NLO) and 0.107 to 0.127 (at NNLO) in steps of 0.001~\cite{mstwpdf}.  These grids will allow users to make $\alpha_S$ determinations from other data sets.  Note that we have implicitly done this for each of the individual data sets in Figs.~\ref{fig:nnloscanasmzA}--\ref{fig:nloscanasmzNLOjets}, where the values of $\alpha_S(M_Z^2)$ at the minima can be read off from the plots, together with an experimental uncertainty for an appropriate choice of $\Delta\chi^2_n\equiv \chi^2_n-\chi^2_{n,{\rm min}}$.  (Generally, the 1-$\sigma$ experimental uncertainty is taken to be the change in $\alpha_S(M_Z^2)$ which gives an increase of one unit in $\chi_n^2$ with respect to the minimum value.)  However, more detailed studies may be desirable, for example, including a theoretical uncertainty in the $\alpha_S(M_Z^2)$ determination from the observed renormalisation scale dependence.  Moreover, $\alpha_S$ extractions from other quantities may be of interest, for example, from the ratio of three-jet to two-jet rates at the Tevatron.

\section{Experimental uncertainty on $\alpha_S(M_Z^2)$}  \label{sec:unc}
Ideally, we would expect the errors on $\alpha_S(M_Z^2)$ to be given by $\Delta\chi^2_{\rm global}\equiv\chi^2_{\rm global}-\chi^2_{\rm min}=1$ or 2.71 for a 68\% or 90\% C.L.~limit respectively.  However, in practice, there are some inconsistencies between the independent fitted data sets, so these ``parameter-fitting'' criteria are not appropriate for global PDF analyses.  Instead, we follow the procedure of Section~6.2 of Ref.~\cite{Martin:2009iq}, where we described how to choose an appropriate value of the tolerance $T=(\Delta\chi^2_{\rm global})^{1/2}$ for each eigenvector of the covariance matrix according to ``hypothesis-testing'' criteria.  Here, we will use the same method to determine the appropriate uncertainty on $\alpha_S(M_Z^2)$.  To summarise, we perform the following steps.

We define the 90\% C.L.~region for each data set $n$ (comprising $N$ data points) by the condition that~\cite{Martin:2009iq}
\begin{equation} \label{eq:90percentCL}
  \chi_n^2 < \left(\frac{\chi_{n,0}^2}{\xi_{50}}\right)\xi_{90},
\end{equation}
where $\xi_{90}$ is the 90th percentile of the $\chi^2$-distribution with $N$ degrees of freedom, and $\xi_{50}\simeq N$ is the most probable value.  (These quantities are defined in detail in Section~6.2 of Ref.~\cite{Martin:2009iq}; see also the example discussed around Eq.~\eqref{eq:68percentCL} below.)  Similarly for the 68\% C.L.~region.  The 90\% and 68\% C.L.~regions determined in this way are shown as the horizontal dashed lines in Figs.~\ref{fig:nnloscanasmzA}--\ref{fig:nloscanasmzNLOjets}.  We then record the values of $\alpha_S(M_Z^2)$ for which the $\chi_n^2$ for each data set $n$ are minimised, together with the 90\% and 68\% C.L.~limits defined by the intercepts of the quadratic curves with the horizontal dashed lines in Figs.~\ref{fig:nnloscanasmzA}--\ref{fig:nloscanasmzNLOjets}.

\begin{figure}
  (a)\\
  \begin{minipage}{\textwidth}
    \centering
    \includegraphics[width=0.8\textwidth]{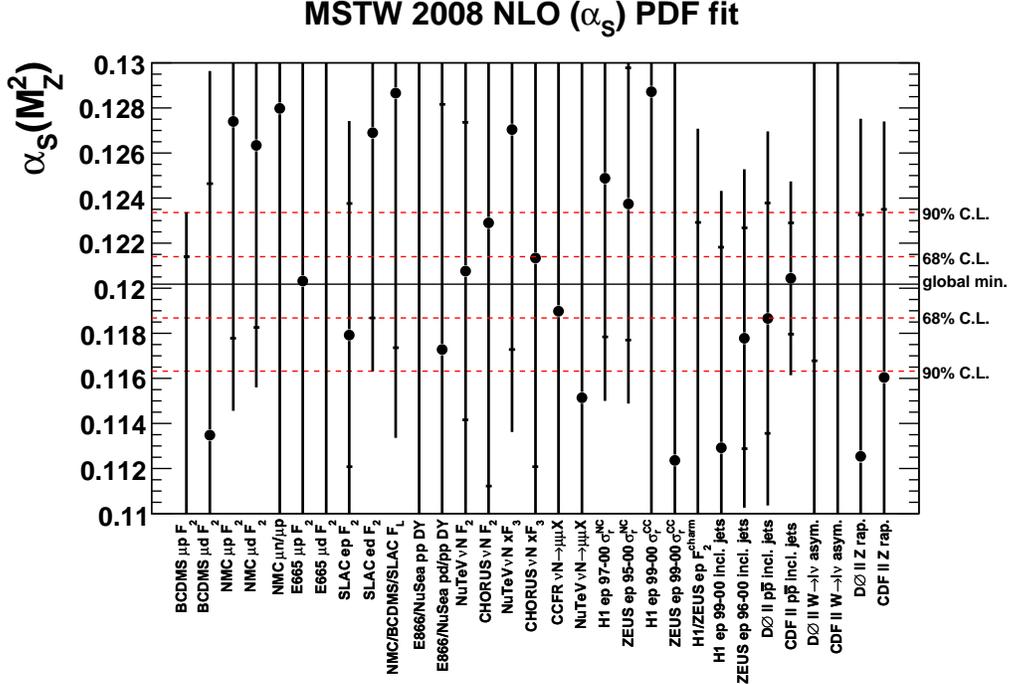}
  \end{minipage}
  (b)\\
  \begin{minipage}{\textwidth}
    \centering
    \includegraphics[width=0.8\textwidth]{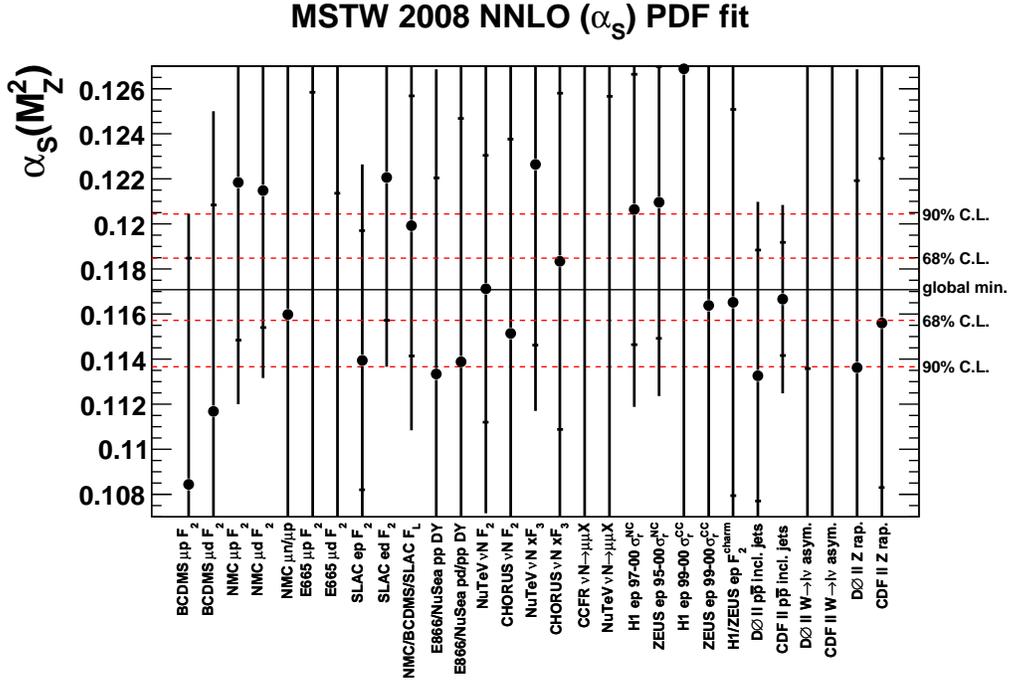}
  \end{minipage}
  \caption{Ranges of $\alpha_S(M_Z^2)$ for which data sets are described within their $90\%$ C.L.~limit (outer error bars) or $68\%$ C.L.~limit (inner error bars) in the (a)~NLO and (b)~NNLO global fits.  The points ($\bullet$) indicate the values of $\alpha_S(M_Z^2)$ favoured by each individual data set $n$, that is, the values for which $\chi_n^2$ is minimised.  The uncertainty on $\alpha_S(M_Z^2)$, indicated by the horizontal dashed lines, is chosen to ensure that all data sets are described within their $68\%$ or $90\%$ C.L.~limits defined by Eq.~\eqref{eq:90percentCL}.}
  \label{fig:rangeasmz}
\end{figure}
In Fig.~\ref{fig:rangeasmz} the points ($\bullet$) indicate these values of $\alpha_S(M_Z^2)$ for which $\chi_n^2$ is minimised, while the inner error bars extend across the 68\% C.L.~region and the outer error bars extend across the 90\% C.L.~region defined by Eq.~\eqref{eq:90percentCL}.  Note that inclusive jet production at the Tevatron is the only process included in the global fit which is proportional to $\alpha_S^2$ at leading-order, therefore these data sets provide a strong constraint on $\alpha_S$.  However, it might initially be considered surprising that the error bars in Fig.~\ref{fig:rangeasmz} are much smaller for CDF than for D{\O}.  Both factors on the right-hand side of Eq.~\eqref{eq:90percentCL} are smaller for CDF than for D{\O}, meaning that the 90\% C.L.~region for $\chi^2_n-\chi^2_{n,0}$, indicated by the horizontal dashed lines in Fig.~\ref{fig:nloscanasmzNLOjets}(b) at NLO (and the corresponding plots in Figs.~\ref{fig:nnloscanasmzC} and \ref{fig:nnloscanasmzD} at NNLO), is smaller by almost a factor of 2 for CDF.  More importantly, the $\chi_n^2$ profiles are significantly steeper for CDF compared to D{\O}, i.e.~the CDF data are more sensitive to the $\alpha_S$ value.  This can largely be explained by the fact that the normalisation of the CDF jet data is tied to the more constraining CDF $Z$ rapidity distribution, due to the common luminosity uncertainty.  This is an interesting example of the interplay between data sets in a global fit.  The normalisation of the D{\O} jet data is independent of the D{\O} $Z$ data, since the latter is presented as a rapidity shape distribution, i.e.~divided by the measured $Z$ total cross section.  The D{\O} jet data therefore has more freedom to compensate for the variation in $\alpha_S$ by changing the normalisation.

We choose the uncertainty on $\alpha_S(M_Z^2)$, indicated by the horizontal dashed lines in Fig.~\ref{fig:rangeasmz}, so that all data sets are described within their 90\% or 68\% C.L.~limits.  It is seen from Fig.~\ref{fig:rangeasmz} that the upper limit on $\alpha_S(M_Z^2)$ is fixed by the deterioration of the quality of the fit to the BCDMS $F_2^{\mu p}$ data (within the context of the full global fit), while the lower limit is provided by the deterioration of the quality of the fit to the SLAC $F_2^{ed}$ data. In each case a number of additional data sets are very close to providing the constraint, i.e.~as with the determination of the tolerance values in Ref.~\cite{Martin:2009iq} there is no particular reliance on any one data set in fixing the limits on $\alpha_S(M_Z^2)$.  The final results are:
\begin{align}
  \text{NLO:}\qquad\alpha_S(M_Z^2) &= 0.1202\quad^{+0.0012}_{-0.0015}\text{ (68\% C.L.)}\quad^{+0.0032}_{-0.0039}\text{ (90\% C.L.)}, \label{eq:asmznlo} \\
  \text{NNLO:}\qquad\alpha_S(M_Z^2) &= 0.1171\quad^{+0.0014}_{-0.0014}\text{ (68\% C.L.)}\quad^{+0.0034}_{-0.0034}\text{ (90\% C.L.)}. \label{eq:asmznnlo}
\end{align}
\begin{figure}
  (a)\\
  \begin{minipage}{\textwidth}
    \centering
    \includegraphics[width=0.8\textwidth]{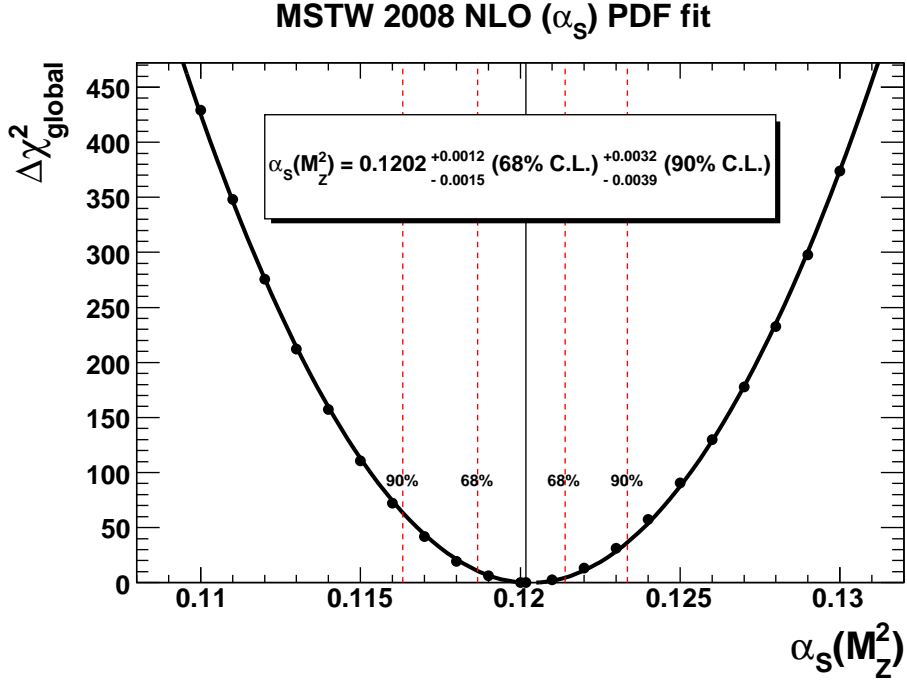}
  \end{minipage}
  (b)\\
  \begin{minipage}{\textwidth}
    \centering
    \includegraphics[width=0.8\textwidth]{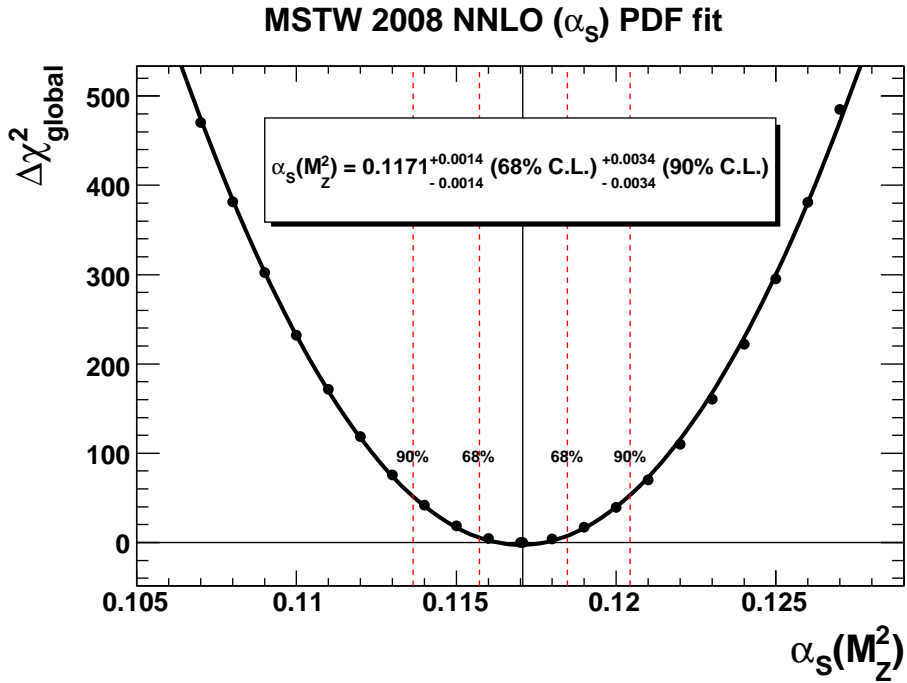}
  \end{minipage}
  \caption{The points ($\bullet$) show $\Delta\chi^2_{\rm global}\equiv\chi^2_{\rm global}-\chi^2_{\rm min}$ as a function of $\alpha_S(M_Z^2)$ for the (a)~NLO and (b)~NNLO global fits.  The solid curve is a fit to a quadratic function of $\alpha_S(M_Z^2)$.  The vertical dashed lines indicate the 68\% and 90\% C.L.~uncertainties on $\alpha_S(M_Z^2)$.}
  \label{fig:chisqglobalasmz}
\end{figure}
In Fig.~\ref{fig:chisqglobalasmz} we show the change in $\chi^2_{\rm global}$ as $\alpha_S(M_Z^2)$ is varied.  The vertical lines indicate the 68\% and 90\% C.L.~uncertainties given by Eqs.~\eqref{eq:asmznlo} and \eqref{eq:asmznnlo}.  The uncertainties on $\alpha_S(M_Z^2)$ at NNLO, determined using the procedure just described, amount to taking $\Delta\chi^2_{\rm global} = ^{+7.6}_{-5.4}$ (68\% C.L.) or $^{+53}_{-51}$ (90\% C.L.).  These are much larger values than the canonical ``parameter-fitting'' values of $\Delta\chi^2_{\rm global} = 1$ (68\% C.L.) or $2.71$ (90\% C.L.).  Conversely, if these standard ``parameter-fitting'' values of $\Delta\chi^2_{\rm global}$ were taken seriously, then the uncertainties on $\alpha_S(M_Z^2)$ would be $^{+0.0008}_{-0.0009}$ (68\% C.L.) or $^{+0.0010}_{-0.0011}$ (90\% C.L.).  The experimental errors on $\alpha_S$ given in Eqs.~\eqref{eq:asmznlo} and \eqref{eq:asmznnlo} are a much refined revision of the previous estimate of $\pm0.002$ from the MRST 2001 analysis~\cite{Martin:2001es} obtained using a fixed $\Delta\chi^2_{\rm global}=20$.

In Table~\ref{tab:asmzcompare} we compare our determination of $\alpha_S(M_Z^2)$ at NLO and NNLO with the results obtained by other PDF fitting groups, showing only the experimental uncertainties.
\begin{table}
  \centering
  \begin{tabular}{l|lll}
    \hline\hline
    \textbf{NLO} & $\alpha_S(M_Z^2)$ (expt.~unc.~only) \\\hline
    MSTW (this work) & $0.1202\quad^{+0.0012}_{-0.0015}$ \\
    CTEQ~\cite{Pumplin:2005rh} & $0.1170\quad\pm0.0047$ \\
    H1~\cite{Adloff:2000qk} & $0.1150\quad\pm0.0017$ \\
    ZEUS~\cite{Chekanov:2005nn} & $0.1183\quad\pm0.0028$ \\
    Alekhin~\cite{Alekhin:2002fv} & $0.1171\quad\pm0.0015$ \\
    BBG~\cite{Blumlein:2006be} & $0.1148\quad\pm0.0019$ \\
    GJR~\cite{Gluck:2007ck} & $0.1145\quad\pm0.0018$ \\
    \hline\hline\multicolumn{2}{l}{}\\\hline\hline
    \textbf{NNLO} & $\alpha_S(M_Z^2)$ (expt.~unc.~only) \\\hline
    MSTW (this work) & $0.1171\quad^{+0.0014}_{-0.0014}$ \\
    AMP~\cite{Alekhin:2006zm} & $0.1128\quad\pm0.0015$ \\
    BBG~\cite{Blumlein:2006be} & $0.1134\quad^{+0.0019}_{-0.0021}$ \\
    ABKM~\cite{Alekhin:2009ni} & $0.1129\quad\pm0.0014$ \\
    JR~\cite{JimenezDelgado:2008hf} & $0.1158\quad\pm0.0035$ \\
    \hline\hline
  \end{tabular}
  \caption{Comparison of the present $\alpha_S$ determination at NLO and NNLO with the values obtained by other PDF fitting groups, showing only the experimental uncertainties.}
  \label{tab:asmzcompare}
\end{table}
The values in Table~\ref{tab:asmzcompare} can be compared with the world average value quoted by the Particle Data Group (PDG) of $\alpha_S(M_Z^2)=0.1176\pm0.002$~\cite{Amsler:2008zzb}, where the PDG world average is dominated by NNLO results, with the NLO results included carrying little weight.  The 2009 world average value of $\alpha_S(M_Z^2)$ obtained by Bethke~\cite{Bethke:2009jm} is $0.1184\pm0.0007$.  The MSTW values are completely consistent with the world average values, but are a little higher than all the others in Table~\ref{tab:asmzcompare}, which are mostly from fits mainly, or exclusively, to structure functions in deep-inelastic scattering.  The NNLO values in Table~\ref{tab:asmzcompare} are generally smaller than the corresponding NLO values from the same fitting group, as is to be expected since NNLO corrections to splitting functions and coefficient functions automatically lead to quicker evolution.  The CTEQ\footnote{The CTEQ analysis uses a definition of $\alpha_S$ at NLO which at low scales lies above other definitions for the same value of $\alpha_S(M_Z^2)$, and tends to result in an extraction of $\alpha_S(M_Z^2)$ about 0.001 lower than the same fit using the other definitions, see Figs.~8 and 9 of Ref.~\cite{Huston:2005jm}.} global analysis~\cite{Pumplin:2005rh} includes additional data on fixed-target Drell--Yan production and Tevatron data on the $W\to\ell\nu$ asymmetry and inclusive jet production.  The H1 analysis~\cite{Adloff:2000qk} includes additional BCDMS data, the ZEUS analysis~\cite{Chekanov:2005nn} includes jet cross sections from deep-inelastic scattering and photoproduction, the GJR analysis~\cite{Gluck:2007ck} includes E866/NuSea Drell--Yan and Tevatron inclusive jet data, and the AMP~\cite{Alekhin:2006zm}, ABKM~\cite{Alekhin:2009ni} and JR~\cite{JimenezDelgado:2008hf} analyses all include fixed-target Drell--Yan data.  The BBG fits~\cite{Blumlein:2006be} are from a non-singlet analysis.  We quote the results of the ``standard'' fits by GJR/JR~\cite{Gluck:2007ck,JimenezDelgado:2008hf} and not the results of the ``dynamical'' fits using a more restricted input parameterisation, which were the focus of these analyses but gave a slightly worse description of data, with $\alpha_S(M_Z^2)$ values lower by around $0.003$  than the values from the corresponding ``standard'' fits (and with smaller uncertainties due to the significant theoretical constraint on the input).  This is another, rather extreme, example of the point raised in Section~\ref{sec:desc}, i.e.~the correlation between a restricted input parameterisation for the gluon distribution and low values of $\alpha_S(M_Z^2)$.  Indeed, this is very likely to be part of the reason for the lower values of all the extractions in Table~\ref{tab:asmzcompare} compared to MSTW.  The exception is BBG~\cite{Blumlein:2006be}, which is a non-singlet analysis, and by definition relies on a much smaller data set and on the data cuts being sufficient that purely non-singlet evolution is applicable.

Indeed, there is a distinct sensitivity to the data sets used.  The fits containing only a limited number of data sets tend to obtain lower values of $\alpha_S$.  This is often because the BCDMS data has a strong influence in the fit, as seen in Fig.~\ref{fig:rangeasmz}.  This tendency for the BCDMS data to bring the value of $\alpha_S(M_Z^2)$ down was well-illustrated in the HERA-LHC ``benchmark'' fits~\cite{Dittmar:2005ed,Dittmar:2009ii}, where both the data sets chosen and the cuts applied led to the BCDMS data carrying a high weight.  It is often thought that the inclusive jet data favour high values of $\alpha_S(M_Z^2)$, and are thus responsible for raising the values significantly in global fits.  The picture is not quite so simple, as the values preferred by these data sets shown in Fig.~\ref{fig:rangeasmz} illustrate, with none lying far above the global best-fit values.  Indeed, in Section 12 of Ref.~\cite{Martin:2009iq} it was shown that removing the Tevatron jet data results in the best-fit value of $\alpha_S(M_Z^2)$ at NLO falling only from 0.1202 to 0.1197.  It is more the case that the inclusion of the jet data can alter the shape of the gluon distribution obtained in a fit to only deep-inelastic scattering data, which indirectly affects the value of $\alpha_S(M_Z^2)$, or can turn a low value of $\alpha_S(M_Z^2)$ with large uncertainty into a much better constrained higher value, as in Ref.~\cite{Chekanov:2005nn}.

In addition to the data fitted, the differences seen in Table~\ref{tab:asmzcompare} between the experimental uncertainties on $\alpha_S(M_Z^2)$ obtained by different groups can be traced partly to the different methods used for error propagation and in particular to the choice of the tolerance, $T=(\Delta\chi^2_{\rm global})^{1/2}$.  For example, the large CTEQ uncertainty corresponds to $T\simeq\sqrt{100/2.71}\simeq 6$ compared to the MSTW values of $T\simeq 2$--$3$, GJR/JR use $T\simeq4.7/4.5$ at NLO/NNLO respectively, while the other groups fit a smaller range of data and use $T=1$.

We shall not address in detail the issue of the theory uncertainty on our $\alpha_S$ determination.  The widely-used method of varying the renormalisation and factorisation scales up and down by a factor of around two is not really sufficient in a global fit, as it misses contributions depending on $\ln(1/x)$ and $\ln(1-x)$ at higher orders and such issues as ambiguities in heavy quark flavour scheme definitions.\footnote{Additionally, scale variation is not currently possible for our fully global fit due to various complications, including the general-mass variable flavour number scheme~\cite{Thorne:2006qt}. However, as one example, if the scale is changed only for inclusive jet production at the Tevatron, from $\mu_R=\mu_F=p_T$ to $\mu_R=\mu_F=p_T/2$, then $\alpha_S(M_Z^2)$ is lowered by $0.001$, at the expense of a slightly worse description of the D{\O} Run II data, as discussed in Section 12 of Ref.~\cite{Martin:2009iq}.}  Alekhin~\cite{Alekhin:2002fv,Alekhin:2002rk} has estimated the theory uncertainty on $\alpha_S(M_Z^2)$ determined from fits to deep-inelastic scattering data due to scale dependence to be $\pm0.0033$ at NLO and $\pm0.0009$ at NNLO.  H1~\cite{Adloff:2000qk} estimated a similar uncertainty on their $\alpha_S(M_Z^2)$ determination from H1 and BCDMS proton data to be $\pm0.005$, and ZEUS~\cite{Chekanov:2005nn} also obtained $\pm0.005$ from varying the scale only for the jet data included in their fit.

A simple estimate of the theory uncertainty can be obtained by taking the difference between the NLO and NNLO determinations of $\alpha_S(M_Z^2)$.  This gives $\pm0.003$, which could be considered to be a minimum theory uncertainty at NLO and a maximum theory uncertainty at NNLO.  It coincides with the estimate at NLO quoted in the MRST 2001 analysis~\cite{Martin:2001es} and the estimate at NLO obtained from scale variation by Alekhin~\cite{Alekhin:2002rk}.  It is also of the same order as changes invoked by introducing models of higher-order corrections or changing the cuts on data at NLO~\cite{Martin:2003sk}, which to us seems to be one of the most reliable estimates.  It is probably an overestimate at NNLO and more quantitative studies are needed.  However, at NNLO the estimates from changing data cuts and introducing models for theoretical corrections~\cite{Martin:2003sk} suggested that variations in $\alpha_S(M_Z^2)$ are distinctly lower than at NLO, and that perhaps $\pm0.001$--$0.002$ would be more appropriate.  Hence, to be conservative we estimate a NNLO theoretical uncertainty of at most $\pm 0.002$, a slightly larger value than that obtained from scale variation at NNLO by Alekhin~\cite{Alekhin:2002rk} to account for the more complicated global fit.

\section{Eigenvector PDF sets with varying $\alpha_S$} \label{sec:eigen}
Following on from our method of determining PDF uncertainties using the Hessian method, we might consider simply letting $\alpha_S(M_Z^2)$ be another free parameter when diagonalising the covariance matrix of the fit, resulting in 42 rather than 40 eigenvector PDF sets, each with a slightly different value of $\alpha_S$.  We do not follow this approach for various reasons.  The first of these is because the coupling does not quite sit on an equal footing with the input PDF parameters.  Let us consider a superposition of different eigenvector PDF sets.  The linear nature of the DGLAP evolution equations means that any linear combination of PDFs, all with a common value of $\alpha_S(M_Z^2)$, is also a well defined PDF set evolving precisely according to the DGLAP evolution equations.  Consequently, the precise linear combination is the same whatever the factorisation scale at which we sample the PDFs.  But if the eigenvector PDF sets in the superposition have differing $\alpha_S(M_Z^2)$, then this picture for linear combinations is not preserved.  A linear combination of PDFs evolving with different $\alpha_S(M_Z^2)$ values do not evolve in precisely the same manner as one single PDF set with a fixed $\alpha_S$, i.e.~they do not follow precisely a real trajectory in PDF space.  This is not a problem in the particular case of using the eigenvector PDF sets to map out the uncertainty band, since each is used independently.  However, it is an issue when rediagonalising the Hessian matrix, such as in Ref.~\cite{Pumplin:2009nm}, where, from a well-defined starting point, the new eigenvector PDF sets would not correspond to a particular PDF set which evolves in \emph{precisely} the manner prescribed by the evolution equations.  Since the whole Hessian method is based on manipulating linear combinations of perturbations in PDF parameters, which to first order in the Taylor expansion is equivalent to manipulating linear combinations of PDF sets, we find this feature troubling, even though it would certainly be a small effect in practice.

There are also more practical reasons why we reject the option of simply including $\alpha_S$ as an extra parameter in the Hessian matrix.  We limit ourselves to 20 eigenvectors because a larger number of free input PDF parameters leads to too large correlations, and to a breakdown of the quadratic behaviour of the global $\chi^2$ distribution in some eigenvector directions~\cite{Martin:2009iq}.  A free $\alpha_S$ would introduce more correlation between parameters, and reduce the stability of the eigenvectors, a feature we prefer to avoid.  Finally, this approach would limit the uncertainty analysis to an expansion about the single best-fit PDF set, with a unique value of $\alpha_S$.  We acknowledge that the user may prefer more flexibility than this, and might wish to utilise PDF sets including uncertainties for a variety of different values of $\alpha_S$.  Hence, we provide eigenvector PDF sets with different, but fixed, $\alpha_S$ values to allow studies of this sort.  In the rest of this section, we give the details of our extraction of PDFs with uncertainties at different $\alpha_S$ values, and examples of our recommended use of them will be given in Section~\ref{sec:cross}.  Here we just note that as we go away from the best-fit value of $\alpha_S(M_Z^2)$, the fit quality is automatically deteriorating, so that the PDFs cannot vary as much before the fit quality becomes unacceptable.  It is an automatic result of our procedure that the PDF uncertainty shrinks as $\alpha_S(M_Z^2)$ deviates from the preferred value determined by the global fit.

We provide eigenvector PDF sets with $\alpha_S(M_Z^2)$ fixed at the limits of the 68\% and 90\% C.L.~uncertainty regions, given by Eqs.~\eqref{eq:asmznlo} and \eqref{eq:asmznnlo}.  We also provide eigenvector PDF sets where $\alpha_S(M_Z^2)$ is fixed at half these limits.  (Further intermediate values of $\alpha_S$ are unnecessary, as we will explain later in Section~\ref{sec:cross}.)  The eigenvector PDF sets $S_k^\pm$ are generated, for each fixed value of $\alpha_S$, from input PDF parameters~\cite{Martin:2009iq}
\begin{equation} \label{eq:skpm}
  a_i(S_k^\pm) = a_i^0 \pm t_k^\pm\,e_{ik},
\end{equation}
where $a_i^0$ are the best-fit input PDF parameters for that value of $\alpha_S$, and the rescaled eigenvectors are $e_{ik}\equiv \sqrt{\lambda_k}\,v_{ik}$.  The covariance (inverse Hessian) matrix has eigenvalues $\lambda_k$ and orthonormal eigenvectors $\boldsymbol{v}_k$ (with components $v_{ik}$).  The distance $t_k^\pm$ along each rescaled eigenvector direction is adjusted to give the desired tolerance $T_k^\pm=(\Delta\chi^2_{\rm global})^{1/2}$.  In determining the tolerance for each of the eigenvector PDF sets using Eq.~\eqref{eq:90percentCL}, and the corresponding equation for the 68\% C.L.~uncertainties, we take $\chi_{n,0}^2$ to be the values at the overall \emph{global} minimum, that is, the values obtained using the best-fit value of $\alpha_S(M_Z^2)$.

Consider the situation when $\alpha_S(M_Z^2)$ is fixed at its \emph{upper} 1-$\sigma$ limit in the NLO fit.  Recall, from Fig.~\ref{fig:rangeasmz}, that this limit is fixed by the BCDMS $F_2^{\mu p}$ data set, for which $\chi_{n,0}^2 = 182.2$ for $N=163$ degrees of freedom.  In this case we have $\xi_{50} = 162.3$ and $\xi_{68} = 171.0$ (while $\xi_{90} = 186.5$); see Fig.~7 of Ref.~\cite{Martin:2009iq}.  For this data set we define the 68\% C.L.~region by the condition that
\begin{equation} \label{eq:68percentCL}
  \chi_n^2 < \left(\frac{\chi_{n,0}^2}{\xi_{50}}\right)\xi_{68}.
\end{equation}
The rescaling factor ($\chi_{n,0}^2/\xi_{50}$) is necessary to take account of the fact that the value of $\chi_{n,0}^2$ at the global minimum is quite far from the most probable value of $\xi_{50}\simeq N$ of this data set $n$.  Indeed, we see that in this case the best-fit value $\chi_{n,0}^2=182.2$ lies outside the strict 68\% C.L.~region $\chi_n^2<\xi_{68}=171.0$.  After applying the rescaling factor, the 68\% C.L.~region is given by $\chi_n^2-\chi_{n,0}^2<9.8$.  That is, the central fit for the BCDMS $F_2^{\mu p}$ data, with $\alpha_S(M_Z^2)$ fixed at its upper 1-$\sigma$ limit, will have $\chi_n^2$ of 9.8 units worse than in the overall best fit.
\begin{figure}
  (a)\hspace{0.5\textwidth}(b)\\
  \includegraphics[width=0.5\textwidth]{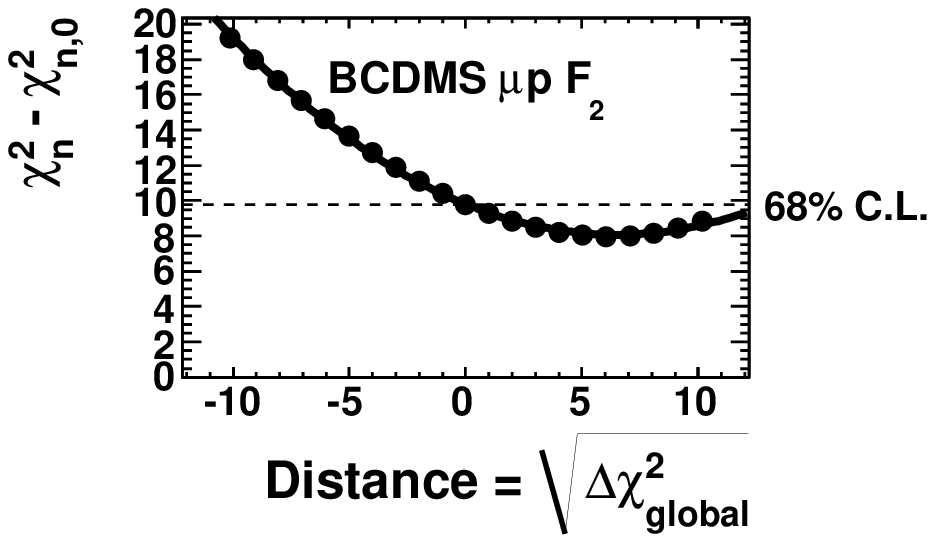}
  \includegraphics[width=0.5\textwidth]{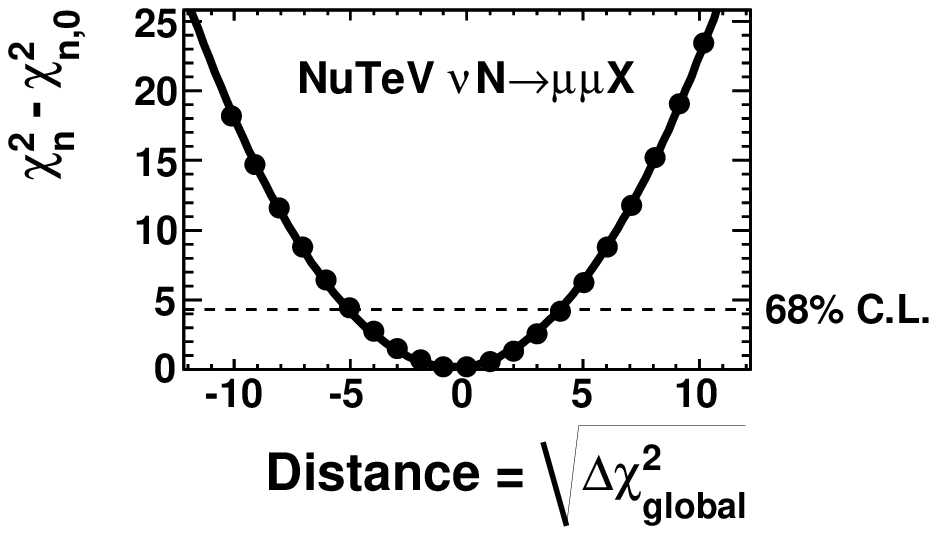}
  (c)\hspace{0.5\textwidth}(d)\\
  \includegraphics[width=0.5\textwidth]{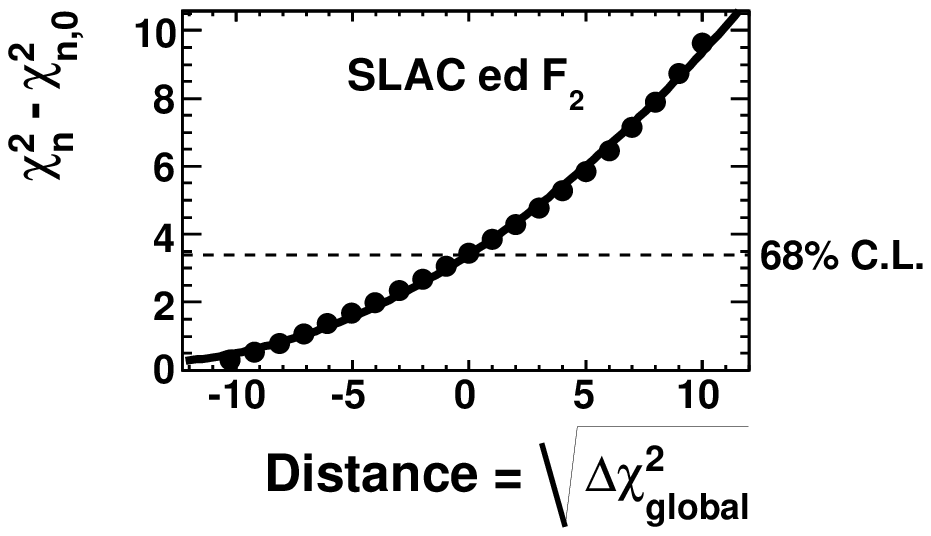}
  \includegraphics[width=0.5\textwidth]{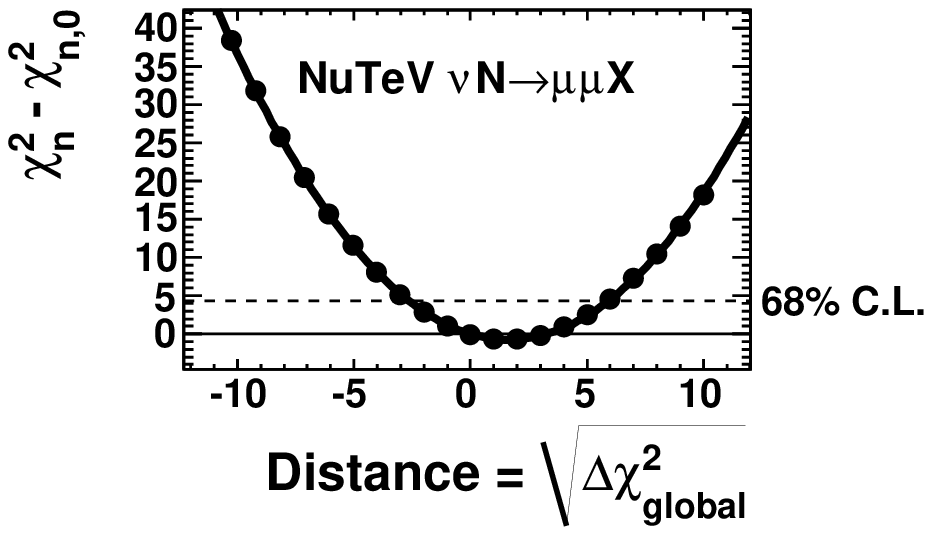}
  \caption{$\chi_n^2$ profiles in the NLO fit when moving along eigenvector 6 for (a)~the BCDMS $F_2^{\mu p}$ data when $\alpha_S(M_Z^2)$ is fixed at its \emph{upper} 1-$\sigma$ limit, (b)~the NuTeV dimuon data when $\alpha_S(M_Z^2)$ is fixed at its \emph{upper} 1-$\sigma$ limit,  (c)~the SLAC $F_2^{ed}$ data when $\alpha_S(M_Z^2)$ is fixed at its \emph{lower} 1-$\sigma$ limit, and (d)~the NuTeV dimuon data when $\alpha_S(M_Z^2)$ is fixed at its \emph{lower} 1-$\sigma$ limit.}
  \label{fig:scan6}
\end{figure}
This can be clearly seen in Fig.~\ref{fig:scan6}(a) where we show $\chi_n^2-\chi_{n,0}^2$ when moving along eigenvector number 6.  At zero distance, $\chi_n^2-\chi_{n,0}^2 = 9.8$, indicated by the horizontal dashed line in Fig.~\ref{fig:scan6}(a).  Moving in the negative direction along this eigenvector, the $\chi_n^2$ increases further, meaning that the tolerance for this eigenvector is \emph{zero} in the negative direction.  In the positive direction, the constraint which fixes the tolerance of 4.0 is provided by another data set, namely the NuTeV dimuon data, as seen in Fig.~\ref{fig:scan6}(b).

In Fig.~\ref{fig:scan6}(c) we show the corresponding plot when $\alpha_S(M_Z^2)$ is fixed at its \emph{lower} 1-$\sigma$ limit in the NLO fit.  Recall from Fig.~\ref{fig:rangeasmz} that this limit is fixed by the SLAC $F_2^{ed}$ data, where $\chi_n^2$ is $3.4$ units worse than the best-fit value of $\chi_{n,0}^2=29.7$.  Here, the tolerance for eigenvector number 6 is zero in the positive direction, while the constraint in the negative direction is again provided by the NuTeV dimuon data, this time giving a tolerance of 2.6, as seen in Fig.~\ref{fig:scan6}(d).

\begin{figure}
  (a)\\
  \begin{minipage}{\textwidth}
    \centering
    \includegraphics[width=0.87\textwidth]{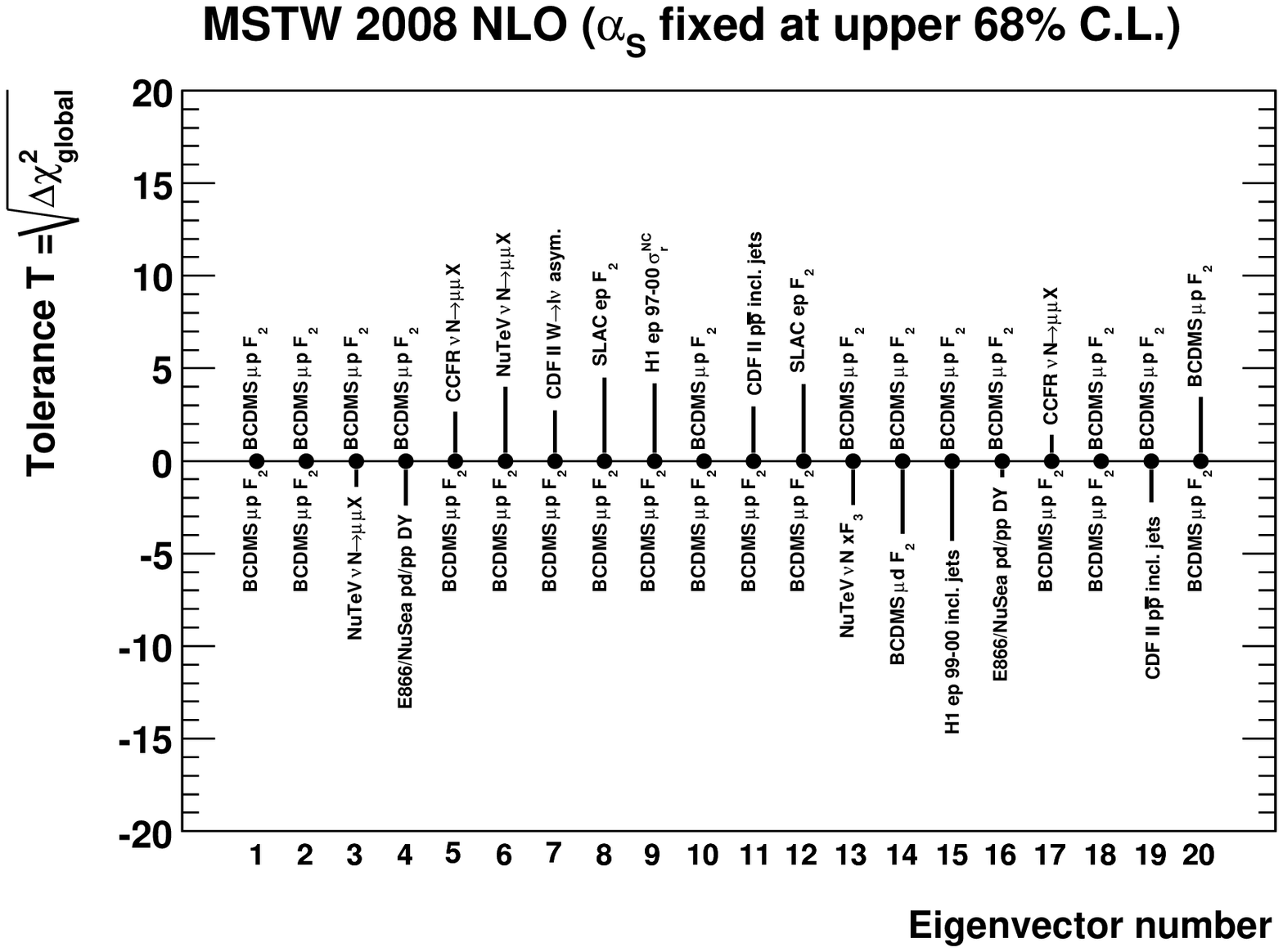}
  \end{minipage}
  (b)\\
  \begin{minipage}{\textwidth}
    \centering
    \includegraphics[width=0.87\textwidth]{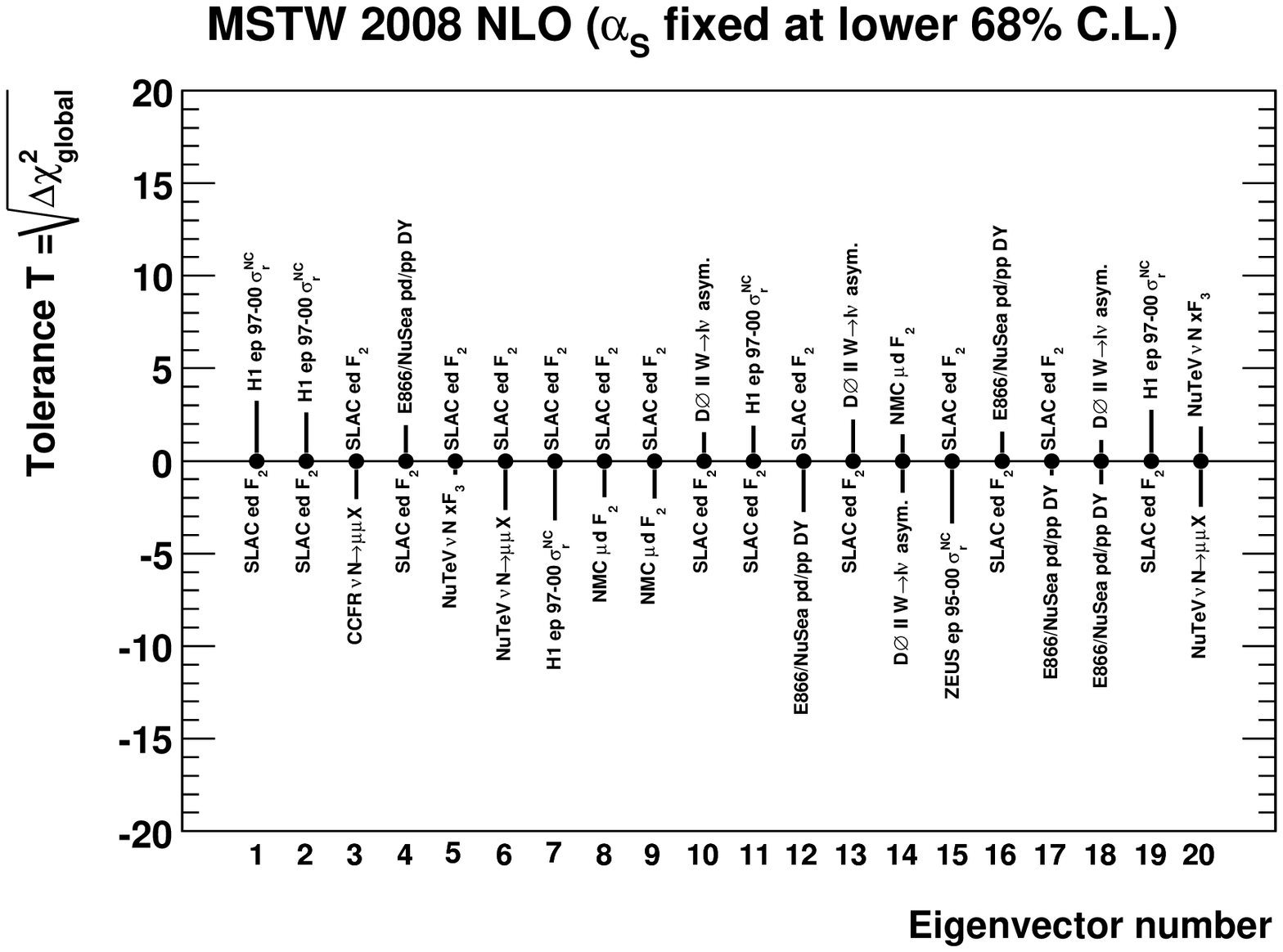}
  \end{minipage}
  \caption{Tolerance values for each eigenvector when $\alpha_S(M_Z^2)$ is fixed to its (a)~upper and (b)~lower 1-$\sigma$ limits in the NLO fit.  The text labels indicate the name of the data set which sets the tolerance constraint on each eigenvector direction.}
  \label{fig:tolerance}
\end{figure}
The tolerance values for each of the eigenvectors, when $\alpha_S(M_Z^2)$ is fixed at each of its 1-$\sigma$ limits in the NLO fit, are shown in Fig.~\ref{fig:tolerance}.  The examples given for eigenvector number 6 were typical.  However, two anomalous situations are seen to arise.  Firstly, when $\alpha_S(M_Z^2)$ is fixed at its upper 1-$\sigma$ limit, then the $\chi_n^2$ for the BCDMS $F_2^{\mu p}$ data when moving along eigenvector numbers 1, 2 and 10 has a minimum (or is almost flat) at zero distance, so the tolerance is zero in both the positive and negative directions.  Secondly, when $\alpha_S(M_Z^2)$ is fixed at its lower 1-$\sigma$ limit, then the $\chi_n^2$ for the SLAC $F_2^{ed}$ data when moving along eigenvector numbers 14, 18 and 20 has a maximum at zero distance.  The tolerance is then non-zero in both the positive and negative directions, since the constraint is provided by data sets other than the SLAC $F_2^{ed}$ data.  Of course, in these two anomalous situations the minima/maxima in $\chi_n^2$ will not occur at \emph{exactly} zero distance, and this is an artifact of working in discrete distance units of $(\Delta\chi^2_{\rm global})^{1/2}$, which is sufficiently accurate for our purposes.

\begin{figure}
  (a)\hspace{0.5\textwidth}(b)\\
  \includegraphics[width=0.5\textwidth]{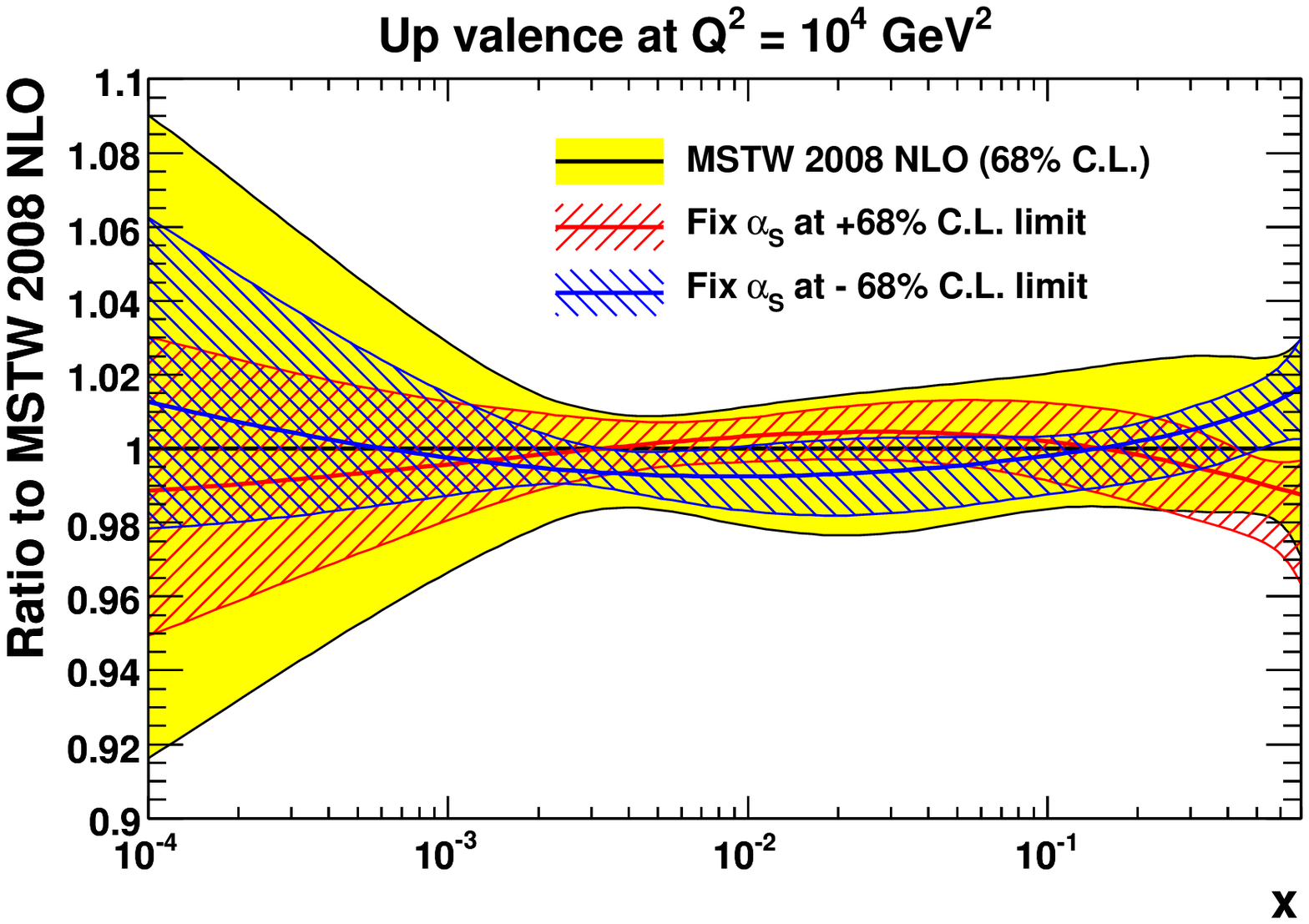}%
  \includegraphics[width=0.5\textwidth]{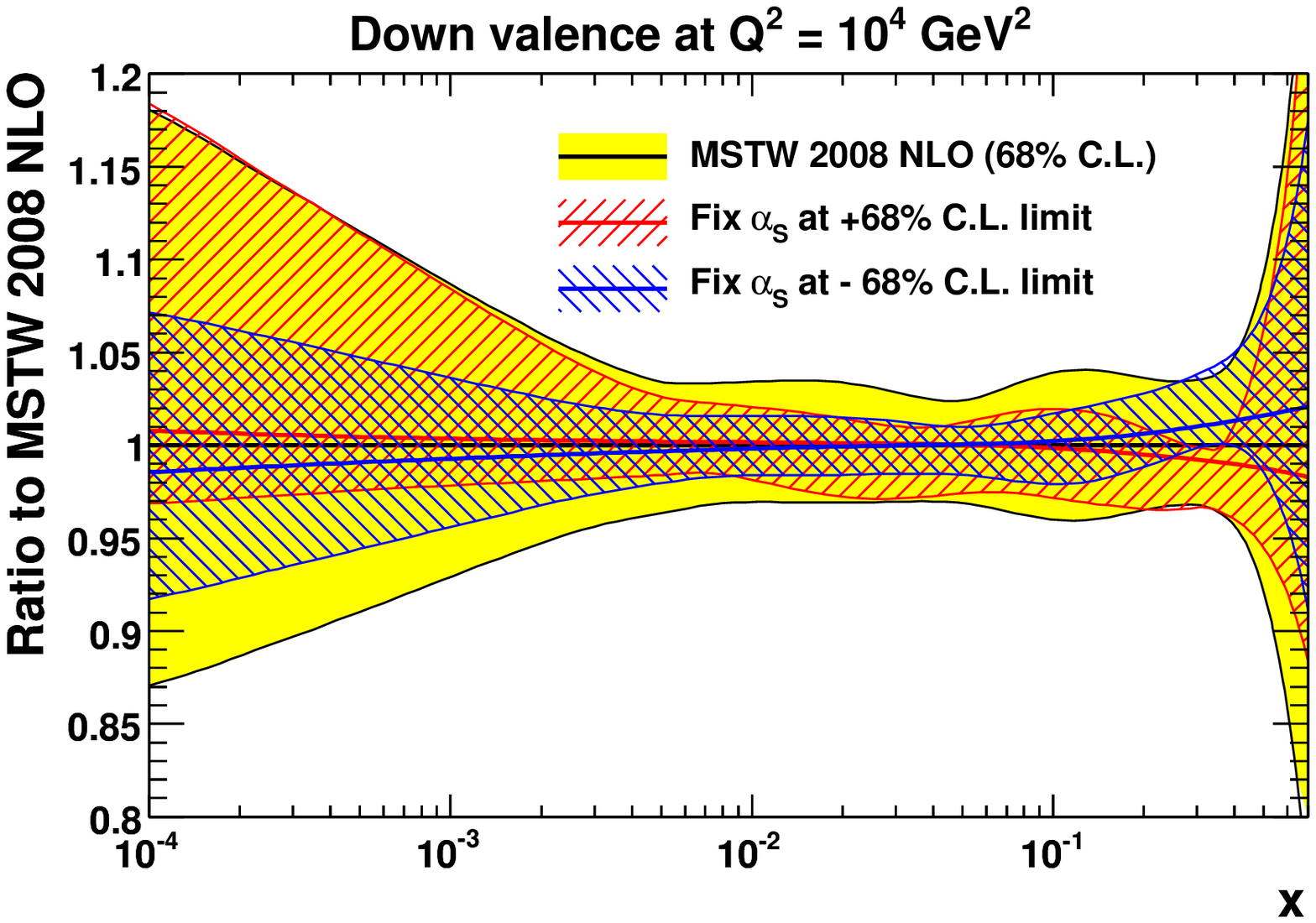}\\
  (c)\hspace{0.5\textwidth}(d)\\
  \includegraphics[width=0.5\textwidth]{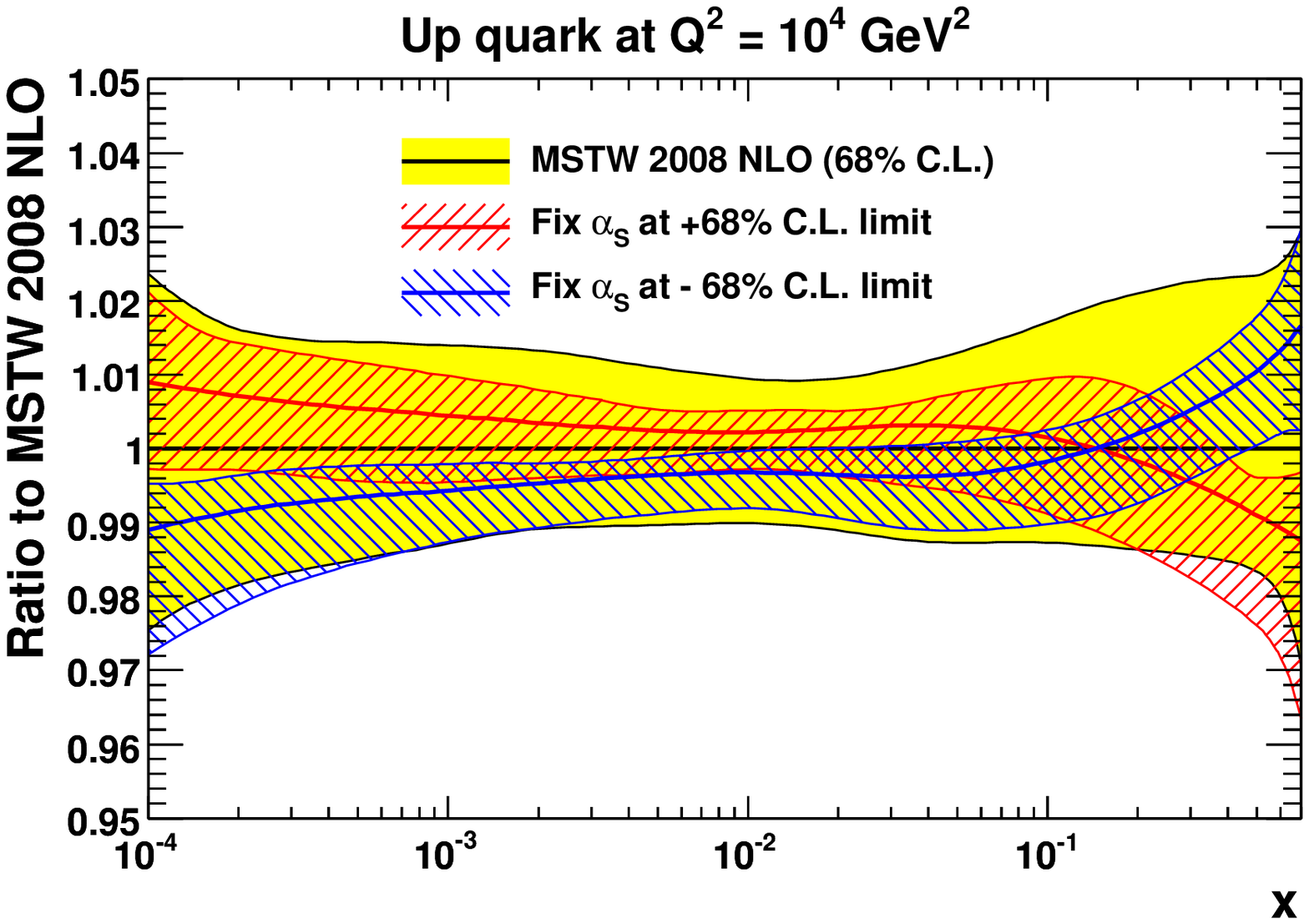}%
  \includegraphics[width=0.5\textwidth]{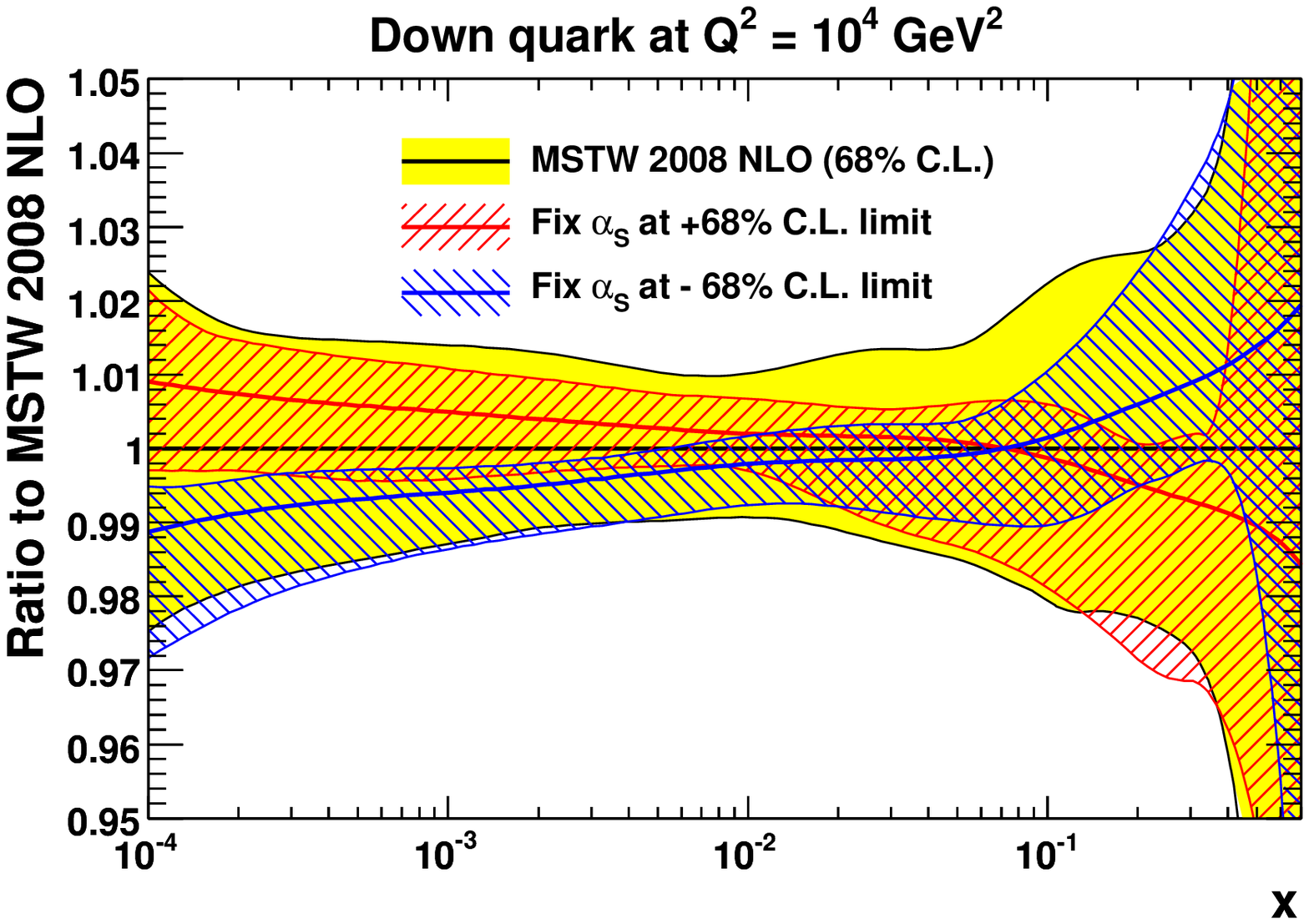}\\
  (e)\hspace{0.5\textwidth}(f)\\
  \includegraphics[width=0.5\textwidth]{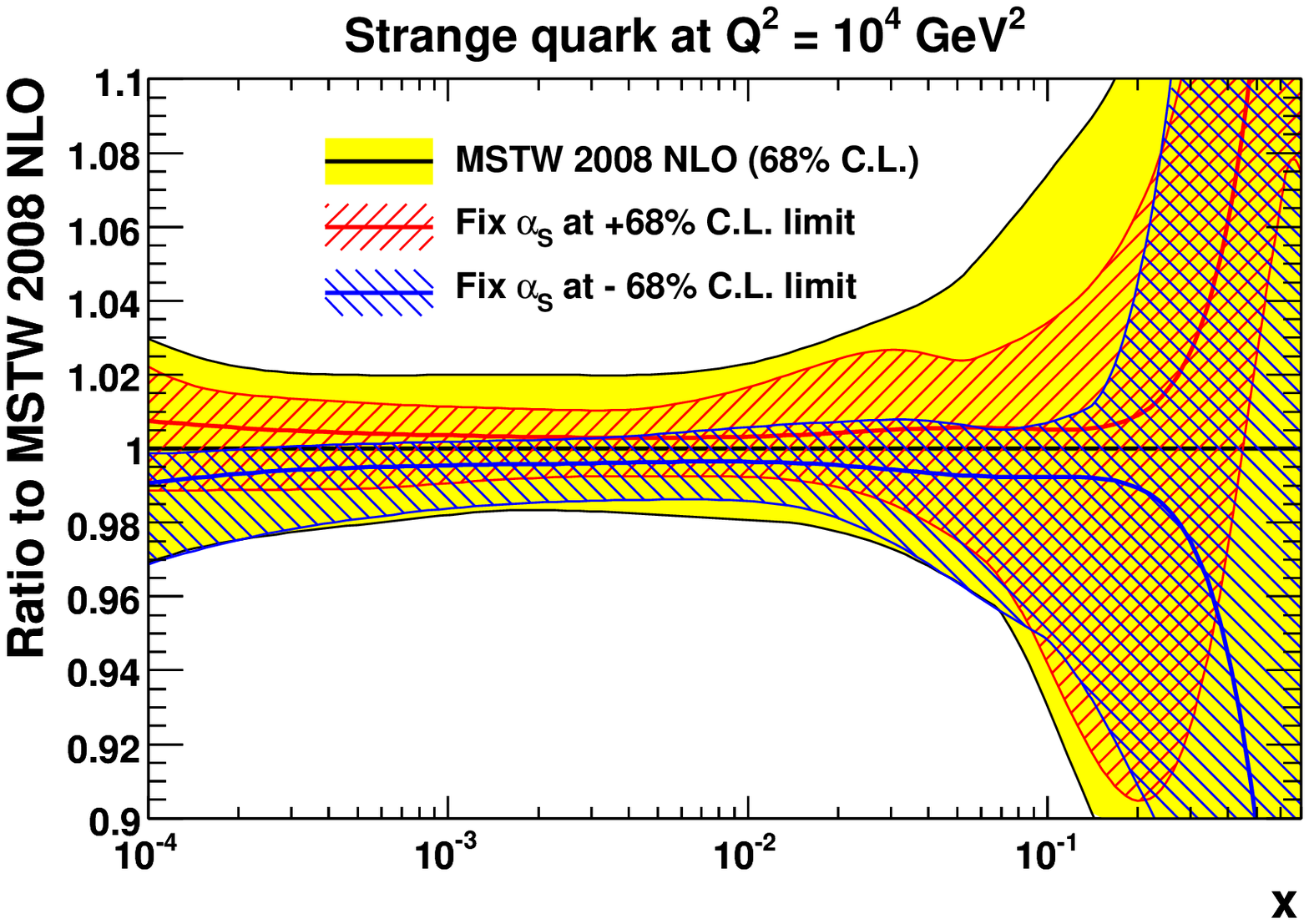}%
  \includegraphics[width=0.5\textwidth]{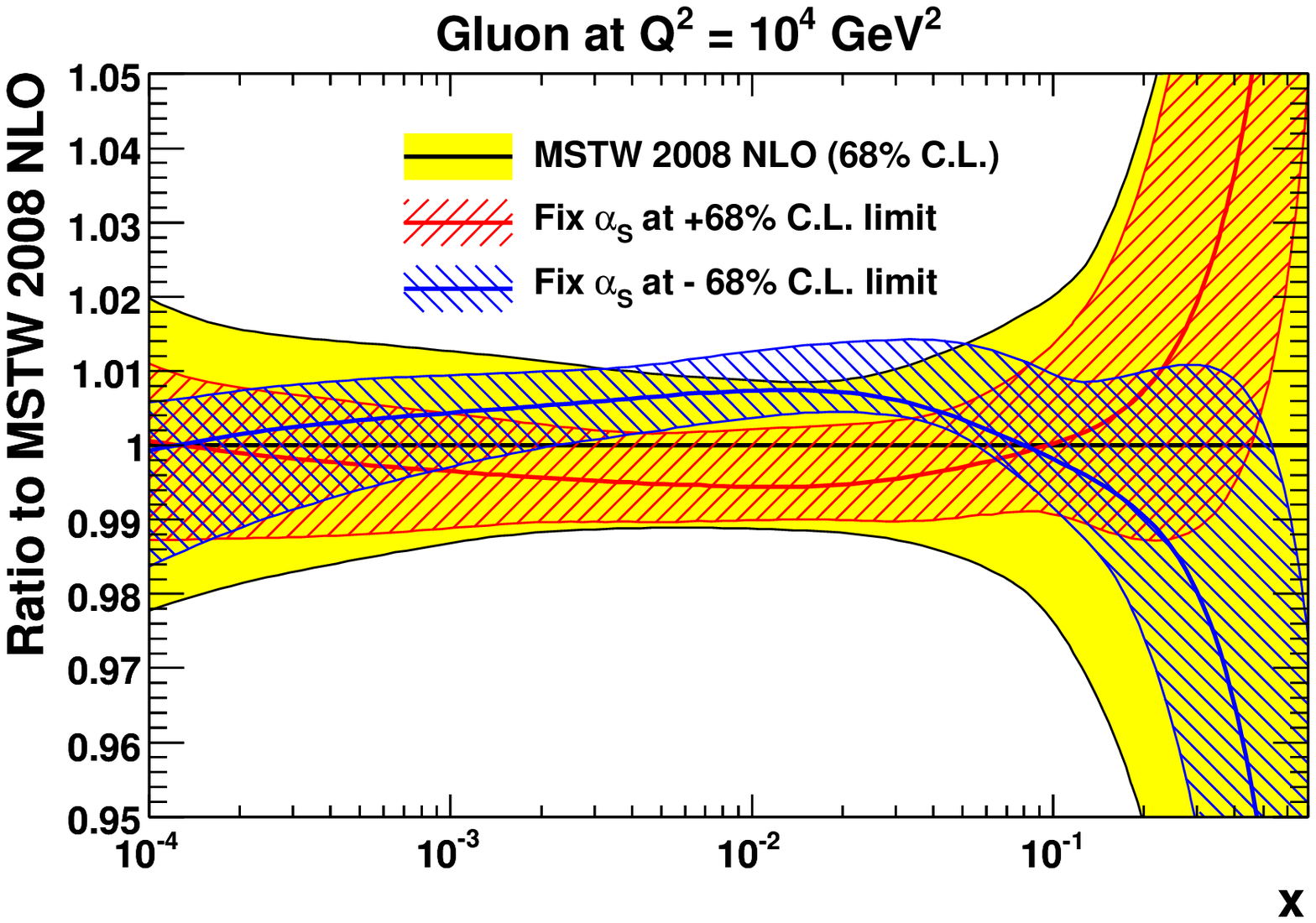}
  \caption{NLO parton distributions at $Q^2=10^4$ GeV$^2$, including the 1-$\sigma$ PDF uncertainty bands, with $\alpha_S$ fixed at either the best-fit value or at each of its 1-$\sigma$ limits.}
  \label{fig:pdferror1}
\end{figure}
Comparing the tolerance plots in Fig.~\ref{fig:tolerance} with those in Fig.~10 of Ref.~\cite{Martin:2009iq}, we see that the PDF uncertainties will be much smaller (and more asymmetric) when $\alpha_S(M_Z^2)$ is fixed at each of its 1-$\sigma$ limits, compared to when it is fixed at the best-fit value.  This statement is confirmed in Fig.~\ref{fig:pdferror1} where we show the PDF uncertainties at $Q^2 = 10^4$ GeV$^2$ for the cases where $\alpha_S(M_Z^2)$ is fixed at the best-fit value or shifted to each of its 1-$\sigma$ limits.  Note that in all cases the uncertainty bands of the PDFs when $\alpha_S$ is at its 1-$\sigma$ limits are at most only slightly outside those for the PDFs when $\alpha_S$ is at its best-fit value.

There are a number of interesting features to note from Fig.~\ref{fig:pdferror1} about the manner in which the central values of the PDFs change as a function of $\alpha_S$.  At high $x$ the valence quarks (and the total up and down quark distributions) are anticorrelated with $\alpha_S$.  This occurs for two reasons.  Firstly, the higher-order coefficient functions for structure functions in deep-inelastic scattering are positive at high $x$.  Increasing $\alpha_S$ therefore means that we increase this contribution and hence require fewer quarks to fit the fixed-target structure function data at relatively low $Q^2$.  Secondly, the increased speed of evolution with larger $\alpha_S$ results in more migration from high $x$ to lower $x$ values.  In absolute terms the effect is similar for up and down quarks, but the greater precision on up quarks means that the proportional effect is greater, and for $x\gtrsim 0.5$ the central value of the default up quark distribution is outside the error bands of the distributions generated with $\alpha_S$ fixed at its 1-$\sigma$ limits.

Another interesting feature, seen in Fig.~\ref{fig:pdferror1}(f), is the confirmation of the anticorrelation between the small-$x$ gluon and $\alpha_S$.  This is seen for $x$ between $10^{-4}$ and $0.1$ at $Q^2=10^4$ GeV$^2$, and is a consequence of maintaining the fit quality to the small-$x$ HERA data, i.e.~the values of $\partial F_2/\partial\ln Q^2\sim \alpha_S\,g$.  From the momentum sum rule this results in a positive correlation of the high-$x$ gluon and $\alpha_S$.  Note that there is some asymmetry in the deviation.  We will return to this point in the next section.  Since the quark distributions at small $x$ are driven by the gluon, this change in the gluon affects the quarks.  However, we note that there is in fact a slight \emph{correlation} between the small-$x$ quark distributions and $\alpha_S$, showing that the increase in evolution from the increased coupling slightly outweighs the effect of the decreased small-$x$ gluon distribution.
\begin{figure}
  (a)\hspace{0.5\textwidth}(b)\\
  \includegraphics[width=0.5\textwidth]{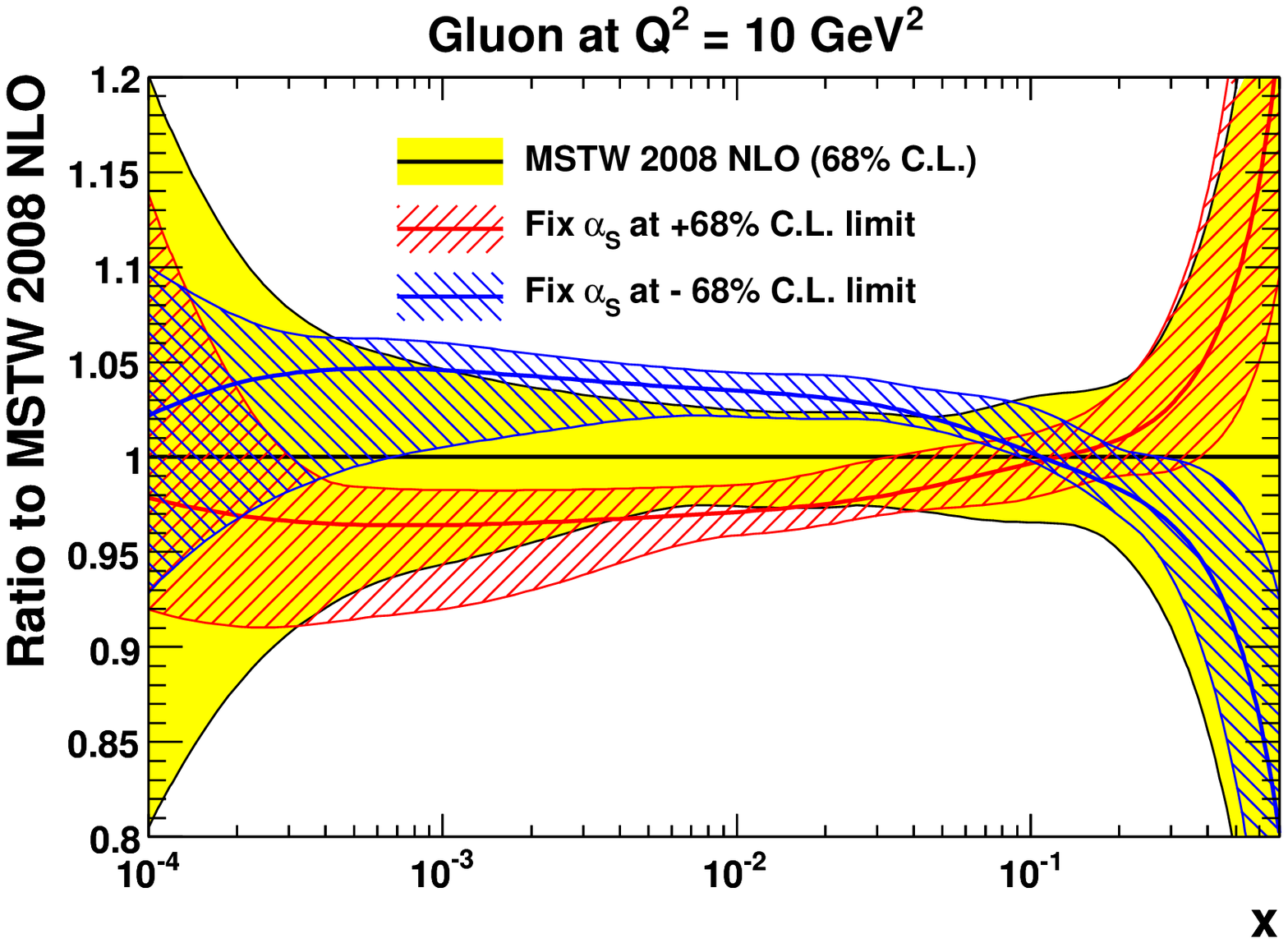}
  \includegraphics[width=0.5\textwidth]{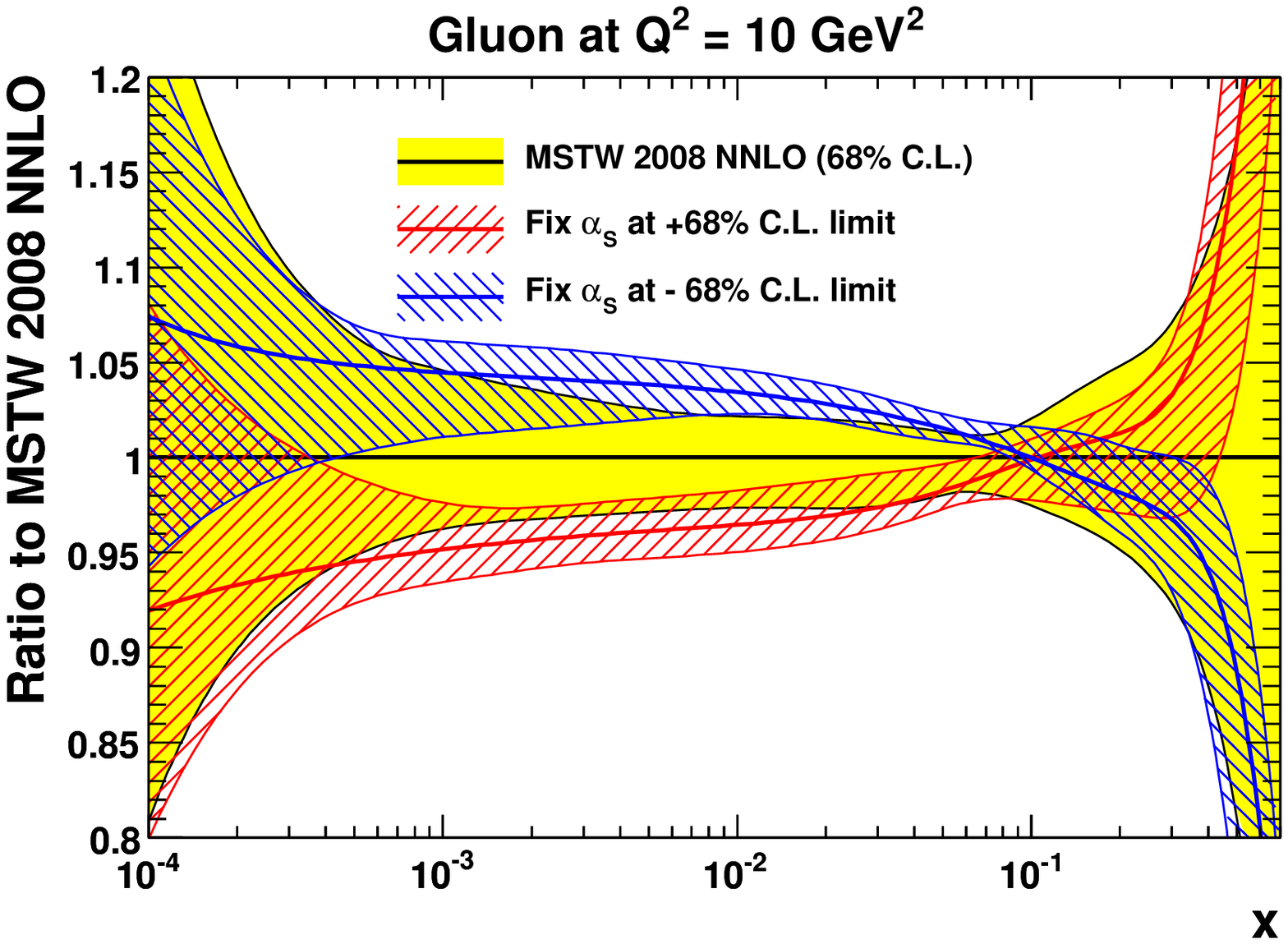}
  \caption{(a)~NLO and (b)~NNLO gluon distribution at $Q^2 = 10$ GeV$^2$, including the 1-$\sigma$ PDF uncertainty bands, with $\alpha_S$ fixed at either the best-fit value or at each of its 1-$\sigma$ limits.}
  \label{fig:gluon10}
\end{figure}
In Fig.~\ref{fig:gluon10} we show similar plots for the gluon distribution at $Q^2 = 10$ GeV$^2$ in both the NLO and NNLO fits.  One can see that, in relative terms, the PDFs at the 1-$\sigma$ limits for $\alpha_S$ are further from the best-fit values at $Q^2 = 10$ GeV$^2$ than at $Q^2 = 10^4$ GeV$^2$, although the error bands always overlap.  This just illustrates that DGLAP evolution drives PDFs together at asymptotic values of $Q^2$.

\section{Implications for cross section calculations} \label{sec:cross}
Each fixed value of $\alpha_S$ is associated with a central PDF set $S_0$ and $2n$ eigenvector PDF sets $S_k^\pm$ defined by Eq.~\eqref{eq:skpm}, where $k=1,\ldots,n$ and $n=20$ (corresponding to 20 input PDF parameters).  An observable PDF-dependent quantity $F$, such as a hadronic cross section, calculated using a particular value of $\alpha_S$, has a central value $F^{\alpha_S}(S_0)$ and asymmetric PDF uncertainties given using the Hessian method by~\cite{Pumplin:2001ct,Nadolsky:2001yg,Martin:2002aw,Campbell:2006wx,Martin:2009iq}
\begin{align}
  (\Delta F_{\textsc{pdf}}^{\alpha_S})_+ &= \sqrt{\sum_{k=1}^n \left\{\max\left[\;F^{\alpha_S}(S_k^+)-F^{\alpha_S}(S_0),\;F^{\alpha_S}(S_k^-)-F^{\alpha_S}(S_0),\;0\right]\right\}^2}, \label{eq:Fp} \\
  (\Delta F_{\textsc{pdf}}^{\alpha_S})_- &= \sqrt{\sum_{k=1}^n \left\{\max\left[\;F^{\alpha_S}(S_0)-F^{\alpha_S}(S_k^+),\;F^{\alpha_S}(S_0)-F^{\alpha_S}(S_k^-),\;0\right]\right\}^2}, \label{eq:Fm}
\end{align}
for a fixed value of $\alpha_S$.

How should this prescription be generalised to calculate an overall ``PDF+$\alpha_S$'' uncertainty on an observable $F$, i.e.~accounting for the additional uncertainty on $F$ due to the uncertainty on $\alpha_S$?  Ideally, we would vary $\alpha_S$ continuously within its experimental uncertainty determined by the global fit.  Each value of $\alpha_S$ would give a central value $F^{\alpha_S}(S_0)$ with PDF uncertainties given by Eqs.~\eqref{eq:Fp} and \eqref{eq:Fm}.  The overall best-fit prediction is then $F^{\alpha_S^0}(S_0)$, where $\alpha_S^0$ is the best-fit $\alpha_S$ value, and the overall ``PDF+$\alpha_S$'' uncertainties are given by the spread in the predictions, including the PDF uncertainty, for each $\alpha_S$ value.  More formally, the ``PDF+$\alpha_S$'' uncertainties are given by
\begin{align}
  (\Delta F_{\textsc{pdf}+\alpha_S})_+ &= \max_{\alpha_S}\left(\left\{F^{\alpha_S}(S_0)+(\Delta F_{\textsc{pdf}}^{\alpha_S})_+\right\}\right) - F^{\alpha_S^0}(S_0), \label{eq:Fpa} \\
  (\Delta F_{\textsc{pdf}+\alpha_S})_- &= F^{\alpha_S^0}(S_0) - \min_{\alpha_S}\left(\left\{F^{\alpha_S}(S_0)-(\Delta F_{\textsc{pdf}}^{\alpha_S})_-\right\}\right), \label{eq:Fma}
\end{align}
where the maximum and minimum are calculated when $\alpha_S$ is varied continuously within, for example, the range $[\alpha_S^0-1\sigma,\;\alpha_S^0+1\sigma]$ to determine the 1-$\sigma$ ``PDF+$\alpha_S$'' uncertainties, and the PDF uncertainties are given by Eqs.~\eqref{eq:Fp} and \eqref{eq:Fm} for each value of $\alpha_S$ in this range.

First suppose that $\alpha_S$ is varied within a given range for the \emph{same} central PDFs and their uncertainties and that $F$ is a monotonic function of $\alpha_S$ (which is usually the case).  Then the extreme values of the observable $F$ would obviously occur when $\alpha_S$ is at either of its limits.  However, in practice, the PDF uncertainty decreases as $\alpha_S$ gets further from its best-fit value.  Moreover, correlations between the relevant PDFs and $\alpha_S$ can enhance the $\alpha_S$ dependence of $F$, while anticorrelations will reduce it.  Therefore, the extreme values of the observable $F$ could, in principle, occur at any intermediate value within a given range of $\alpha_S$.  For reasons of economy, we provide PDF sets with uncertainties only for five different fixed values of $\alpha_S$ (i.e.~the best-fit $\alpha_S$, $\alpha_S$ fixed at the two limits, and $\alpha_S$ at half these two limits).  In fact we find that in the majority of the processes we consider, the extreme values of the observable $F$ come from $\alpha_S$ fixed at either the best-fit value or one of the two limits.  Even in the rare cases where the extreme values of $F$ come from $\alpha_S$ fixed at half the limit, omitting these intermediate $\alpha_S$ values would not significantly reduce the overall ``PDF+$\alpha_S$'' uncertainty.  Hence we do not consider it necessary to provide PDF uncertainty sets at further intermediate $\alpha_S$ values.

Since this prescription might seem quite complicated at first sight, we will give a few concrete examples of its application and consequences in the following subsections.\footnote{We also provide a Fortran example program at Ref.~\cite{mstwpdf}.}

\subsection{$W$ and $Z$ total cross sections}
\begin{figure}
  (a)\\
  \begin{minipage}{\textwidth}
    \centering
    \includegraphics[width=0.8\textwidth]{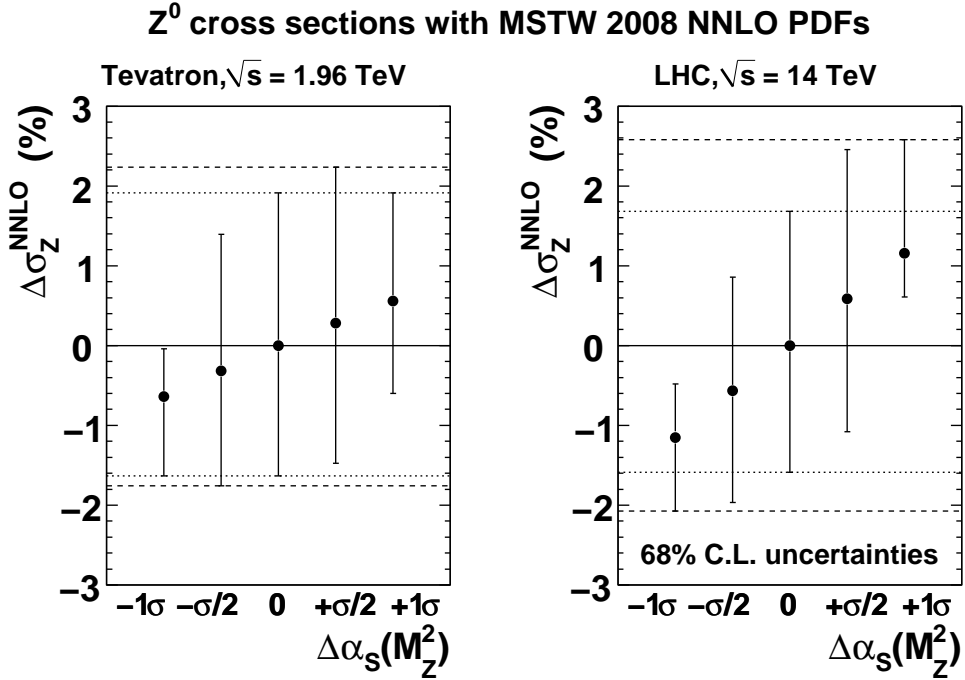}
  \end{minipage}
  (b)\\
  \begin{minipage}{\textwidth}
    \centering
    \includegraphics[width=0.7\textwidth]{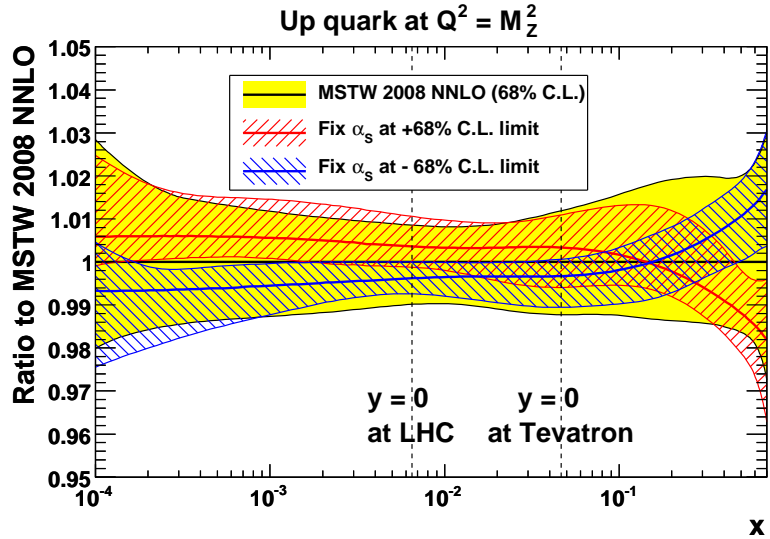}
  \end{minipage}
  \caption{(a)~PDF uncertainties on the $Z$ total cross sections for each of the five PDF sets obtained from global fits performed with different fixed values of $\alpha_S(M_Z^2)$ around the best-fit value $\Delta\alpha_S(M_Z^2)=0$.  The results are shown as the percentage difference from the overall best-fit value.  The horizontal dotted (dashed) lines indicate the PDF (PDF+$\alpha_S$) 68\% C.L.~percentage uncertainty on $\sigma_Z$.  The $W^{\pm}$ case is very similar.  (b)~NNLO up quark distribution at $Q^2=M_Z^2$.  The values of $x=M_Z/\sqrt{s}$ relevant for central production (assuming $p_T^Z=0$) at the Tevatron and LHC are indicated.}
  \label{fig:ztot_asmz}
\end{figure}
In Fig.~\ref{fig:ztot_asmz}(a) we show the PDF uncertainties on the $Z$ total cross section at the Tevatron and LHC for each of the five sets with different fixed $\alpha_S(M_Z^2)$ values.  (The situation is similar for the $W^\pm$ cross sections.)  The cross sections are calculated as described in Section~15 of Ref.~\cite{Martin:2009iq}, e.g.~using the PDG 2008~\cite{Amsler:2008zzb} electroweak parameters.  The increase of the $Z$ cross section with increasing $\alpha_S(M_Z^2)$ is due to a combination of two effects.  Firstly, there is the effect of the $\alpha_S$ dependence in the (positive) higher-order corrections to the partonic cross section.  Secondly, there is the $\alpha_S$ dependence of the (predominantly) quark distributions.  In Fig.~\ref{fig:ztot_asmz}(b) we show the up quark distribution at $Q^2 = M_Z^2$.  The momentum fractions $x=M_Z/\sqrt{s}$ (assuming $p_T^Z=0$, i.e.~LO kinematics) probed at the Tevatron and LHC at central rapidity ($y=0$) are indicated.  As previously noted, the small-$x$ up quark distribution is slightly correlated with the value of $\alpha_S$ (see also Fig.~\ref{fig:pdferror1} for other parton flavours).  However, when integrating over rapidity to obtain the total cross section, the PDFs will also be sampled at larger (and smaller) values of $x$.  From Fig.~\ref{fig:ztot_asmz}(b), the up quark distribution is \emph{anti}correlated with $\alpha_S$ in the large $x$ region, which will be sampled more at the Tevatron than at the LHC, effectively cancelling out some of the correlation with $\alpha_S$ arising from the sampling of smaller $x$ values.  This explains why there is less dependence on $\alpha_S$ at the Tevatron compared to the LHC in Fig.~\ref{fig:ztot_asmz}(a).

We also indicate in Fig.~\ref{fig:ztot_asmz}(a) how the spread of the five individual predictions can be used to give an overall uncertainty, which is larger than that obtained when $\alpha_S(M_Z^2)$ is fixed at the global best-fit value (as is usually done).  Note that the extreme values for the $Z$ cross section at the Tevatron arise when $\alpha_S(M_Z^2)$ is shifted to half its 1-$\sigma$ limit.  In general, there exists a $X\in[0,1]$ such that the extreme value of a given observable (or a PDF itself) occurs when $\alpha_S(M_Z^2)$ is shifted by $X$-$\sigma$ from the best-fit value.  Often, perhaps in most cases, $X=1$, but in other cases $X<1$, as we have here for $Z$ production at the Tevatron.  However, we see from Fig.~\ref{fig:ztot_asmz}(a) that the extreme values of the $\pm1\sigma$ results are not far from the extreme values of the $\pm\sigma/2$ results.

Note that at the LHC there is an asymmetry in the extra uncertainty generated by allowing $\alpha_S$ to vary, with more increase in the upwards direction than downwards.  This is seen to be a consequence of a similar asymmetry in the PDF uncertainty of the up quark distribution when $\alpha_S$ is fixed at its upper limit, shown in Fig.~\ref{fig:ztot_asmz}(b), with the PDF uncertainty in the relevant $x$ region giving more freedom for upwards movement compared to the best fit than downwards.  This is due to the fact that the HERA structure function data at small $x$ would prefer a little more evolution than in the best global fit at NNLO.  The HERA data therefore allow more freedom for extra evolution when $\alpha_S$ increases than for reduced evolution when $\alpha_S$ decreases.\footnote{This effect is obscured slightly in the NLO quark distribution plots shown in Fig.~\ref{fig:pdferror1}, because the $\pm1\sigma$ limits for $\alpha_S(M_Z^2)$ itself, mainly determined by fixed-target data, are asymmetric at NLO and larger in the downwards direction, see Eq.~\eqref{eq:asmznlo}, but are practically symmetric at NNLO, see Eq.~\eqref{eq:asmznnlo}.}

\begin{table}
  \centering
  \begin{tabular}{l|c|c|c}
    \hline\hline
    Tevatron, $\sqrt{s} = 1.96$ TeV & $B_{\ell\nu} \cdot \sigma_W$ (nb) & $B_{\ell^+\ell^-}\cdot\sigma_Z$ (nb) & $R_{WZ}$ \\
    \hline
    NNLO (PDF unc.~only) & $2.747^{+0.049}_{-0.042}$ $\left(^{+1.8\%}_{-1.5\%}\right)$ & $0.2507^{+0.0048}_{-0.0041}$ $\left(^{+1.9\%}_{-1.6\%}\right)$ & $10.96^{+0.03}_{-0.03}$ $\left(^{+0.2\%}_{-0.2\%}\right)$ \\
    NNLO (PDF+$\alpha_S$ unc.) & $2.747^{+0.060}_{-0.047}$ $\left(^{+2.2\%}_{-1.7\%}\right)$ & $0.2507^{+0.0056}_{-0.0044}$ $\left(^{+2.2\%}_{-1.8\%}\right)$ & $10.96^{+0.03}_{-0.03}$ $\left(^{+0.3\%}_{-0.3\%}\right)$ \\
    \hline\hline\multicolumn{4}{c}{}\\\hline\hline
    LHC, $\sqrt{s} = 7$ TeV & $B_{\ell\nu} \cdot \sigma_W$ (nb) & $B_{\ell^+\ell^-}\cdot\sigma_Z$ (nb) & $R_{WZ}$ \\
    \hline
    NNLO (PDF unc.~only) & $10.47^{+0.18}_{-0.17}$ $\left(^{+1.7\%}_{-1.6\%}\right)$ & $0.958^{+0.017}_{-0.015}$ $\left(^{+1.7\%}_{-1.5\%}\right)$ & $10.92^{+0.02}_{-0.02}$ $\left(^{+0.2\%}_{-0.2\%}\right)$ \\
    NNLO (PDF+$\alpha_S$ unc.) & $10.47^{+0.27}_{-0.20}$ $\left(^{+2.5\%}_{-1.9\%}\right)$ & $0.958^{+0.024}_{-0.018}$ $\left(^{+2.5\%}_{-1.9\%}\right)$ & $10.92^{+0.03}_{-0.02}$ $\left(^{+0.3\%}_{-0.2\%}\right)$ \\
    \hline\hline\multicolumn{4}{c}{}\\\hline\hline
    LHC, $\sqrt{s} = 10$ TeV & $B_{\ell\nu} \cdot \sigma_W$ (nb) & $B_{\ell^+\ell^-}\cdot\sigma_Z$ (nb) & $R_{WZ}$ \\
    \hline
    NNLO (PDF unc.~only) & $15.35^{+0.26}_{-0.25}$ $\left(^{+1.7\%}_{-1.6\%}\right)$ & $1.429^{+0.024}_{-0.022}$ $\left(^{+1.7\%}_{-1.6\%}\right)$ & $10.74^{+0.02}_{-0.02}$ $\left(^{+0.2\%}_{-0.2\%}\right)$ \\
    NNLO (PDF+$\alpha_S$ unc.) & $15.35^{+0.39}_{-0.31}$ $\left(^{+2.6\%}_{-2.0\%}\right)$ & $1.429^{+0.037}_{-0.027}$ $\left(^{+2.6\%}_{-1.9\%}\right)$ & $10.74^{+0.03}_{-0.03}$ $\left(^{+0.3\%}_{-0.3\%}\right)$ \\
    \hline\hline\multicolumn{4}{c}{}\\\hline\hline
    LHC, $\sqrt{s} = 14$ TeV & $B_{\ell\nu} \cdot \sigma_W$ (nb) & $B_{\ell^+\ell^-}\cdot\sigma_Z$ (nb) & $R_{WZ}$ \\
    \hline
    NNLO (PDF unc.~only) & $21.72^{+0.36}_{-0.36}$ $\left(^{+1.7\%}_{-1.7\%}\right)$ & $2.051^{+0.035}_{-0.033}$ $\left(^{+1.7\%}_{-1.6\%}\right)$ & $10.59^{+0.02}_{-0.03}$ $\left(^{+0.2\%}_{-0.3\%}\right)$ \\
    NNLO (PDF+$\alpha_S$ unc.) & $21.72^{+0.56}_{-0.48}$ $\left(^{+2.6\%}_{-2.2\%}\right)$ & $2.051^{+0.053}_{-0.043}$ $\left(^{+2.6\%}_{-2.1\%}\right)$ & $10.59^{+0.03}_{-0.03}$ $\left(^{+0.3\%}_{-0.3\%}\right)$ \\
    \hline\hline
  \end{tabular}
  \caption{Predictions for $W\equiv W^++W^-$ and $Z$ total cross sections at the Tevatron and LHC, and their ratio $R_{WZ}$, with PDF uncertainties only~\cite{Martin:2009iq} and with the combined ``PDF+$\alpha_S$'' uncertainty.  The 68\% C.L.~uncertainties are given in all cases.  We take $\mu_R=\mu_F=M_{W,Z}$.}
  \label{tab:wztot}
\end{table}
\begin{table}
  \centering
  \begin{tabular}{l|c|c|c}
    \hline\hline
    LHC, $\sqrt{s} = 7$ TeV & $B_{\ell\nu} \cdot \sigma_{W^+}$ (nb) & $B_{\ell\nu}\cdot\sigma_{W^-}$ (nb) & $R_\pm$ \\
    \hline
    NNLO (PDF unc.~only) & $6.16^{+0.11}_{-0.10}$ $\left(^{+1.8\%}_{-1.6\%}\right)$ & $4.31^{+0.08}_{-0.07}$ $\left(^{+1.8\%}_{-1.6\%}\right)$ & $1.429^{+0.013}_{-0.012}$ $\left(^{+0.9\%}_{-0.8\%}\right)$ \\
    NNLO (PDF+$\alpha_S$ unc.) & $6.16^{+0.16}_{-0.12}$ $\left(^{+2.6\%}_{-2.0\%}\right)$ & $4.31^{+0.11}_{-0.08}$ $\left(^{+2.5\%}_{-2.0\%}\right)$ & $1.429^{+0.015}_{-0.012}$ $\left(^{+1.1\%}_{-0.8\%}\right)$ \\
    \hline\hline\multicolumn{4}{c}{}\\\hline\hline
    LHC, $\sqrt{s} = 10$ TeV & $B_{\ell\nu} \cdot \sigma_{W^+}$ (nb) & $B_{\ell\nu}\cdot\sigma_{W^-}$ (nb) & $R_\pm$ \\
    \hline
    NNLO (PDF unc.~only) & $8.88^{+0.15}_{-0.15}$ $\left(^{+1.7\%}_{-1.6\%}\right)$ & $6.47^{+0.11}_{-0.11}$ $\left(^{+1.7\%}_{-1.6\%}\right)$ & $1.373^{+0.012}_{-0.010}$ $\left(^{+0.8\%}_{-0.7\%}\right)$ \\
    NNLO (PDF+$\alpha_S$ unc.) & $8.88^{+0.23}_{-0.19}$ $\left(^{+2.6\%}_{-2.1\%}\right)$ & $6.47^{+0.16}_{-0.13}$ $\left(^{+2.5\%}_{-2.0\%}\right)$ & $1.373^{+0.013}_{-0.010}$ $\left(^{+0.9\%}_{-0.7\%}\right)$ \\
    \hline\hline\multicolumn{4}{c}{}\\\hline\hline
    LHC, $\sqrt{s} = 14$ TeV & $B_{\ell\nu} \cdot \sigma_{W^+}$ (nb) & $B_{\ell\nu}\cdot\sigma_{W^-}$ (nb) & $R_\pm$ \\
    \hline
    NNLO (PDF unc.~only) & $12.39^{+0.22}_{-0.21}$ $\left(^{+1.8\%}_{-1.7\%}\right)$ & $9.33^{+0.16}_{-0.16}$ $\left(^{+1.7\%}_{-1.7\%}\right)$ & $1.328^{+0.011}_{-0.009}$ $\left(^{+0.8\%}_{-0.7\%}\right)$ \\
    NNLO (PDF+$\alpha_S$ unc.) & $12.39^{+0.32}_{-0.28}$ $\left(^{+2.6\%}_{-2.3\%}\right)$ & $9.33^{+0.24}_{-0.20}$ $\left(^{+2.6\%}_{-2.1\%}\right)$ & $1.328^{+0.011}_{-0.009}$ $\left(^{+0.9\%}_{-0.7\%}\right)$ \\
    \hline\hline
  \end{tabular}
  \caption{Predictions for $W^+$ and $W^-$ total cross sections at the LHC, and their ratio $R_\pm$, with PDF uncertainties only~\cite{Martin:2009iq} and with the combined ``PDF+$\alpha_S$'' uncertainty.  The 68\% C.L.~uncertainties are given in all cases.  We take $\mu_R=\mu_F=M_W$.}
  \label{tab:w+w-tot}
\end{table}
In Tables \ref{tab:wztot} and \ref{tab:w+w-tot} we update the NNLO predictions for $W^\pm$ and $Z$ production at the Tevatron and LHC that were given in Ref.~\cite{Martin:2009iq}, to allow for the enlarged uncertainty when the variation of $\alpha_S$ is taken into account.\footnote{In addition to the cross sections presented in Ref.~\cite{Martin:2009iq}, we give results for $\sqrt{s} = 7$~TeV at the LHC in light of the recently announced centre-of-mass energy at start-up~\cite{CERNnews}.}  For the NNLO total cross sections, we see that the combined ``PDF+$\alpha_S$'' uncertainty is about $2\%$ at the Tevatron and $2.5\%$ at the LHC.  This is larger than the estimate of the theoretical uncertainty obtained from scale variation which gives below $1\%$~\cite{Anastasiou:2003ds}.  However, at the LHC, the sensitivity to parton distributions at $x$ values $\lesssim 10^{-3}$ could lead to additional uncertainties which are difficult to estimate, but variations of PDFs from resummed fits~\cite{White:2006yh} and contributions from small-$x$ resummation to the Drell--Yan cross section~\cite{Marzani:2008uh}, suggest that a few percent is quite possible.

\subsection{Higgs boson total cross sections}
\begin{figure}
  (a)\\
  \begin{minipage}{\textwidth}
    \centering
    \includegraphics[width=0.8\textwidth]{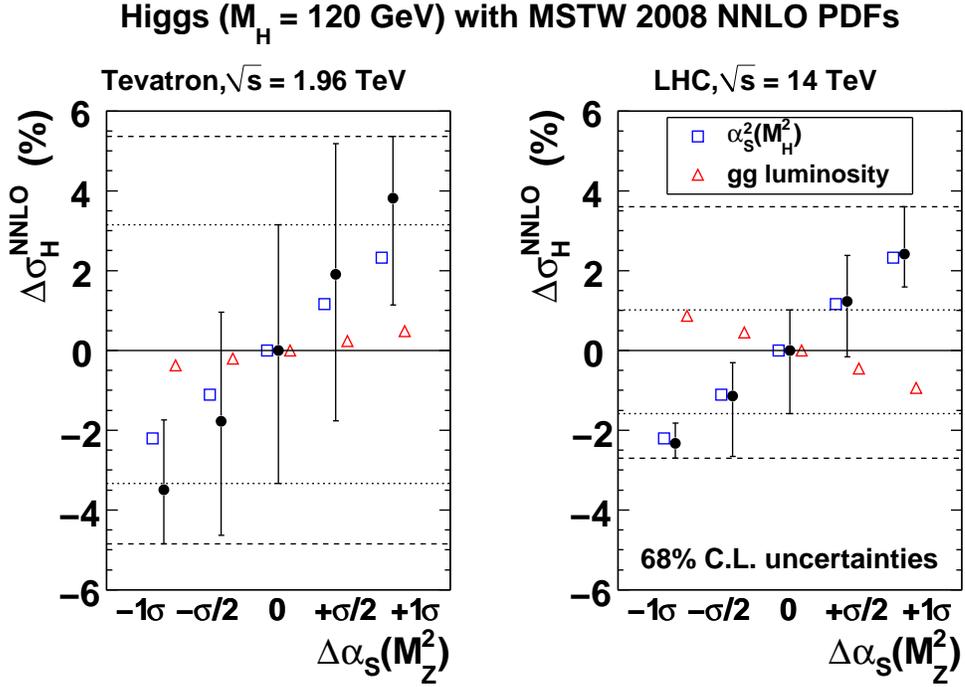}
  \end{minipage}
  (b)\\
  \begin{minipage}{\textwidth}
    \centering
    \includegraphics[width=0.7\textwidth]{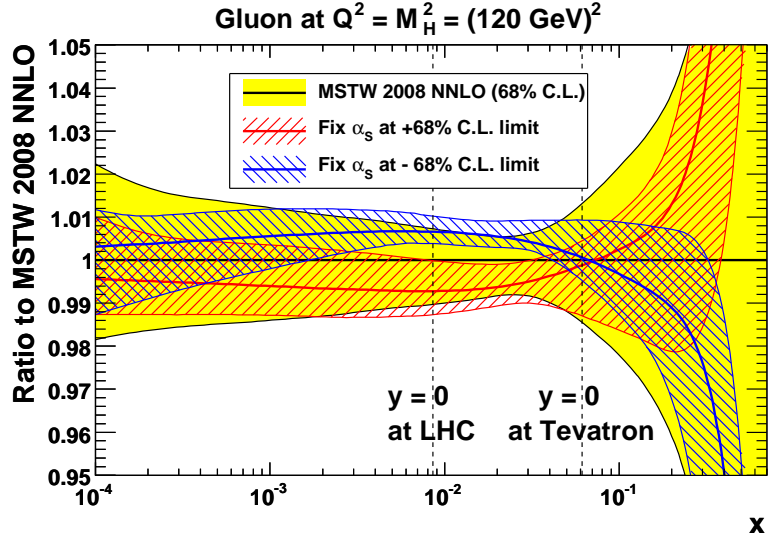}
  \end{minipage}
  \caption{(a)~PDF uncertainties on the Higgs total cross sections ($M_H=120$ GeV) for each of the five PDF sets with different fixed values of $\alpha_S(M_Z^2)$.  The results are shown as the percentage difference from the overall best-fit value.  We also indicate the $\alpha_S$ dependence of the trivial $\alpha_S^2$ factor and the $gg$ luminosity (by open squares and triangles respectively, slightly offset for clarity), again shown as the percentage difference from the best-fit value.  (b)~NNLO gluon distribution at $Q^2=M_H^2=(120\,{\rm GeV})^2$.  The values of $x=M_H/\sqrt{s}$ relevant for central production (assuming $p_T^H=0$) at the Tevatron and LHC are indicated.}
  \label{fig:htot_asmz}
\end{figure}

In Fig.~\ref{fig:htot_asmz}(a) we show a similar plot for Higgs boson production via gluon--gluon fusion with $M_H=120$ GeV.  
The Higgs cross sections are calculated using the LO subprocess $gg\to H$ via a top-quark loop (with $m_t=175$ GeV).  Higher-order corrections are included in the $m_t\to\infty$ limit up to NNLO, with only terms up to $(1-z)^1$ (where $z\equiv M_H^2/\hat{s}$) included in the expansion of the ``hard'' pieces of the NNLO partonic cross section around the kinematic point $z=1$ in powers of $(1-z)$~\cite{Harlander:2002wh} (this approximation is accurate to better than 1\%~\cite{Harlander:2002wh}).

At LO, the PDF dependence of the Higgs cross section is given simply by the effective gluon--gluon luminosity, defined as
\begin{equation} \label{eq:gglum}
  \frac{\partial {\cal L}_{gg}}{\partial M_H^2} = \frac{1}{s} \int_\tau^1\frac{{\rm d}x_1}{x_1}\;f_g(x_1,M_H^2)\;f_g(x_2=\tau/x_1,M_H^2),\quad \tau = \frac{M_H^2}{s},
\end{equation}
while the $\alpha_S$ dependence of the partonic cross section is given by an overall $\alpha_S^2$ factor.  In Fig.~\ref{fig:htot_asmz}(a) we show the percentage difference of the trivial $\alpha_S^2$ factor and the gluon--gluon luminosity, with respect to the values for the best-fit $\alpha_S$, for each of the five PDF sets.  (The percentage PDF uncertainty on the $gg$ luminosity for each of the five PDF sets, not shown in Fig.~\ref{fig:htot_asmz}(a), is very similar to the percentage PDF uncertainty on the NNLO total cross section.)  The dependence of the $gg$ luminosity on $\alpha_S$ can be understood by looking at Fig.~\ref{fig:htot_asmz}(b), where we show the relevant gluon distribution as a function of momentum fraction $x$.  The momentum fractions probed at the Tevatron and LHC at central rapidity ($y=0$), assuming LO kinematics where $p_T^H=0$ and $x_1=x_2=M_H/\sqrt{s}$, are indicated.  At the Tevatron, $y=0$ corresponds almost exactly to the crossing point of the three gluon distributions shown in Fig.~\ref{fig:htot_asmz}(b).  However, integrating over $x_1$ (or equivalently, rapidity $y$) in Eq.~\eqref{eq:gglum} leads to some contribution from larger $x$ values, leading to a slight overall correlation between the $gg$ luminosity and $\alpha_S$ at the Tevatron.  At the LHC, we see from Fig.~\ref{fig:htot_asmz}(b) that $y=0$ corresponds to the point of almost maximal anticorrelation between the gluon distribution and $\alpha_S$.  Again, there will be contributions from both larger and smaller $x$ values in the integral over $x_1$ in Eq.~\eqref{eq:gglum}, but there is still a substantial anticorrelation between the $gg$ luminosity and $\alpha_S$ shown in Fig.~\ref{fig:htot_asmz}(a).  At LO, for each fixed $\alpha_S$ value, the percentage difference of the Higgs total cross section, with respect to the value obtained with the best-fit $\alpha_S$, would be approximately given by simply adding the percentage differences of the $\alpha_S^2$ factor and the $gg$ luminosity.  However, higher-order corrections will significantly increase the $\alpha_S$ dependence of the partonic cross section beyond the LO $\alpha_S^2$ factor.  We see from Fig.~\ref{fig:htot_asmz}(a) that at the Tevatron both higher-order corrections and the correlation of the $gg$ luminosity with $\alpha_S$ increase the $\alpha_S$ dependence of the Higgs total cross section.  On the other hand, at the LHC, higher-order corrections compensate almost exactly for the anticorrelation of the $gg$ luminosity with $\alpha_S$, meaning that the $\alpha_S$ dependence of the total Higgs cross section is surprisingly almost the same as the trivial $\alpha_S^2$ factor.

In Fig.~\ref{fig:htot_asmz}(a) we also show that the ``PDF+$\alpha_S$'' uncertainty, indicated by the dashed lines, is much enhanced compared to the ``PDF only'' uncertainty indicated by the dotted lines.  Note that there is an even more marked asymmetry in the $\alpha_S$ dependence of the total uncertainty for Higgs production than for $Z$ production, with much more of an increase in the upwards direction.  This is due to the same source.  The HERA data demand more of an increase in the gluon (with tighter bands) when $\alpha_S$ decreases than a decrease in the gluon when $\alpha_S$ increases; see Fig.~\ref{fig:htot_asmz}(b).  This asymmetry would be absent to first order if the default fit to the HERA structure function data was perfect.

\begin{figure}
  (a)\\
  \begin{minipage}{\textwidth}
    \centering
    \includegraphics[width=0.8\textwidth]{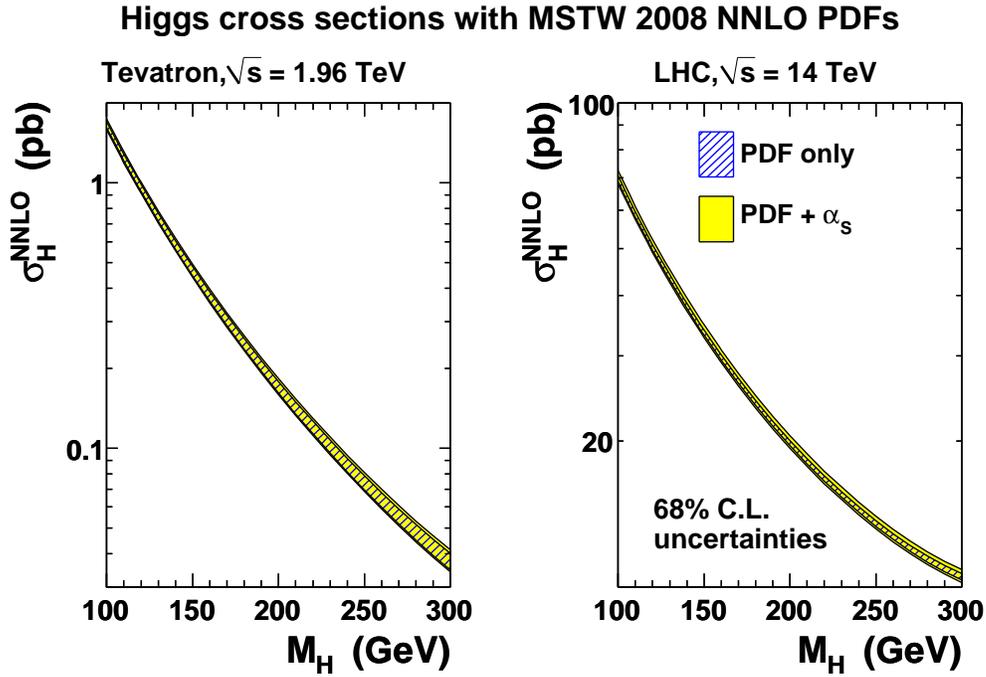}\\
  \end{minipage}
  (b)\\
  \begin{minipage}{\textwidth}
    \centering
    \includegraphics[width=0.8\textwidth]{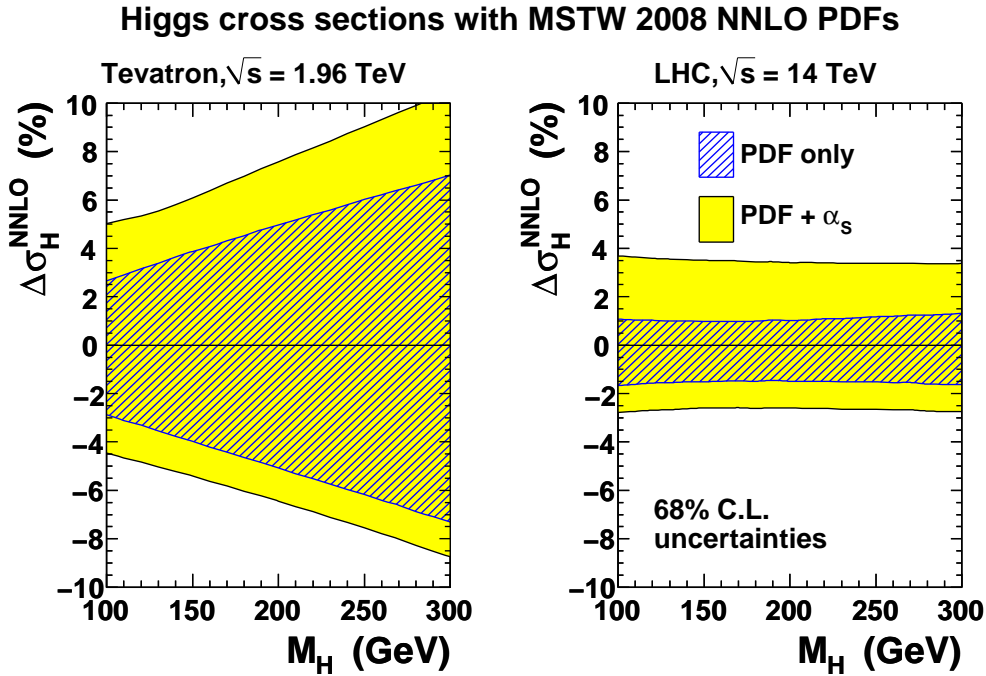}
  \end{minipage}
  \caption{(a)~Higgs total cross sections as a function of the Higgs mass at the Tevatron and LHC.  (b)~Percentage uncertainty in the Higgs total cross sections when accounting simultaneously for PDF and $\alpha_S$ uncertainties (outer error bands) as compared to that due to the PDF uncertainty alone (inner error bands).}
  \label{fig:hmass_asmz}
\end{figure}
In Fig.~\ref{fig:hmass_asmz} we show, as a function of the Higgs mass, how the uncertainty increases when variation of $\alpha_S$ is included.  Both uncertainty bands increase in size for increasing Higgs mass at the Tevatron as we become more sensitive to the less well-determined high-$x$ gluon distribution.  At the LHC, the size of the uncertainties is largely independent of the Higgs mass in the range considered here, because we are always dominated by the gluon in the region of $x\sim 10^{-2}$.  The total ``PDF+$\alpha_S$'' uncertainty can become comparable to the theory uncertainty estimated from scale variation (most recently updated in Refs.~\cite{Anastasiou:2008tj,deFlorian:2009hc}) of $\sim\pm 10\%$ for high-mass Higgs production at the Tevatron, while the uncertainty from unknown higher-order QCD corrections dominates at the LHC where the combined ``PDF+$\alpha_S$'' uncertainty is still relatively small.

\subsection{Inclusive jet production}
\begin{figure}
  (a)\\
  \begin{minipage}{\textwidth}
    \centering
    \includegraphics[width=0.8\textwidth]{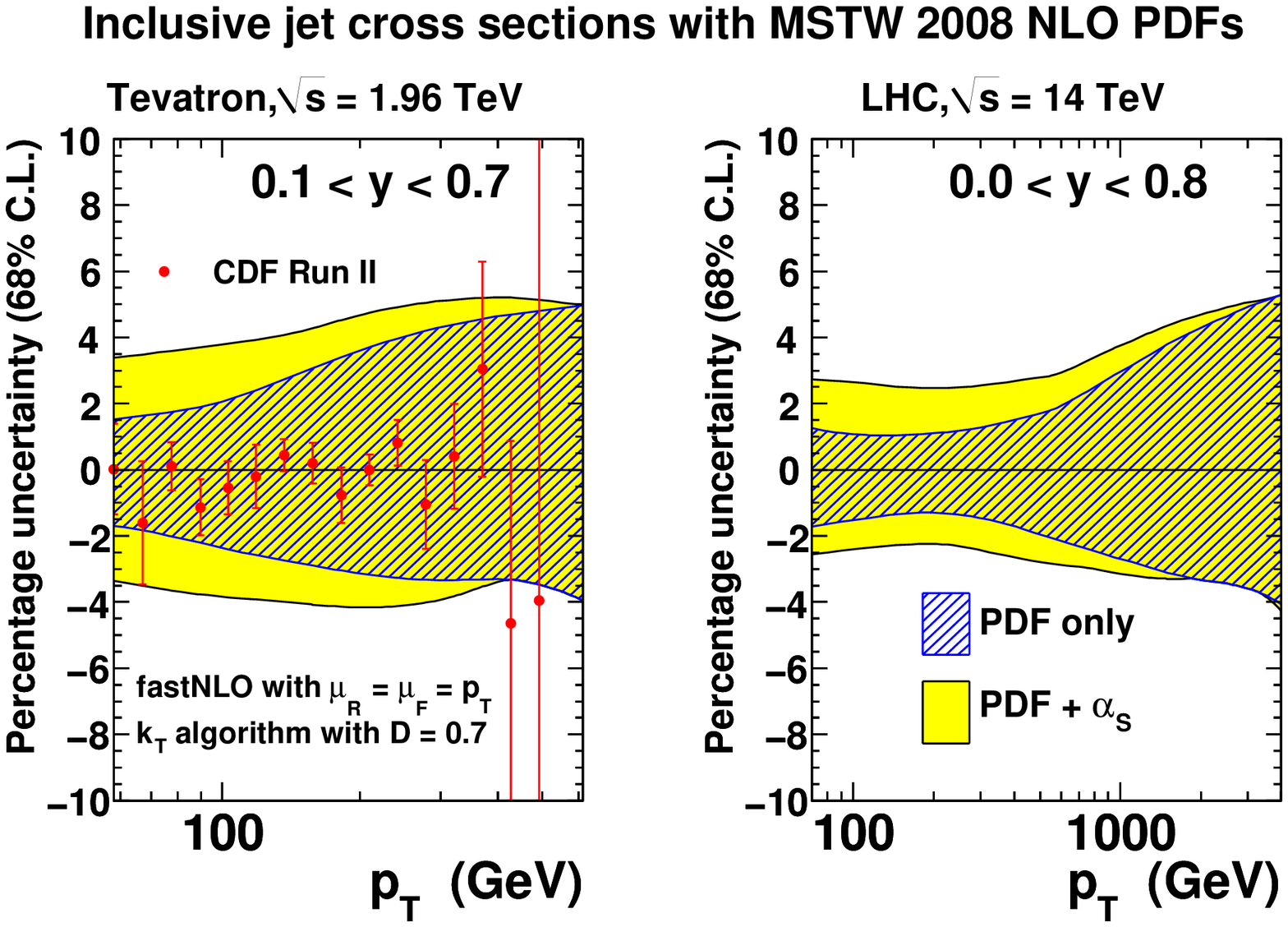}\\
  \end{minipage}
  (b)\\
  \begin{minipage}{\textwidth}
    \centering
    \includegraphics[width=0.8\textwidth]{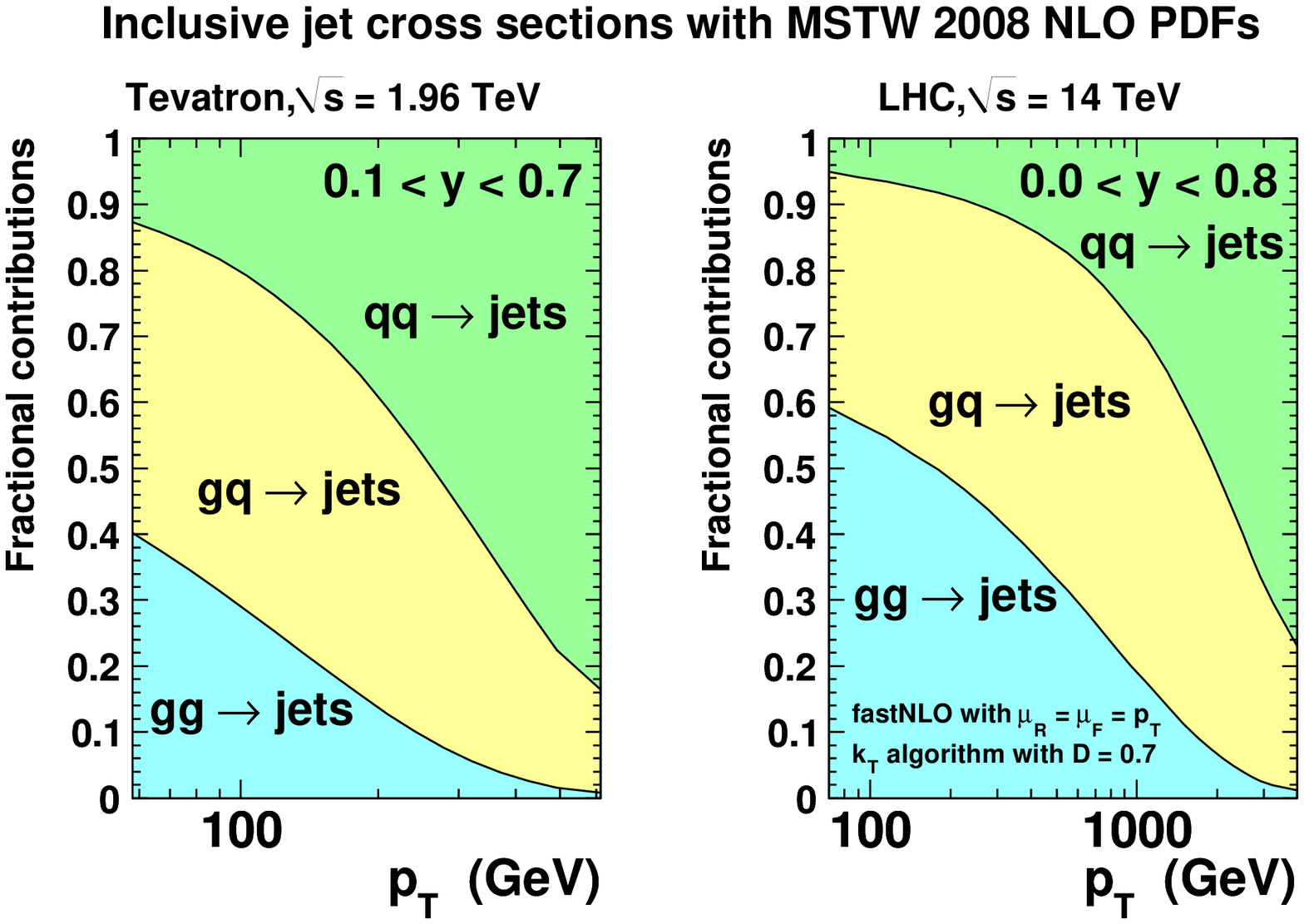}
  \end{minipage}
  \caption{(a)~``PDF+$\alpha_S$'' uncertainty compared to the ``PDF only'' uncertainty as a function of $p_T$ for inclusive jet production in a central rapidity region at the Tevatron and LHC.  At the Tevatron we also show the CDF Run II data points~\cite{Abulencia:2007ez} (statistical uncertainties only).  (b)~Fractional contributions of the $gg$-, $gq$- and $qq$-initiated processes as a function of $p_T$.}
  \label{fig:jets_asmz}
\end{figure}
In Fig.~\ref{fig:jets_asmz}(a) we show the ``PDF+$\alpha_S$'' uncertainty compared to the ``PDF only'' uncertainty as a function of $p_T$ for inclusive jet production in a central rapidity region at the Tevatron and LHC.  The inclusive jet cross sections are calculated using the \textsc{fastnlo} package~\cite{Kluge:2006xs}, based on \textsc{nlojet++}~\cite{Nagy:2001fj,Nagy:2003tz}, with the $k_T$ jet algorithm.  At the Tevatron we show the data-to-theory ratio for the CDF Run II data~\cite{Abulencia:2007ez}, where the error bars represent only the statistical uncertainties on the data, and the data points have been shifted by the correlated systematic uncertainties determined by the best-fit MSTW 2008 NLO PDFs; see Eq.~(38) and Fig.~48 of Ref.~\cite{Martin:2009iq}.  Note that the systematic uncertainties on the Tevatron data on inclusive jet production are generally very much larger than the statistical uncertainties.  The $x$ value probed in the PDFs is approximately $x_T=2p_T/\sqrt{s}$, and hence the Tevatron and LHC plots show a similar trend if comparing the same $x_T$ rather than the same $p_T$.  The fractional contributions of the $gg$-, $gq$- and $qq$-initiated processes to the cross section are shown in Fig.~\ref{fig:jets_asmz}(b).  Here, $gq$ implicitly includes $qg$ contributions, and $q$ stands for both quarks and antiquarks.  At smaller values of $p_T$ the additional uncertainty on $\alpha_S$ leads to an increase in the overall uncertainty which is due to the largely gluon-initiated cross section being correlated (or at worst uncorrelated) with $\alpha_S$.  At larger values of $p_T$, where the $x$ probed $\gtrsim 0.3$--$0.4$, the cross section is dominated by $qq$ scattering.  As seen in Section~\ref{sec:eigen}, the high-$x$ quarks are highly anticorrelated with $\alpha_S$, and hence the uncertainty on $\alpha_S$ does not lead to a further enhancement of the PDF uncertainty.  In the transition region of intermediate $p_T$, the upper ``PDF+$\alpha_S$'' uncertainty is enhanced very slightly by the presence of the PDF sets with $\alpha_S$ shifted upwards by $\sigma/2$, but in all other regions the extreme values are provided by either the best-fit $\alpha_S$ or the shifts of $\alpha_S$ by $\pm1\sigma$.  The situation is similar in other rapidity bins.  The total ``PDF+$\alpha_S$'' uncertainty is smaller than the theory uncertainty at NLO estimated from scale variation, which can be $\sim\pm 10\%$.  However, the PDF uncertainty tends to allow more variation in shape as a function of $p_T$, whereas scale variation affects mainly the normalisation (for low rapidities).

\section{Conclusions} \label{sec:conclusions}
Parton distribution functions (PDFs) must be used together with the appropriate value of $\alpha_S$, since both the input PDFs and $\alpha_S(M_Z^2)$ are determined simultaneously from global fits to deep-inelastic and related hard-scattering data within the framework of leading-twist fixed-order collinear factorisation (in the $\overline{\rm MS}$ scheme).  In a previous paper~\cite{Martin:2009iq} we determined the best-fit values of $\alpha_S(M_Z^2)$ at leading-order (LO), next-to-leading order (NLO) and next-to-next-to-leading order (NNLO).  In this paper we have determined the experimental uncertainties on $\alpha_S(M_Z^2)$, finding that at
\begin{align}
  \text{NLO:}\qquad\alpha_S(M_Z^2) &= 0.1202\quad^{+0.0012}_{-0.0015}\text{ (68\% C.L.)}\quad^{+0.0032}_{-0.0039}\text{ (90\% C.L.)},\\
  \text{NNLO:}\qquad\alpha_S(M_Z^2) &= 0.1171\quad^{+0.0014}_{-0.0014}\text{ (68\% C.L.)}\quad^{+0.0034}_{-0.0034}\text{ (90\% C.L.)}.
\end{align}
The experimental errors on $\alpha_S$ quoted here were obtained using an extension of the ``dynamic tolerance'' method~\cite{Martin:2009iq}, which gives a much refined revision on the previous estimate of $\pm0.002$ from the MRST 2001 analysis~\cite{Martin:2001es} obtained using a fixed $\Delta\chi^2_{\rm global}=20$.  We did not address in detail the important issue of the additional theory uncertainty on our $\alpha_S$ determination, but an estimate was given of $\pm0.003$ at NLO and at most $\pm0.002$ at NNLO.

In the process of determining the experimental uncertainty on $\alpha_S(M_Z^2)$ we performed global fits with different fixed values of $\alpha_S(M_Z^2)$ in steps of 0.001 for a range of 0.110--0.130 at NLO and 0.107--0.127 at NNLO.  Public grids for these PDF sets are available from Ref.~\cite{mstwpdf}.  These PDF sets will be invaluable for $\alpha_S$ determinations by other groups.

We then considered, for the first time, the correlation between PDF uncertainties and uncertainties on $\alpha_S(M_Z^2)$.  By an extension of the ``dynamic tolerance'' method~\cite{Martin:2009iq} we showed how to consistently account for the effect of both these sources of uncertainty in cross section calculations.  When $\alpha_S(M_Z^2)$ is fixed at either of its 1-$\sigma$ limits, then the tolerance for each eigenvector PDF set will be one-sided, i.e.~zero in one direction and non-zero in the other direction.  Therefore, the PDF uncertainties will be much smaller (and more asymmetric) when $\alpha_S(M_Z^2)$ is fixed at either of its 1-$\sigma$ limits than when $\alpha_S(M_Z^2)$ is fixed at its best-fit value.  We also provide ``in-between'' PDF sets with $\alpha_S(M_Z^2)$ fixed at \emph{half} the 1-$\sigma$ limits, where the size of the PDF uncertainty is generally smaller than when $\alpha_S(M_Z^2)$ is at its best-fit value, but larger than when $\alpha_S(M_Z^2)$ is at its 1-$\sigma$ limits.  The PDF uncertainty for physical observables, such as cross sections, should be calculated separately for each of the five sets (each comprising the best-fit PDF set and 40 eigenvector PDF sets) with $\alpha_S(M_Z^2)$ displaced by $\{-1\sigma,\,-\sigma/2,\,0,\,+\sigma/2,\,+1\sigma\}$ from its best-fit value.  Then the combined ``PDF+$\alpha_S$'' uncertainty is given by the envelope of these five predictions.  We also provide similar 90\% confidence-level (C.L.) PDF sets, where $\alpha_S(M_Z^2)$ is displaced to each of its 90\% C.L.~limits, and to half these limits.  Public grids for all these PDF sets, for use either with the standalone MSTW interpolation code or via the \textsc{lhapdf} interface~\cite{Whalley:2005nh} (from version 5.7.0), are available from Ref.~\cite{mstwpdf}.

As examples, we calculated the total cross sections for production of $W$, $Z$ and Higgs bosons at the Tevatron and LHC.  For $W$ and $Z$ production, where the LO subprocess is $\mathcal{O}(\alpha_S^0)$ and is quark-initiated, there is not a significant enhancement due to the combined ``PDF+$\alpha_S$'' uncertainty as compared to the PDF-only uncertainty with a fixed $\alpha_S$.  However, the additional uncertainty due to $\alpha_S$ is more important for Higgs boson production via gluon--gluon fusion, where the LO subprocess is $\mathcal{O}(\alpha_S^2)$.  This is particularly the case at the LHC, where the PDF uncertainty (for fixed $\alpha_S$) is very small.  Finally, we considered the combined ``PDF+$\alpha_S$'' uncertainty on the cross sections for inclusive jet production at the Tevatron and LHC as a function of $p_T$.  The additional uncertainty from $\alpha_S$ enhances the PDF-only uncertainty at low $p_T$, where the largely gluon-initiated cross section is correlated with $\alpha_S$, but not at high $p_T$, where the largely quark-initiated cross section is anticorrelated with $\alpha_S$.  Having discussed these few basic processes, the general implications for the ``PDF+$\alpha_S$'' uncertainty on new physics production (for example, $Z^\prime$ or supersymmetric particles) should be clear by considerations such as powers of $\alpha_S$, the relevant initiating partons and the approximate $x$ values.  For any particular process the details of the uncertainty can be explicitly calculated in a straightforward way using the tools we have provided in this paper.

\newpage
\appendix
\setcounter{equation}{0}
\renewcommand{\theequation}{A.\arabic{equation}}
\section{Appendix: Definition of $\alpha_S(Q^2)$}
There is more than one definition of the strong coupling $\alpha_S(Q^2)$ beyond LO commonly used in QCD phenomenology.  The various prescriptions are all formally equivalent since they differ only at higher orders.  The strong coupling $\alpha_S(Q^2)$ runs according to the renormalisation group equation (RGE):
\begin{equation} \label{eq:rge}
  \frac{{\rm d}}{{\rm d}\ln Q^2}\left(\frac{\alpha_S}{4\pi}\right) = -\beta_0\left(\frac{\alpha_S}{4\pi}\right)^2-\beta_1\left(\frac{\alpha_S}{4\pi}\right)^3-\beta_2\left(\frac{\alpha_S}{4\pi}\right)^4-\ldots,
\end{equation}
where the $\beta$-function coefficients up to NNLO are
\begin{equation}
  \beta_0(n_f) = 11 - \frac{2}{3}n_f, \qquad \beta_1(n_f) = 102 - \frac{38}{3}n_f, \qquad \beta^{\overline {\rm MS}}_2(n_f) = \frac{2857}{2}-\frac{5033}{18}n_f+\frac{325}{54}n_f^2.
\end{equation}
The form of $\alpha_S(Q^2)$ used in all previous MRST fits is given by
\begin{equation} \label{eq:invalphas}
  \alpha_S^{-1}(Q^2) =
  \begin{cases}
    \alpha_S^{-1}(Q^2,3)+\alpha_S^{-1}(m_c^2,4)-\alpha_S^{-1}(m_c^2,3)
    &:\quad Q^2 < m_c^2\\
    \alpha_S^{-1}(Q^2,4) &:\quad m_c^2 \le Q^2 \le m_b^2\\
    \alpha_S^{-1}(Q^2,5)+\alpha_S^{-1}(m_b^2,4)-\alpha_S^{-1}(m_b^2,5)
    &:\quad Q^2 > m_b^2
  \end{cases},
\end{equation}
where $\alpha_S(Q^2,n_f)$ is defined as the solution of the RGE, Eq.~\eqref{eq:rge}, which can be rewritten as
\begin{equation}
  \frac{{\rm d} a}{{\rm d}\ln Q^2} = -a^2-b\,a^3-c\,a^4-\ldots,
\end{equation}
where $a\equiv\beta_0(n_f)\alpha_S(Q^2,n_f)/(4\pi)$, $b\equiv \beta_1(n_f)/\beta_0^2(n_f)$ and $c\equiv \beta_2(n_f)/\beta_0^3(n_f)$.  The solution of this equation in terms of an input parameter $\Lambda$ is~\cite{Roberts:1990ww}
\begin{equation} \label{eq:deflambda}
  \ln\left(\frac{Q^2}{\Lambda^2}\right) =
  \begin{cases}
    \frac{1}{a}&:\quad\text{LO}\\
    \frac{1}{a}-b\ln\left(\frac{1}{a}+b\right)&:\quad\text{NLO}\\
    \frac{1}{a}-\frac{b^2-2c}{\sqrt{4c-b^2}}\tan^{-1}\left(\frac{b+2ac}{\sqrt{4c-b^2}}\right)-b\ln\left(\frac{\sqrt{1+ab+a^2c}}{a}\right)&:\quad\text{NNLO}
  \end{cases}.
\end{equation}
Note that $\alpha_S(Q^2)$ defined by Eq.~\eqref{eq:invalphas} is continuous across the flavour thresholds, and that the fitted parameter $\Lambda$ used in solving Eq.~\eqref{eq:deflambda} is independent of the number of active flavours $n_f$.  This is different from other common prescriptions where it is necessary to choose $\Lambda$ to be $n_f$-dependent to ensure the continuity of $\alpha_S(Q^2)$.  The MRST prescription, Eq.~\eqref{eq:invalphas}, for $\alpha_S(Q^2)$ is numerically very similar at NLO to the other two principal definitions in use, provided that the same input value $\alpha_S(M_Z^2)$ is taken~\cite{Huston:2005jm}.  However, the value of $\Lambda$ determined using one prescription for $\alpha_S(Q^2)$ should not be used in a different prescription.  For example, the MRST relation between $\Lambda$ and $\alpha_S(Q^2)$ differs from the PDG relation~\cite{Amsler:2008zzb}, which also differs from the CTEQ relation~\cite{Huston:2005jm,Pumplin:2005rh}.

By differentiating Eq.~\eqref{eq:invalphas} it can be shown that $\alpha_S(Q^2)$ at LO satisfies the RGE exactly, and at NLO there is an additional higher-order term of $\mathcal{O}(\alpha_S^4)$, which is sufficiently small to be beyond the desired accuracy.  Therefore, the MRST definition Eq.~\eqref{eq:invalphas} is formally valid at LO and NLO, but not at NNLO.  In addition, at NNLO, the strong coupling in the $\overline{\rm MS}$ scheme is discontinuous at the heavy flavour thresholds, $Q^2 = m_H^2$, where $m_H$ is the pole mass of the heavy quarks, i.e.
\begin{equation} \label{eq:asdisc}
  \alpha_S(m_H^2,n_f+1) = \alpha_S(m_H^2,n_f) + \frac{14}{3}\left(\frac{\alpha_S(m_H^2,n_f)}{4\pi}\right)^3.
\end{equation}
These discontinuities in $\alpha_S$ were not taken into account in any of the MRST NNLO fits, including the MRST 2006 NNLO analysis~\cite{Martin:2007bv} where the corresponding discontinuities in the PDFs were included for the first time.  (It is theoretically possible to modify Eq.~\eqref{eq:invalphas} to be correct up to NNLO, but only via a much more cumbersome and unattractive relationship.)  Moreover, rather than use the solution of Eq.~\eqref{eq:deflambda} directly, the MRST NNLO fits used a parameterisation (called \texttt{qwikalf}) of the solution as a degree-5 polynomial in $\sqrt{\ln(Q^2/\Lambda)}$, with the coefficients fitted.

\begin{figure}
  (a)\hspace{0.5\textwidth}(b)\\
  \includegraphics[width=0.5\textwidth]{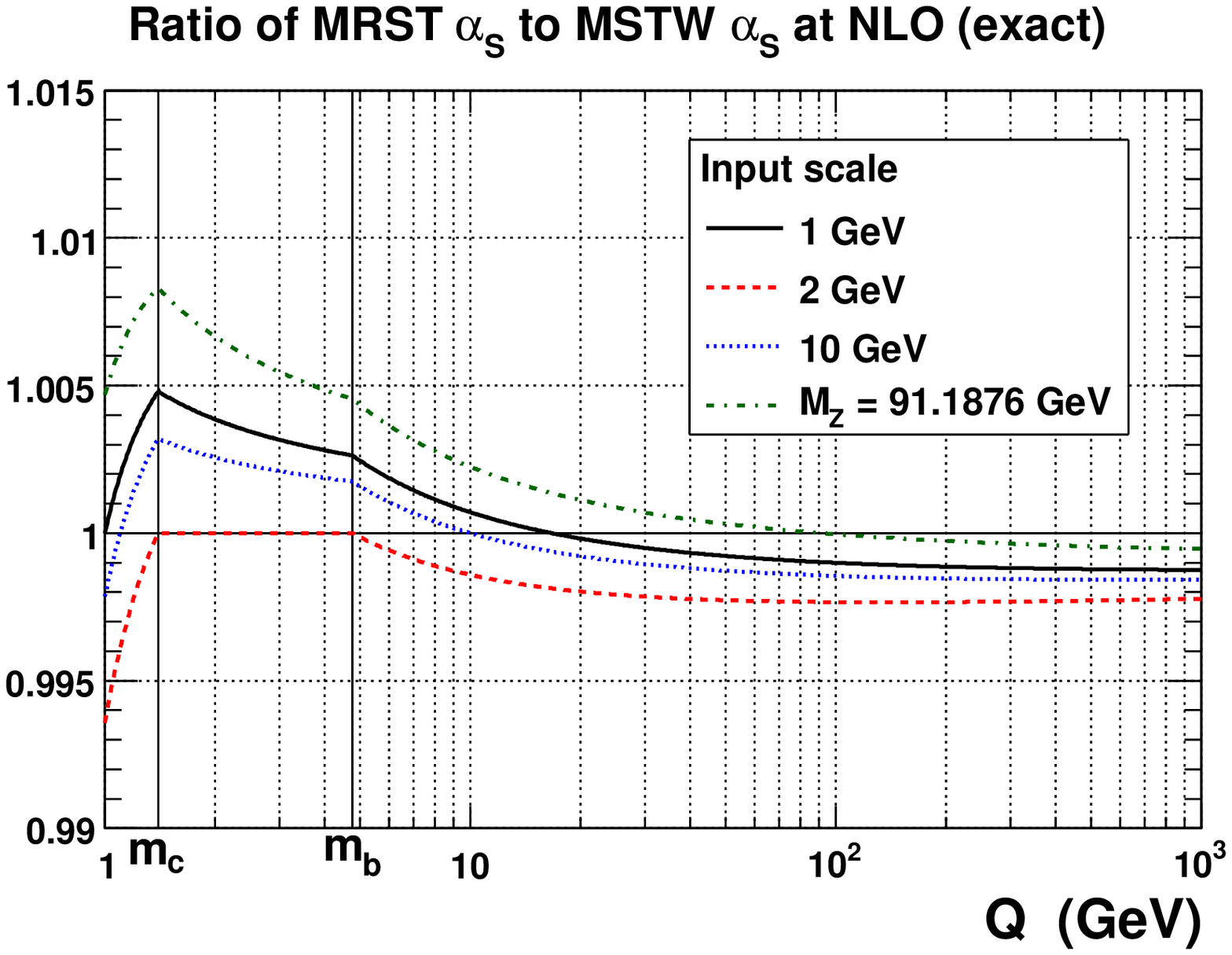}%
  \includegraphics[width=0.5\textwidth]{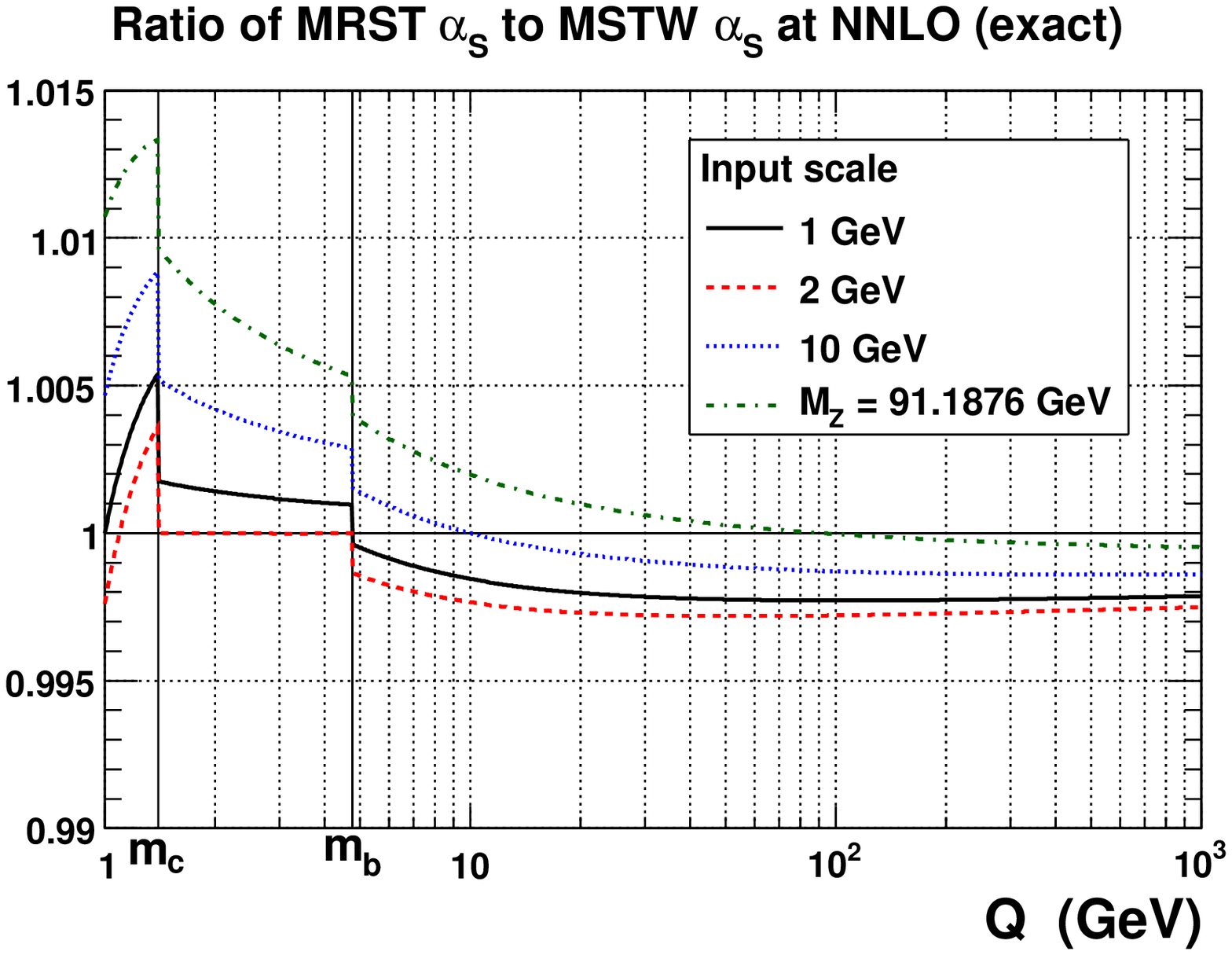}\\
  (c)\hspace{0.5\textwidth}(d)\\
  \includegraphics[width=0.5\textwidth]{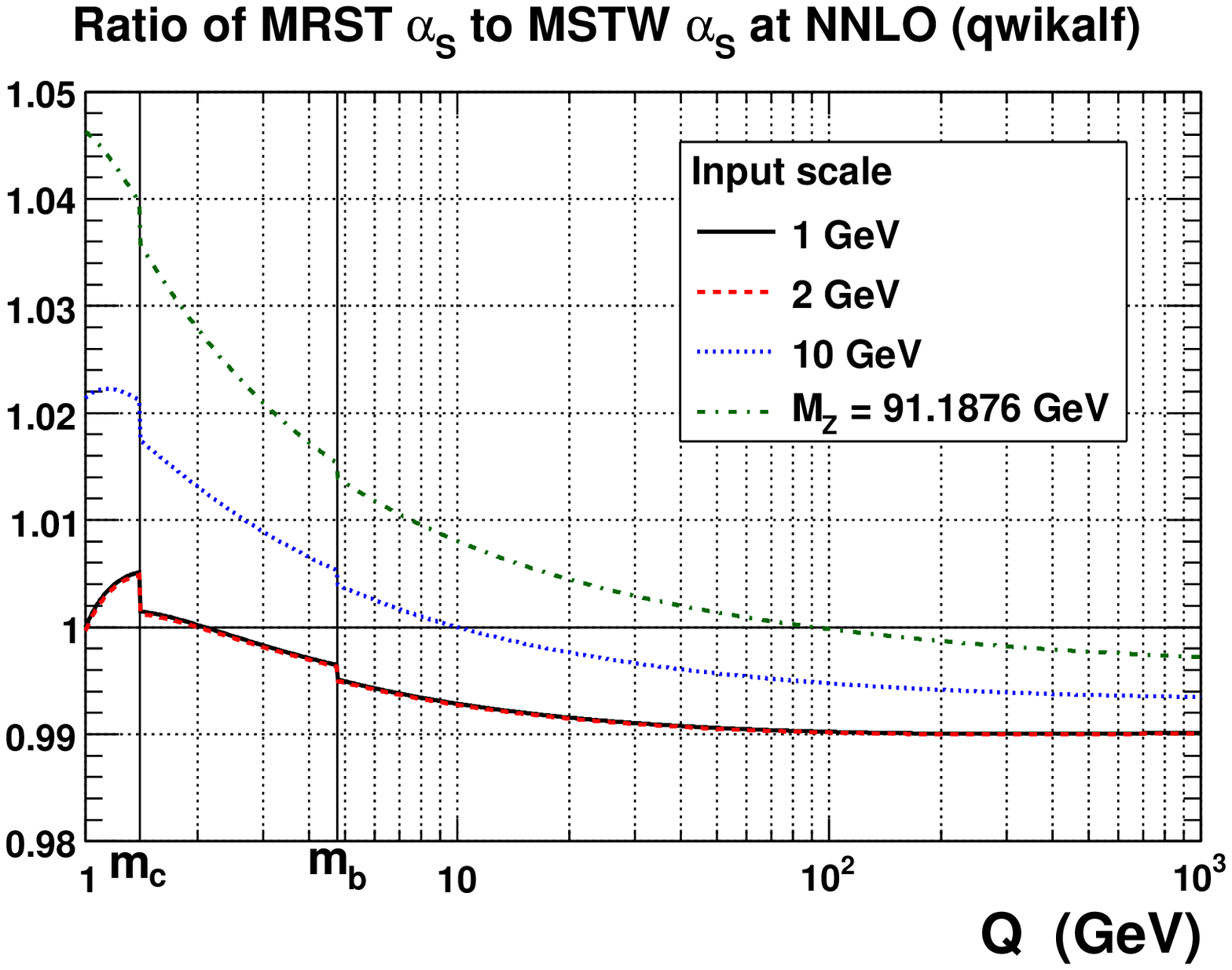}%
  \includegraphics[width=0.5\textwidth]{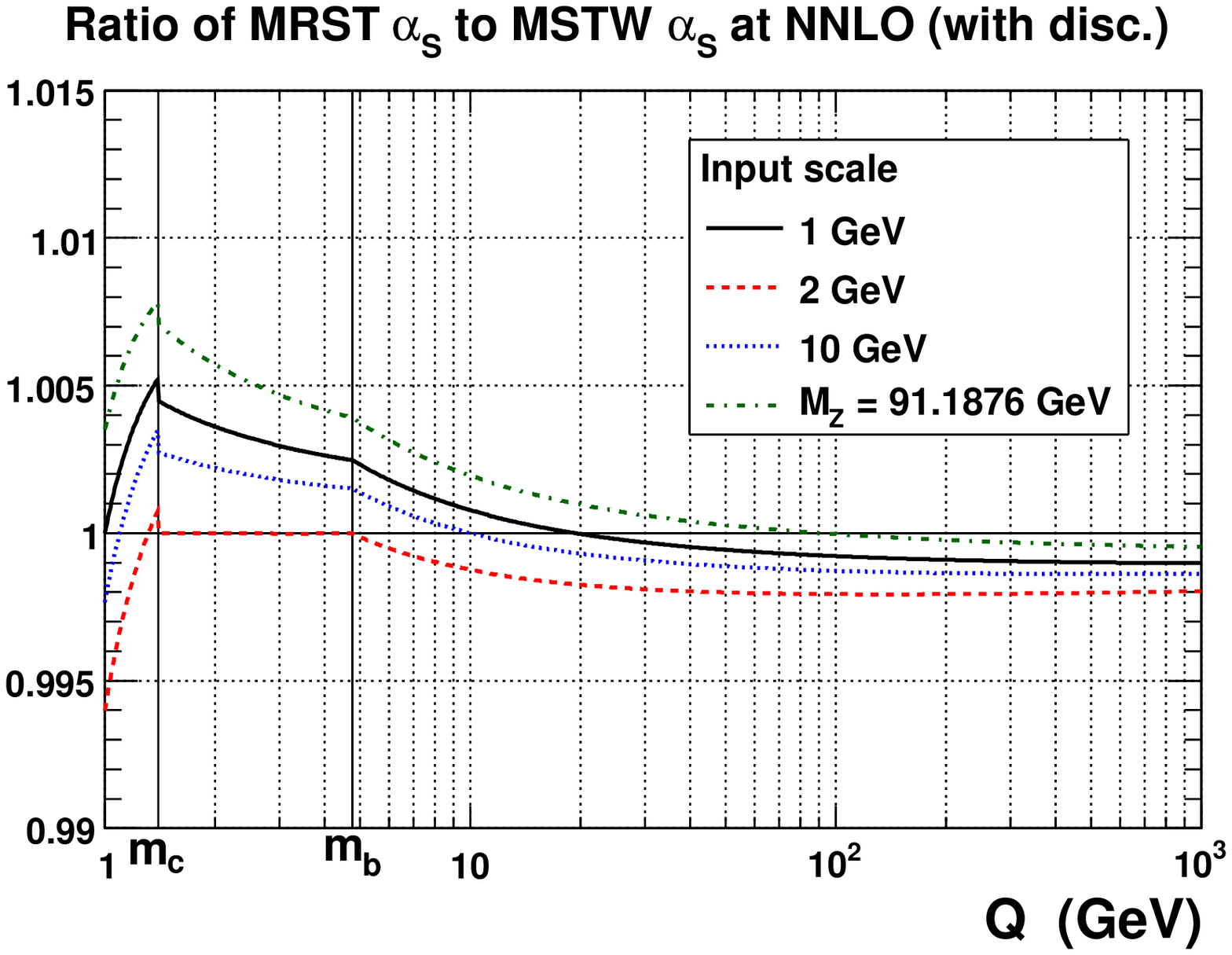}
  \caption{Ratio of the old MRST to the new MSTW $\alpha_S$ definitions when the common input is taken at different scales ($Q=1$ GeV, $2$ GeV, $10$ GeV or $M_Z$), at (a)~NLO and (b)~NNLO.  (At LO, the two definitions are identical.)  The input values are taken from the MSTW 2008 best fits at the corresponding orders.  The heavy flavour thresholds are at $m_c = 1.40$ GeV and $m_b = 4.75$ GeV.  (c)~The MRST \texttt{qwikalf} is a parameterisation of the exact result, which is clearly rather less accurate than the exact NNLO result (notice the different axis scale), but was used in all the MRST NNLO fits and is used in \textsc{lhapdf}~\cite{Whalley:2005nh} with the MRST NNLO sets.  (d)~In principle, it is possible to add discontinuities to the MRST NNLO definition, which leads to an improvement compared to the MSTW definition.}
  \label{fig:asratio}
\end{figure}
We have changed the definition of $\alpha_S$ in the recent MSTW analyses~\cite{Martin:2009iq} to match the definition used in public evolution codes such as \textsc{pegasus}~\cite{Vogt:2004ns} and \textsc{hoppet}~\cite{Salam:2008qg}, that is, we use the exact solution of Eq.~\eqref{eq:rge} with flavour matching using Eq.~\eqref{eq:asdisc} at NNLO.  (Both the MRST and MSTW definitions differ from the truncated form used by CTEQ~\cite{Huston:2005jm,Pumplin:2005rh} and also by the PDG~\cite{Amsler:2008zzb}.)  This change of $\alpha_S$ definition allowed our evolution code to be checked against the results from these two independent public evolution codes for the first time~\cite{Martin:2009iq}.  Here, for completeness, we numerically compare the MRST definition of $\alpha_S$ with the MSTW definition, taking the same input values of $\alpha_S$ at 1 GeV, 2 GeV, 10 GeV and $M_Z$.  The input values at these scales are taken from the best-fit MSTW 2008 analyses~\cite{Martin:2009iq} at the respective order, and the heavy flavour thresholds are set to $m_c=1.40$ GeV and $m_b=4.75$ GeV, evolving $\alpha_S$ with a maximum of five flavours.  In Fig.~\ref{fig:asratio} we show the ratio $\alpha_S^{\rm MRST}/\alpha_S^{\rm MSTW}$ as a function of the renormalisation scale $\mu_R=Q$, using, in turn, each of the four input scales.  Note that $\alpha_S^{\rm MRST} = \alpha_S^{\rm MSTW}$ only for four flavours if the input scale is also taken in the four-flavour region.  Otherwise, there are non-negligible differences.  The parameterisation \texttt{qwikalf} used in the MRST NNLO fits leads to sizable discrepancies compared to the exact result given by the solution of Eq.~\eqref{eq:deflambda}.  We see that if the MRST and MSTW analyses are forced to have the same value of $\alpha_S$ in the region containing the most data $(Q^2 \sim 20~{\rm GeV}^2)$ then by $Q^2=M_Z^2$ a discrepancy of more than 0.5$\%$ can occur.

\end{document}